
\magnification=\magstep1
\overfullrule=0pt

\font\HUGE=cmbx12 scaled \magstep4
\font\Huge=cmbx10 scaled \magstep4
\font\Large=cmr12 scaled \magstep3

{
\nopagenumbers
\centerline{\HUGE Universit\"at Bonn}
\vskip 5pt
\centerline{\Huge Physikalisches Institut}
\vskip 1.0cm
\centerline{\Large Fusion Algebras and Characters of}
\vskip 3pt
\centerline{\Large Rational Conformal Field Theories}
\vskip 7pt
\vskip 0.7cm
\centerline{by}
\centerline{Wolfgang \ Eholzer}
\vskip 1.0cm
\centerline{\bf Abstract}
\vskip 0.2cm
We introduce the notion of (nondegenerate) strongly-modular fusion algebras.
Here strongly-modular means that the fusion algebra is induced via Verlinde's
formula by a representation of the modular group whose kernel
contains a congruence subgroup. Furthermore, nondegenerate means that
the conformal dimensions of possibly underlying rational conformal field
theories do not differ by integers.
Our first main result is the classification of all strongly-modular
fusion algebras
of dimension two, three and four and the classification of all nondegenerate
strongly-modular fusion algebras of dimension less than~24.
\smallskip
Secondly, we show that the conformal characters
of various rational models of $\Cal W$-algebras can
be determined from the mere knowledge of the central charge
and the set of conformal
dimensions. We also describe how to actually
construct conformal characters by using theta series
associated to certain lattices.
On our way we develop several tools
for studying representations of the modular group
on spaces of modular functions. These methods,
applied here only to certain rational conformal field
theories, are in general useful for the analysis rational models.
\noindent
\vskip 1cm
\noindent
\settabs \+&  \hskip 110mm & \phantom{XXXXXXXXXXX} & \cr
\+ & Post  address:                        & hep-th/9502160  & \cr
\+ & Nu{\ss}allee 12                       & BONN-IR-95-10   & \cr
\+ & D-53115 Bonn                          & Bonn University & \cr
\+ & Germany                               & February 1995   & \cr
\+ & e-mail:                               & ISSN-0172-8733  & \cr
\+ & eholzer\@avzw01.physik.uni-bonn.de     &                 & \cr
\vfil\eject
\centerline{\HUGE Universit\"at Bonn}
\vskip 5pt
\centerline{\Huge Physikalisches Institut}
\vskip 2.0cm
\centerline{\Large Fusion Algebras and  Characters of}
\vskip 8pt
\centerline{\Large Rational Conformal Field Theories}
\vskip 11pt
\vskip 2.5cm
\centerline{von}
\centerline{W. Eholzer }
\vskip 2.0cm
\vfill
\noindent
Dieser Forschungsbericht wurde als Dissertation von
der Mathematisch-Natur\-wis\-sen\-schaft\-lich\-en Fakult\"at
der Universit\"at Bonn angenommen.
\vskip 1.0cm\noindent
\halign{\noindent # \hfill & # \hfill\cr
Angenommen am: & 9.2.1995 \cr
Referent: & Prof. Dr. W. Nahm  \cr
Korreferent: & Prof. Dr. R. Flume \cr}
\vfil\eject
}

{
\nopagenumbers
\phantom{n}
\vskip 2cm
\centerline{ \bf Meinen Eltern }
\vskip 6.5cm
\it{
\leftline{ \hskip 7cm Das gro{\ss}e Lalul$\bar a$}
\vskip 0.5cm
\leftline{ \hskip 7cm Kroklokwafzi? Se$\bar m$eme$\bar m$i! }
\leftline{ \hskip 7cm Seiokrontro - prafriplo: }
\leftline{ \hskip 7cm Bifzi, bafzi; hulale$\bar m$i: }
\leftline{ \hskip 7cm quasti basti bo ... }
\leftline{ \hskip 7cm Lalu lalu lalu lalu la! }
\vskip 0.3cm
\leftline{ \hskip 7cm  Hontraruru miromente}
\leftline{ \hskip 7cm  zasku zes r\"u r\"u?}
\leftline{ \hskip 7cm  Entepente, leiolente}
\leftline{ \hskip 7cm  klekwapufzi l\"u}
\leftline{ \hskip 7cm  Lalu lalu lalu lalu la! }
\vskip 0.3cm
\leftline{ \hskip 7cm  Simarar kos malzipempu}
\leftline{ \hskip 7cm  silzuzankunkrei (;)!}
\leftline{ \hskip 7cm  Majomar dos: Quempu Lempu}
\leftline{ \hskip 7cm  Siri Suri Sei $[]$!}
\leftline{ \hskip 7cm  Lalu lalu lalu lalu la! }
}
\vskip 1cm
\leftline{\hskip 8cm \rm Christian Morgenstern}
\vfil\eject
}


\pageno=1
\magnification=1200
\documentstyle{amsppt}
\Monograph
\language=0


\def\N{\Bbb N}                \def\Z{\Bbb Z}
\def\Q{\Bbb Q}                \def\R{\Bbb R}
\def\C{\Bbb C}
\def\F{{\Bbb F}}              \def\FF{\F{(l^2)}}

\def\SLZ{\operatorname{SL}(2,\Z)}
\def\CSLZ{{\operatorname{DSL}(2,\Z)}}
\def\SLZp{\operatorname{SL}(2,\Z_{p})}
\def\SLZpl{\operatorname{SL}(2,\Z_{p^\lambda})}
\def\Spl#1#2{\operatorname{SL}(2,\Z_{#1^{#2}})}
\def\SL{\Gamma}
\def\CSL{D\Gamma}
\def\GL#1{\operatorname{GL}(#1,\C)}
\def\GLK#1{\operatorname{GL}(#1,K)}
\def\MG#1{\operatorname{SL}(2,\Z/{#1}\Z)}

\def\Im{\operatorname{Im}}            \def\Re{\operatorname{Re}}
\def\min{\operatorname{min}}          \def\diag{\operatorname{diag}}
\def\tr{\operatorname{tr}}            \def\diag{\operatorname{diag}}
\def\ad{\operatorname{ad}}            \def\norm{\operatorname{n}}
\def\e{\operatorname{e}}
\def\LO{{\Cal O}}
\def\Hom{\operatorname{Hom}}          \def\End{\operatorname{End}}
\def\id{{\text 1 \kern-2.8pt {\text I} }}
\def\aproxeq{{\buildrel \approx \over \to }}
\def\Res{\operatorname{Res}}

\def\O{{\frak O}}		\def\o{{\frak o}}
\def\p{{\frak p}} 		

\def\PH{\frak H}		\def\w{{\Cal W}}
\def\rhot{\widetilde\rho}       \def\B{\Bbb B}
\def\QF{{\Cal Q}}               \def\P{\Bbb P}

\def\mn{\medskip\smallskip\noindent}
\def\sn{\smallskip\smallskip\noindent}
\font\Large=cmr12 scaled \magstep3

\centerline{\Large  Contents}
\mn\mn

\topmatter
\toc
\head
1. Introduction
\page{3}
\endhead

\head
2. Rational conformal field theories and fusion algebras
\page{8}
\endhead
  \subhead
  2.1 Vertex operator algebras, $\w$-algebras and rational models
\page{8}
  \endsubhead
  \subhead
  2.2 Definition of fusion algebras
\page{19}
  \endsubhead
  \subhead
  2.3 Some simple properties of modular fusion algebras
\page{22}
  \endsubhead

\head
3. On the classification of modular fusion algebras
\page{24}
\endhead
  \subhead
  3.1 Results on the classification of strongly-modular fusion algebras
\page{24}
  \endsubhead
  \subhead
  3.2 Realization of strongly-modular fusion algebras in RCFTs and data
      of certain rational models
\page{25}
  \endsubhead
  \subhead
  3.3 Some theorems on level $N$ representations of $\SLZ$
\page{28}
  \endsubhead
  \subhead
  3.4 Weil representations associated to binary quadratic forms
\page{30}
  \endsubhead
  \subhead
  3.5 The irreducible representations of $\SLZpl$ for $p\ne2$
\page{33}
  \endsubhead
  \subhead
  3.6 The irreducible representations of $\Spl{2}{\lambda}$
\page{34}
  \endsubhead
  \subhead
  3.7 Proof of the classification of the strongly-modular fusion algebras
      of dimension less than or equal to four
\page{37}
  \endsubhead
  \subhead
  3.8 Proof of a Lemma on diophantic equations
\page{45}
  \endsubhead
  \subhead
  3.9 Proof of the classification of the nondegenerate strongly-modular
      fusion algebras of dimension less than 24
\page{46}
  \endsubhead

\head
4. Uniqueness of conformal characters
\page{48}
\endhead
  \subhead
  4.1 Results on the uniqueness of conformal characters of certain
      rational models
\page{48}
  \endsubhead
  \subhead
  4.2 A dimension formula for vector valued modular forms
\page{50}
  \endsubhead
  \subhead
  4.3 Three basic lemmas on representations of $\SLZ$
\page{52}
  \endsubhead
  \subhead
  4.4 Proof of the theorem on the uniqueness of conformal characters
      of certain rational models
\page{53}
  \endsubhead

\head
5. Construction of conformal characters
\page{59}
\endhead
  \subhead
  5.1  The general construction:
       Realization of modular representations by theta series
\page{59}
  \endsubhead
  \subhead
  5.2 An example (I):
     Theta series associated to quaternion algebras and certain
     conformal characters
\page{62}
  \endsubhead
  \subhead
  5.3 An example (II): Comparison to formulas derivable from the
      representation theory of Casimir $\w$-algebras
\page{67}
  \endsubhead

\head
6. Conclusion and outlook
\page{69}
\endhead

\head
7. Appendix
\page{72}
\endhead
  \subhead
  7.1 The irreducible level $p^\lambda$ representations
      of dimension $\le 4$
\page{72}
  \endsubhead
  \subhead
  7.2 The strongly-modular fusion algebras of dimension $\le 4$
\page{76}
  \endsubhead
  \subhead
  7.3 The strongly-modular fusion algebras of dimension less than 24:
      Representations, fusion matrices and graphs
\page{78}
  \endsubhead
  \subhead
  7.4 Minimal models of Casimir $\w$-algebras
\page{81}
  \endsubhead

\head
{} References
\page{82}
\endhead
\endtoc
\endtopmatter

\vfill\eject

\topmatter
\title  List of Tables \endtitle
\endtopmatter

\toc
\head
{} Table 3.2a:  Central charges and conformal dimensions related to
             simple strongly-modular fusion algebras of dimension $\le 4$
\page{26}
\endhead
\head
{} Table 3.2b: Data of certain rational $\w$-algebras related to
            the $ADE$-classification
\page{27}
\endhead
\head
{} Table 3.2c:  Data of the rational models
\page{28}
\endhead
\head
{} Table 3.5a: Irreducible representations of $\SLZp$ for $p\not=2$
\page{33}
\endhead
\head
{} Table 3.5b: Irreducible representations of $\SLZpl$ for $p\not=2$ and
            $\lambda>1$
\page{34}
\endhead
\head
{} Table 3.6a: Irreducible representations of $\Spl{2}{}$
\page{35}
\endhead
\head
{} Table 3.6b: Irreducible representations of $\Spl{2}{2}$
\page{35}
\endhead
\head
{} Table 3.6c: Irreducible representations of $\Spl{2}{3}$
\page{35}
\endhead
\head
{} Table 3.6d: Irreducible representations of $\Spl{2}{4}$
\page{35}
\endhead
\head
{} Table 3.6e: Irreducible representations of $\Spl{2}{5}$
\page{36}
\endhead
\head
{} Table 3.6f: Irreducible representations of $\Spl{2}{\lambda}$ for $\lambda >
5$
\page{36}
\endhead
\head
{} Table 4.4: Representations of $\SL$ and weights related to certain
           rational models
\page{55}
\endhead
\head
{} Table 5.2: Certain data related to five rational models
\page{62}
\endhead
\head
{} Table 7.1a: Two dimensional irreducible level $p^\lambda$ representations
\page{72}
\endhead
\head
{} Table 7.1b: Three dimensional irreducible level $p^\lambda$  representations
\page{73}
\endhead
\head
{} Table 7.1c: Four dimensional irreducible level $p^\lambda$ representations
\page{74}
\endhead
\head
{} Table 7.2a: Two and three dimensional strongly-modular fusion algebras
\page{76}
\endhead
\head
{} Table 7.2b: Four dimensional simple strongly-modular fusion algebras
\page{77}
\endhead
\head
{} Table 7.3: Simple nondegenerate strongly-modular fusion algebras
           of dimension less than 24
\page{78}
\endhead
\head
{} Table 7.4: Values of $m_i,m_i^\vee$ for all simple Lie algebras
\page{81}
\endhead
\endtoc

\vfill\eject

\head
1. Introduction
\endhead

One of the most successful insights in modern physics is  that
symmetry is fundamental.
Perhaps, this is most apparent in gauge theories:
The standard model -formulated as a gauge field theory-
constitutes the most prominent example of a quantum field theory.
It provides a description of all known fundamental forces besides gravity.
Furthermore, some of its predictions have been checked experimentally and
are by far the most accurate ones in the history of physics.
However, there are still a lot of unsolved problems concerning the
standard model itself and, even more, concerning the unification
of the standard model with general relativity to a theory of everything.
Many of the problems of the standard model are mathematical in nature
and only a few can be treated in a mathematically rigorous way.
For example, Minkowskian theories typically formulated in the language of path
integrals seem not to make sense mathematically.
Many different lines of research have been developed
to overcome these problems. We only want to mention very shortly
two of them and their relation to conformal field theories (CFTs),
the topic of this thesis.
\mn
Firstly, algebraic quantum field theory (AQFT)
(developed by Haag, Kastler and Borchers in the fifties and sixties
(see e.g\. \cite{Ha} and references therein)) starts
at a very fundamental level and  encodes the basic features of
quantum field theories in a very clear and mathematically rigorous way.
So far, however, the treatment of `realisitic' quantum field theories
on the basis of algebraic quantum field theory is not possible.
The best understood examples of quantum field theories that can -at least
to some extend- be described within this framework are euclidean
two dimensional conformal field theories.
\mn
Secondly, string theories, originally developed to describe strong
interactions, are considered nowadays as one of the most promising candidates
for theories of everything unifying the standard model
and general relativity (for a review see e.g\. \cite{GSW}).
Although a lot of progress has been made in string theory in the last
two decades, the description of `realistic' states of matter and
something like a derivation of the standard model from string theory
are far from being solved.
Formulating field theory on the `world sheet' of the strings
gives rise to a two dimensional conformal field theory.
\mn
In this thesis we will be concerned with two dimensional chiral conformal
field theories. In the last ten years two dimensional CFTs
have played a profound role in theoretical physics as well as in
mathematics. Starting with the work of A.A. Belavin, A.M. Polyakov
and  A.B. Zamolodchikov \cite{BPZ} in 1984 it was shown that
all correlation functions of chiral rational conformal field theories (RCFTs),
i.e\. conformal field field theories depending only on one of the two light
cone coordinates and having only finitely many primary fields,
are determined by the symmetry of the theory and can
-at least in principle- be calculated.
Using conformal field theory many new results connecting statistical
mechanics and string theory with the theory of topological invariants
of 3-manifolds or with number theory were found (see e.g\ \cite{Wi,C}).
\mn\mn\mn
The classification of RCFTs became one of the important problems
in mathematical physics.
However, a complete classification seems to be an impossible task
since, for example, all self dual double even lattices lead to RCFTs
and there is at this stage no hope to classify all such lattices of
rank greater than 24. Nevertheless, it might be possible to classify
all RCFTs with `small' effective central charge $\tilde c$.
(The effective central charge is given by the difference of the central
charge and 24 times the smallest conformal dimension of the rational model
under consideration.) In particular, for $\tilde c \le 1$ a classification
of RCFTs can be obtained by using a theorem of Serre-Stark describing all
modular forms of weight $1/2$ on congruence subgroups if one assumes that
the corresponding conformal characters are modular functions on
a congruence subgroup.
\mn
With this thesis we want to contribute to the classification
program of RCFTs with only a few primary fields and for low values of
the effective central charge. Our investigations concern mainly two
different directions:
\mn
Firstly, we investigate the structure of modular fusion algebras
associated to RCFTs using the known classification of the irreducible
representations of the finite groups $\Spl{p}{\lambda}$.
For $\tilde c>1$ only partial results have been obtained so far.
One of the possibilities is to look at RCFTs where the corresponding
fusion algebra has a `small' dimension.
In the special case of a trivial fusion algebra
the RCFT has only one superselection sector and
a classification of the corresponding modular invariant
partition functions for unitary theories with $c\le24$ has
been obtained \cite{Sche}.
As a next step in the classification one can try to classify the
nontrivial fusion algebras of low dimension first and then
investigate corresponding RCFTs. Indeed, the modular fusion algebras of
dimension less than or equal to three satisfying the so-called
Fuchs conditions have been classified  (see e.g.\ \cite{MMS,CPR}).
In this thesis we develop several tools, following the ideas of
references  \cite{E$2$, E$3$}, which enable us to classify
all strongly-modular fusion algebras of dimension less than or equal to four
(for a definition of strongly-modular fusion algebras see \S2.2).
Our approach is based on the known classification of the irreducible
representations of the groups $\Spl{p}{\lambda}$ \cite{NW}.

Another possibility is to investigate theories where the corresponding
fusion algebra has a certain structure but may have arbitrary or
`big' dimension.
Here, a classification of all selfconjugate fusion algebras which
are isomorphic to a polynomial ring in one variable, where the distinguished
basis has a certain form and where the structure
constants are less than or equal to one, has been obtained (see e.g.\
\cite{CPR}\footnotemark~).\footnotetext{
More precisely, in \cite{CPR} all selfconjugate modular fusion algebras
with $N_{i j}^k\le 1$, which are isomorphic to $\Q[x]/<P(x)>$ and
$\Phi_0 \cong 1, \Phi_1 \cong x, \Phi_j \cong p_j(x)\ (j=2,\dots,n-1)$
for some polynomials $P$ and $p_j$ and where the degree of $P$ is $n$
and the degree of the $p_j$ is $j$, have been classified
(the assumption on the degree of $p_j$ was used implicitly in loc.\ cit.\ ).}
Furthermore, a classification of all fusion algebras which are isomorphic
to a polynomial ring in one variable and where the quantum dimension
of the elementary field is smaller or equal to 2 is known
(this classification contains the fusion algebras occurring
in the classification of ref. \cite{CPR};
for a review  see e.g.\ \cite{F}).
With the tools developed in this thesis we obtain another partial
classification, namely of those strongly-modular fusion algebras
of dimension less than 24 where the corresponding representation $\rho$
of the modular group is such that $\rho(T)$ has nondegenerate eigenvalues.
The nondegeneracy of the eigenvalues of $\rho(T)$ means that the difference
of any two conformal dimensions of a possibly underlying RCFT is not
an integer.
The restriction on the dimension is of purely technical nature so that
it should  be possible to obtain a complete classification of all
nondegenerate strongly-modular fusion algebras with the methods
described in this thesis by systematical use of Galois theory.

\mn
Secondly, we discuss properties of conformal characters related to rational
models which are an important tool in the study of rational models of
$\w$-algebras.
These conformal characters $\chi_h$ form a finite set of modular functions
satisfying a transformation law
$$\chi_h(A\tau)=\sum_{h'} \rho(A)_{h,h'}\chi_{h'}(\tau).$$
Here $A$ runs through the full modular group $\SL=\SLZ$ or through a
certain subgroup $G(2)$ (if the underlying $\w$-algebra is
fermionic), and $\rho$ is a matrix representation of
$\Gamma$ or $G(2)$, which depends on the rational model
under consideration.

It already has been noticed in the literature that conformal characters
are very
distinguished modular functions: First of all,
similar to the $j$-function, their Fourier coefficients are nonnegative
integers and they have no poles in the upper half plane. They
sometimes admit interesting sum formulas: These formulas, which
allow an interpretation as
generating functions of the spectrum of certain quasi-particles, can be
used to deduce dilogarithm-identities (see e.g\. \cite{NRT,KRV}).
In some cases the conformal characters have simple product expansions.
If one has both, sum and product expansions, the
resulting identities are what is known in combinatorics as Rogers-Ramanujan
type identities.

In this thesis we add one more piece to this theme.
We show that in a number of cases the
conformal characters of some RCFT are uniquely determined by
the corresponding central charge and set of conformal dimensions.
More precisely, we shall state a few  general
and simple axioms which are satisfied by the conformal characters of all
known rational models of $\w$-algebras. These axioms state essentially
not more than the $\SLZ$-invariance of the space of functions spanned by
 the conformal characters, the rationality of their Fourier coefficients
and an upper bound for the order of their poles. The only data of
the underlying rational model occurring in these axioms are the central
charge and the conformal dimensions, which give the upper bound for the
pole orders and a certain restriction on the $\SLZ$-invariance.
We then prove that, for various sets of central charges and conformal
dimensions, there is at most one set of modular  functions which satisfies
these axioms (cf\. the main theorem 3 in \S4.1).

Finally, we describe a mean which can be used to construct conformal characters
using theta series associated to certain lattices.
In particular, we shall apply our method to
the case of five special rational models. The reason for the choice of these
models is that the $\SLZ$-representations on their
conformal characters can be treated  in some generality, and that
the conformal characters of one of these models
(of type $\w(2,8)$ with central charge $c=-\frac{3164}{23}$)
could not be computed explicitly by the so far known methods.

\vfil\eject
This thesis is organized as follows: In section 2 we give a short
introduction into the theory of vertex operator algebras and present
basic (working) definitions of $\w$-algebras and RCFTs or rational models.
Furthermore, this section contains the abstract definition of fusion algebras
and some of their basic properties.
Section 3 contains two of our main results: The classification of the
strongly-modular fusion algebras of dimension less than or equal to four and
the
classification of the nondegenerate strongly-modular fusion algebras of
dimension less than 24. In the other subsections we prove our results
and comment on the realization of the fusion algebras occurring in
our classifications contained in \S3.1.
In \S4 we present and prove another main result of this thesis, namely
theorem 3 on uniqueness of conformal characters which states
that, for several rational models,  the central charge and the set of
conformal dimensions together with a set of axioms fulfilled by all known
RCFTs uniquely determine the conformal characters of the rational model
under consideration. In order to prove the main theorem 3 we develop
in \S4.2 and \S4.3 some mathematical tools  which may be of independent
interest.
In the next section, we describe how one can actually construct
conformal characters transforming under a certain congruence representation
of the modular group. After presenting a general construction procedure
we discuss concrete examples by constructing explicitly
the conformal characters of certain rational models.
Finally, we draw some conclusions and discuss open questions in~\S6.
\mn\mn\mn
Parts of this thesis have already been published:


\vfill\eject
\subhead
Notation
\endsubhead
We use $\Z_N$ for $\Z/N\Z$,
$\PH$ for the complex upper half plane,
$\tau$ as a variable in $\PH$,
$q= e^{2\pi i \tau}$,
$q^\delta=e^{2\pi i \delta\tau}$,
$T=\left(\smallmatrix 1&1\\0&1\endsmallmatrix\right)$,
$S=\left(\smallmatrix 0&-1\\1&0\endsmallmatrix\right)$,
$\SL$ for the group $\SLZ$, and
$$\Gamma(n) = \{ A \in \SLZ \ \vert \ A \equiv \id \pmod n \}$$
for the principal congruence subgroup  of $\SLZ$ of level $n$.
Recall that a congruence subgroup of $\SL$ is a subgroup containing
$\Gamma(n)$ for some $n$.
We use  $\eta$ for the Dedekind eta function
$$\eta(\tau)=\e^{\pi i\tau/12}\prod_{n\ge 1}(1-q^n).$$
The group $\SL$ acts on $\PH$ by
$$A\tau=\frac{a\tau+b}{c\tau+d}\qquad
(A=\pmatrix a&b\\c& d\endpmatrix).$$
For a complex vector valued function $F(\tau)$ on $\PH$, and for an
integer $k$  we write $F|_kA$ for the function defined by
$$(F|_kA)(\tau)=(c\tau+d)^{-k}F(A\tau).$$
Finally, for a matrix representation $\rho\colon\SL\to\GL{n}$ and
an integer $k$ we use $M_k(\rho)$ for the vector space of all holomorphic
maps $F\colon\PH\to\C^n$ (= column vectors) which satisfy $F|_kA=\rho(A)F$
for all $A\in\SL$, and which are bounded in any region $\Im(\tau)\ge r>0$.
Thus, if $\rho$ is the trivial representation, then $M_k(\rho)$ is the
space of ordinary modular forms on $\Gamma$ and of weight $k$.

\vfill\eject
\head
2. Rational conformal field theories and fusion algebras
\endhead

In order to proceed towards a classification of RCFTs one needs
precise definitions of the objects under consideration.
One attempt to formulate the axioms of RCFTs mathematically rigorous
starts with the definition of vertex operator algebras (see e.g\. \cite{FHL}).
We summarize some basic facts about vertex operator algebras,
their representations, intertwining operators and fusion algebras
in this section. In particular, we concentrate on those aspects which
are closely related to conformal field theory. We do not give all the
mathematical details but rather try to describe the basic structures
one needs for dealing with conformal field theory problems in the language
of vertex operator algebras. The results in section \S2.1 serve as
(mathematically) motivating introduction and are not really needed in
the following.

This section is organized as follows:
the three parts of section 2.1 contain the definition of vertex operator
algebras (VOAs), their representations and intertwining operators.
Furthermore, we give working definitions of $\w$-algebras and rational
models and review some basic theorems.
In \S2.2 we define various types of fusion algebras
and comment on the relation of abstract fusion algebras
to fusion algebras associated to RCFTs.
Finally, in \S2.3 we state and prove some basic lemmas on modular fusion
algebras which we need in \S3.

\subhead
2.1 Vertex operator algebras, $\w$-algebras and rational models
\endsubhead

$\w$-algebras are a special kind of vertex operator algebras.
For the reader's convenience we repeat the definition of vertex
operator algebras and their representations
(see e.g.\ \cite{FHL,FZ}) and comment on their relation to
conformal field theory.

\subhead
$\underline{ \text{Vertex algebras and vertex operator algebras}}$
\endsubhead

Let us first comment on some basic properties of conformal field theories
motivating the definition of vertex algebras below.
Conformal field theories in two space time dimensions consist of
fields $\phi(z,\bar z)$ which are parameterized by coordinates
$z$ and $\bar z$. These theories live on a cylinder with
time coordinate $t$ and space coordinate $x$ which is periodic with
period $2\pi$. The coordinates $z,\bar z$
are given by $z = e^{t+ix}$ and $\bar z = e^{t-ix}$, respectively.
The fact that conformal field theories describe massless phenomena and that
they live in  two space time dimensions allows to consider right and
left movers (i.e.\ holomorphic and antiholomorphic fields) separately
(the corresponding fields are called chiral). We will concentrate in
the following only on holomorphic fields.
Holomorphy on the cylinder implies that a
field $\phi(z)$ (corresponding to the formal power series
$Y(\phi,z)$ below) has a Laurent series expansion
$$ \phi(z) = \sum_{n\in\Z} z^{-n-1} \phi_n $$
where the `modes' $\phi_n$ are given by
$\phi_n = \Res_z\left( z^n\phi(z) \right)$.
In addition to the holomorphic chiral fields there exists the
vacuum state (denoted by $\id$) such that the map
$\phi:=\phi(0)\id \to \phi(z)$ is injective.
Translational covariance is implemented by
the generator $L_{-1}$ which acts via
$$ \left(L_{-1} \phi\right) (z) = \frac{d}{dz} \phi(z)$$
on the fields.
Furthermore, locality implies that
$$ \phi(z) \psi = \sum_{m\in\Z} z^{-m-1} \chi^m  $$
where $\chi^m$ is given by
$ \chi^m = \Res_z\left( z^m \phi(z)\psi\right)$.
Using translation invariance of the vacuum we find that
$$ R(\phi(z)\psi(w)) \id= \sum_{m\in\Z}(z-w)^{-m-1} \chi^m(w) \id  $$
where the left hand side as to be understood as the radial ordered
product of the two fields
$$  R(\phi(z)\psi(w)) := \cases \phi(z)\psi(w) &\vert z\vert > \vert w \vert \\
                               \psi(w)\phi(z) &\vert z\vert < \vert w \vert
\endcases $$
(for details see e.g.\ [Gi]).
The fields $\chi^m(w)$ are given by
$$ \chi^m(w) = \Res_{z-w}\left( (z-w)^m R(\phi(z)\psi(w))\right). $$
Using Cauchy integration one can easily calculate the $n$-th mode of
$\chi^k(w)$ and thus obtains the so-called Jacobi identity, i.e.\ the formula
in axiom (3) in the definition of a vertex algebra below.
For $k\le -1$ the field $\chi^k(w)$ is (up to normalization) the
`normal ordered product of $\left(\frac{d}{dz}\right)^{-m-1}\ \phi(z)$
and $\psi(z)$' and usually denoted by
$\frac1{(-m-1)!}N(\partial^{-m-1}\phi,\psi)(z)$ in the physical literature.
Finally, time translations are implemented by the energy generator $L_0$
which gives rise to a grading with respect to the energy. Covariance with
respect to this grading and the full conformal covariance, i.e.\ a
representation of the Virasoro algebra on the space of fields, completes
the properties of conformal field theories which motivate
the mathematically rigorous definition of vertex algebras.

\definition{Definition (Vertex algebra)}
A vertex algebra is a complex $\Z$-graded vector space
$$ V = \bigoplus_{n\in\Z} V_n$$
(an element $\phi\in V_n$ is said to be of dimension $n$),
together with a linear map
$$V \to (\End V)[[z,z^{-1}]],\qquad
  \phi\mapsto Y(\phi,z) = \sum_{n\in\Z} \phi_n\ z^{-n-1}, $$
(the elements of the image are called vertex operators), and two distinguished
elements $1 \in V_0$ (called the vacuum)
and $\omega\in V_2$ (called the Virasoro element)
satisfying the following axioms:
\roster
\item
The map $\phi \mapsto Y(\phi,z)$ is injective.
\item
For all $\phi,\psi\in V$ there exists an $n_0$ such that
$\phi_n \psi = 0$ for all $n\ge n_0$.
\item
For all $\phi,\psi\in V$ and $m,n\in\Z$ one has
$$
  (\phi_m \psi)_n = \sum_{i\ge 0} (-1)^i \binom{m}{i}
   \left( \phi_{m-i} \psi_{n+i} -
          (-1)^m \psi_{m+n-i} \phi_{i} \right). $$
(For $m<0$ the sum on the right hand side is infinite; in this case this
identity has to be read argumentwise, i.e.\ it has to be understood in the
sense that the left hand side applied to an arbitrary element of $V$ equals the
right hand side applied to the same element: Note that this makes sense since
by (2) in the
sum on the right hand side all but a finite number of terms become 0 when
evaluated at an element of $V$.)
\item
$Y(1,z) = \id_V$.
\item
Writing $Y(\omega,z) = \sum_{n\in\Z} L_{n}z^{-n-2}$ one has
$$ L_0{\vert}_{V_n} = n\,\id_{V_n}, $$
$$ Y(L_{-1}\phi,z) = \frac{d}{dz}Y(\phi,z),$$
$$ [L_m,L_n] = (m-n)L_{m+n} +
   \delta_{m+n,0}\, (m^3-m) \frac{c}{12} \id_V,$$
for all $n,m\in\Z$, $\phi\in V$, where $c$
is a complex constant (called the central charge or rank).
\endroster
\enddefinition

\remark{Remarks}
\roster
\item
For $m\ge 0$ property (3) is equivalent to
$$ [\psi_m,\phi_n] = \sum_{i\ge 0} \binom{m}{i}
                                   (\psi_i\phi)_{m+n-i}. $$
where the left hand side denotes the ordinary commutator
of endomorphisms.
\item
This commutator identity implies in particular
$[L_0,\phi_n]=(L_{-1}\phi)_{n+1}+(L_{0}\phi)_n$, hence
$[L_0,\phi_n]=(d-1-n)\phi_n$
for $\phi\in V_d$ (here we used $(L_{-1}\phi)_{n+1}=(-n-1)\phi_n$
from axiom~(5)). From this one obtains
$$\phi_nV_m\subseteq V_{m+d-n-1}.$$
\item
Although elements of negative dimension do not turn up directly
in physical applications, ghosts (fields of negative dimension)
quite often serve as an important tool in free field constructions.
The corresponding structures can e.g.\ be described in terms of
vertex algebras.
\endroster
\endremark


Symmetry algebras of conformal field theories have additional
properties motivating the
\definition{Definition (Vertex operator algebra)}
A vertex algebra is called a vertex operator  algebra (VOA) if
\roster
\item the spectrum of $L_0$ is bounded from below by $0$, and
\item the graded components $V_n$ of $V$ are finite dimensional.
\endroster
\enddefinition

Of particular interest are special elements of VOAs which
are lowest weights with respect to the $sl(2)$ or Virasoro
algebra inside the VOA.
\definition{Definition ((Quasi-)primary elements of a VOA)}
An element $\psi\in V_d$ of a VOA $V$ is called
quasi-primary of dimension $d$ if $L_1\psi = 0$, i.e.\ $\psi\in\ker(L_1)$,
and primary of dimension $d$ if $L_n\psi = 0$ for all $n>0$.
\enddefinition
\remark{Remark}
Note that for a quasi-primary element $\psi$ of dimension $d$ one has
$\psi_n\cdot 1 = 0 $ for $n \ge 0$ and
$[L_m,\psi_n]=((d-1)(m+1)-n)\psi_{n+m}$ ($m=0,\pm 1$).
For primary elements this formula holds true without any restrictions on $m$.
\endremark

All known symmetry algebras arising in conformal field theory
are generated by quasi-primary elements.
\definition{Definition (Quasi-primary generated VOA)}
A vertex operator algebra $V$ is called quasi-primary generated if
$$ V = \oplus_{n=0}^\infty (L_{-1})^n \ker(L_1). $$
\enddefinition
\remark{Remark}
All homogeneous elements of a quasi-primary generated VOA
are linear combinations of terms of the form $\psi_n \cdot 1$
for some quasi-primary $\psi$.
\endremark

In order to make closer contact with physics we need the notion
of an `invariant' bilinear form on VOAs.
\definition{Definition (Invariant bilinear form of a VOA)}
A bilinear form $(\cdot,\cdot)$ on a VOA $V$ is said to be
invariant
if it satisfies the condition:
$$ (\psi_n u, v) = (-1)^d\ \sum_{m\ge0} \frac{1}{m!} \,
    (u,(L_1^m \psi)_{-n-m-2(d-1)}v) $$
for all $u,v\in V$ and $\psi\in V_d$.
\enddefinition
\remark{Remark}
Note that for $\psi\in V_d$ quasi-primary the invariance condition reads
$$ (\psi_n u,v) = (u,\psi_{-n-2(d-1)}v). $$
Therefore, one defines for a quasi-primary element $\psi\in V_d$
$$\psi^\dagger_n := \psi_{-n-2(d-1)},$$
in particular $L_n^\dagger = L_{-n}$.
\endremark

For quasi-primary generated VOAs there always exists a `natural'
invariant bilinear form.
\proclaim{Theorem (Existence of an invariant bilinear form of a VOA \cite{L}) }
Let $V$ be a quasi-primary generated simple VOA. Then one has
\roster
\item  $\dim(V_0)=1$, and
\item  the invariant bilinear form $(\cdot,\cdot)$ on $V$ defined by:
       $$(V_n,V_m) = 0,\qquad n\not=m$$
       and
       $$(\psi_m\cdot 1, u) \cdot 1 = \psi^\dagger_m u,\qquad
          u,\psi_m\cdot 1 \in V_n\ \text{and quasi-primary}\ \psi$$
       is nondegenerate.
\endroster
\endproclaim

Of course we also need the notion of
\vfill\eject

\subhead
$\underline{\text{Modules of vertex operator algebras and intertwining
                  operators}}$
\endsubhead
\mn
\definition{Definition (Representation of a VOA)}
A representation of a VOA $V$  is a linear map
$$\rho\colon V \to (\End M)[[z,z^{-1}]],\qquad
   \phi\mapsto Y_M(\phi,z) = \sum_{n\in\Z} \rho(\phi)_n z^{-n-1}, $$
where $M$ is an
$\N$-graded complex vector space
$$ M = \bigoplus_{n\in\N} M_n,$$
such that the following axioms are satisfied:
\roster
\item
For all $\phi\in V_d$ and $m,n$ one has
$\rho(\phi)_n M_m \subset M_{m-n-1+d}.$
\item
For all $\phi\in V$ and $v\in M$ there exists an $n_0$ such that
$\rho(\phi)_n v = 0$
for all $n\ge n_0$.
\item
For all $\phi, \psi\in V$ and all $m,n\in\Z$ one has
$$ \rho(\phi_m \psi)_n = \sum_{i\ge 0} (-1)^i \binom{m}{i}
   \left( \rho(\phi)_{m-i} \rho(\psi)_{n+i} -
          (-1)^m \rho(\psi)_{m+n-i} \rho(\phi)_{i} \right), $$
where again this identity has to be read
argumentwise.
\item
$Y_M(1,z) =  \id_{M}.$
\item
Using $Y_M(\omega,z) = \sum_{n\in \Z} \rho(L)_n z^{-n-2}$, i.e.\
$\rho(L)_n=\rho(\omega)_{n+1}$ (note that this equality is not an identity
involving some special $L\in V$, but introduces only a suggestive abbreviation
for the right hand side), one has
$$ Y_M(L_{-1}\phi,z) = \frac{d}{dz}Y_M(\phi,z),$$
$$ [\rho(L)_m,\rho(L)_n] = (m-n)\rho(L)_{m+n} +
   \delta_{m+n,0}\, (m^3-m) \frac{c}{12} \id_M,$$
for all $n,m\in\Z$, $\phi\in V$,
where $c$ is the central charge of $V$.
\endroster
The representation $\rho$ is called irreducible if there is
no nontrivial subspace of $M$ which is invariant under all
$\rho(\phi)_n$.
\enddefinition

In the following we shall occasionally use the term
$V$-module $M$ instead of representation $\rho\:V\to\End(M)[[z,z^{-1}]]$.

\remark{Remarks}
\roster
\item
Note that a vertex operator algebra $V$ is a $V$-module itself via $\phi\mapsto
Y(\phi,z)$
(use remark~(2) after the definition of vertex operator algebra
for verifying axiom~(1) of a representation).
\item
A VOA is called simple if it is irreducible as a module of itself.
\item
For a given module $M$ of a VOA $V$ there exists a dual module $M'$
of $V$ given by:
$$ M' = \oplus_{n\in\N}M'_n := \oplus_{n\in\N}M^*_n $$
and
$$ <Y_{M'}(\phi,z)w',w> =
   <w',Y_{M}(e^{zL_1}(-z^{-2})^{L_0}\phi,z^{-1})w>
$$
where $<\cdot,\cdot >$ is the natural pairing between $M'$ and $M$.
Furthermore, $(Y_{M},M)$ is irreducible if and only if
$(Y_{M'},M')$ is irreducible and $(Y_{M'},M')$ is isomorphic to
$(Y_{M},M)$ \cite{FHL}.
\endroster
\endremark

As a simple consequence of the definition we have the
\proclaim{Lemma}
Let $\rho\colon V\to\End(M)[[z,z^{-1}]]$ be an irreducible representation
of VOA with $\dim(M_n)<\infty$ $(n\in\N)$. Then there exists a complex constant
$h_m$ such that $$ \rho(L)_0 {\vert}_{M_n} = (h_M+n)\,\id_{M_n}$$
for all $n\in\N$.
\endproclaim
\demo{Proof}
By axiom (1) of a vertex operator algebra representation we have that
$\rho(L)_0M_0\subseteq M_0$.
Hence, since $M_0$ is finite dimensional, there exists an eigenvector $v$ of
$\rho(L)_0$ in $M_0$. Let $h_M$ be the corresponding eigenvalue.
Since $\rho$ is irreducible the vector space $M$ is generated by the vectors
$\rho(\phi)_nv$ ($\phi\i V_d$, $d\in\N$, $n\in\Z$); for proving this, note that
the subspace spanned by the latter vectors is invariant under all
$\rho(\phi)_n$ as can be deduced from axiom~(3)). For $m\in\N$ let $M_m^\prime$
be the subspace generated by all $\rho(\phi)_nv$ with $\phi\in M_d$ and
$d-n-1=m$.
By axiom (1) we have $M_m^\prime\subseteq M_m$, and since $M$ is the sum of all
the $M_m^\prime$ we conclude $M_m^\prime=M_m$.

On the other hand, one has $[\rho(L)_0,\rho(\phi)_n]=(d-n-1)\phi_n$ for all $n$
and all $\phi\in V_d$ (similarly as in remark~(2) after the definition of
vertex operator algebras). From this we obtain
$\rho(L){\vert}_{M_n^\prime}=(h_M+n)\,\id_{M_n^\prime}$. This proves the lemma.
\qed\enddemo

The lemma suggests the following
\definition{ Definition (Character of a VOA module)}
Let $M$ be an irreducible module of the vertex
operator algebra $V$ (with respect to the representation $\rho$)
and assume that $\dim(M_n)<\infty$ ($n\in\N$).
Then the character $\chi_M$ of $M$ is
the formal power series defined by
$$ \chi_M(q) := \tr_{M}( q^{\rho(L)_0-c/24} )
           := q^{h_M-c/24}\sum_{n\in\N} \dim(M_n) q^{n}
$$
where $c$ is the central charge of $V$ and $h_M$ the
conformal dimension of $M$.
\enddefinition

The most important class of VOAs is
given by `rational' vertex operator algebras:
\definition{Definition (Rationality of a VOA)}
A vertex operator algebra $V$ is called rational if
the following axioms are satisfied:
\roster
\item   $V$ has only finitely many inequivalent irreducible representations
$M$.
\item   For all inequivalent irreducible representations $M$ one has
        $\dim(M_n)<\infty$  ($n\in\N$).
\item   Every finitely generated representation of $V$ is equivalent
        to a direct sum of finitely many irreducible representations.
\endroster
\enddefinition

Here the notions equivalence and direct sum
are to be understood in the obvious sense.
Furthermore, finitely generated means that there exists a finite dimensional
subspace $V'$  of $V$ such that the smallest vectorspace containing
$V'$ which is invariant under all $\rho(\psi)_n$ ($n\in \Z , \psi\in V'$)
equals $V$ (this should not be confused with the (different) notion of
finitely generated $\w$-algebras cf\. below).

The importance of the rational
algebras becomes clear by the following theorem:
\proclaim{ Theorem (Zhu \cite{Zh})}
Let $M_i$ ($i=0,\dots,n-1$) be a complete set of inequivalent irreducible
modules of the rational vertex operator algebra $V$.
Assume, furthermore, that Zhu's  finiteness condition
is satisfied, i.e.\
$$ \dim( V/(V)_{-2}V ) < \infty $$
where $(V)_{-2}V \subset V$ is defined by
$(V)_{-2}V := \{ \phi_{-2} \psi \vert \phi,\psi\in V\}.$
Then the conformal characters $\chi_{M_i}$ become holomorphic
functions on the upper complex half plane $\PH$ by setting
$q = e^{2\pi i\tau}$ with $\tau\in\PH$.
Furthermore, the space spanned by the conformal characters
$\chi_{M_i}$ ($i=0,\dots,n-1$) is invariant under the natural
action $(\chi(\tau),A) \mapsto \chi(A\tau)$
of the modular group $\SLZ$.
\endproclaim

Naively one would like to talk about the multiplicity of a certain
representation of a VOA in the tensor product of two VOA representations.
However, the tensor product of two
representations does in general not carry the structure of a
VOA representation. Instead, we use the notion of intertwining
operators and fusion rules.

\definition{Definition (Intertwining operator)}
An intertwining operator $\Cal I$ of three irreducible modules
$(\rho^i,M^i),(\rho^j,M^j),(\rho^k,M^k)$
of a VOA $V$ satisfying $\dim(M^\alpha_n) <\infty$ ($n\in\N; \alpha = i,j,k$)
is a linear map
$$\align
  &{\Cal I} \colon M^i \to z^{-h_i-h_j+h_k}\ \Hom(M^j,M^k)[[z,z^{-1}]], \\
  &v \mapsto I(v,z) = z^{-h_i-h_j+h_k}\  \sum_{n\in\Z} I(v)_n z^{-n-1},
\endalign
$$
such that the following axioms are satisfied:
\roster
\item
For all $v\in M^i_d$ and $m,n$ one has
$I(v)_n M^j_m \subset M^k_{m-n-1+d}.$
\item
For all $\phi\in V$, $v\in M^i$ and all $m,n\in\Z$ one has
$$ I(\rho^i(\phi)_m v)_n = \sum_{l\ge 0} (-1)^l \binom{m}{l}
   \left( \rho^k(\phi)_{m-l} I(v)_{n+l} -
          (-1)^m I(v)_{m+n-l} \rho^j(\phi)_{l} \right), $$
where again this identity has to read argumentwise.
\item
For all $v\in M^i$ one has
$ I(L_{-1}v,z) = \frac{d}{dz}I(v,z)$.
\endroster
We call $(M^i,M^j,M^k)$ the type of the intertwining operator $\Cal I$.
\enddefinition
\remark{Remark}
Note that an irreducible representation $\rho$ of a simple VOA is
an intertwining operator of type $(V,M,M)$.
\endremark

We are now able to define the fusion rule coefficients which will be
the starting point of our results on the classification of
fusion algebras in \S3.
\definition{Definition (Fusion rule coefficient)}
The fusion rule coefficient $N_{i,j}^k$ of three irreducible modules
$(\rho^i,M^i),(\rho^j,M^j),(\rho^k,M^k)$
of a VOA $V$ which satisfy $\dim(M^\alpha_n) <\infty$
($n\in\N; \alpha = i,j,k$) is the dimension of the space of the
corresponding intertwining operators.
\enddefinition
This definition can be viewed as a natural generalization
of the situation for simple Lie algebras.
In the case of simple Lie algebras the dimension of the space
of intertwing operators between three irreducible representations
gives exactly the multiplicity of the third representation in
the tensor product of the first two representations.
This also provides us with a motivation for calling property
(3) in the definition of vertex algebras and in the definition of
representation of VOAs and property (2) in the definition of
intertwining operators `Jacobi identity'.

\remark{Remark}
Let $M^i$ ($i=0,\dots,n-1$) be a complete set of inequivalent irreducible
modules of a simple rational VOA and assume that all fusion coefficients
are finite, i.e\. $N_{i,j}^k<\infty$. It is then proven -under certain
further assumptions- that (for details see \cite{HL,Hu}):
\roster
\item
The representation $\rho^0$ of the VOA acting on itself is isomorphic
to its dual representation $\rho^{0'}$.
\item
The following equalities for the fusion coefficients hold true
$$ \align
   N_{0,i}^j = \delta_{i,j},\qquad N_{i,j}^0 &= \delta_{i,j'},\qquad
   N_{i,j}^k = N_{j,i}^k,\qquad N_{i,j}^k = N_{i',j'}^{k'}, \\
   &\sum_{k=0}^n N_{i,j}^k N_{k,l}^m = \sum_{k=0}^n N_{i,k}^m N_{j,l}^k,
\endalign
 $$
where $i,j,l,m$ run from $0$ to $n-1$.
\endroster
In this case one can interpret the fusion coefficients as structure
constants of a unital associative commutative algebra, the fusion
algebra (see \S2.2 for a abstract definition of fusion algebras).
\endremark

\subhead
$\underline{\w\text{-algebras and rational models}}$
\endsubhead
\mn
One of our aims in this section is to make the notion of
$\w$-algebras and rational models mathematically precise.
Note, however, that the definitions below just collect the
basic properties of the objects called `$\w$-algebras' and
`rational models' in the physical literature. We would like to
stress therefore that our definitions can only serve as working
definitions and it might be necessary to change them in the future.
Nevertheless, we think that the definitions below will among others
help to clarify notions.

\definition{Definition ($\w$-algebra)}
A simple vertex operator algebra $V$ is called $\w$-algebra
if it is quasi-primary generated.
\enddefinition

All known $\w$-algebras which define rational models are `finitely
generated'. We make the notion of being `finitely generated' precise as
follows.
For any subspace $W\subset V$ of a $\w$-algebra $V$ denote by $U(W)$ the
smallest subspace of $V$ which is invariant under $\psi_m$
($\psi\in W; m\le -1$) and
contains $1$. A subspace $V'$ of a $\w$-algebra $V$ is called a generating
subspace if $V'\subset\ker(L_1)$ and $V= U(V')$.

\definition{Definition (Finitely generated $\w$-algebra)}
A $\w$-algebra $V$ is called finitely generated if
there exists a finite dimensional subspace $V'\subset \ker(L_1)$
generating $V$, i.e\. $V$ is the smallest vectorspace which contains $1$ and
is invariant under all $\psi_m$ ($\psi\in V'; m\le -1$).
\enddefinition

\remark{Remark}
A generating subspace $V$ is called minimally generating if
$$V_n \cap U(\oplus_{i<n}V_i) = \{ 0 \}\qquad (n\in\N).$$
For any generating subspace $V$ there obviously exists a minimal
generating subspace $\hat V$ contained in $V$.
For a generating subspace $V$ with
$V = \oplus_{i=1}^n V_{d_i}$ and $\dim(V_{d_i}) = k_i$
define the type of $V$ by
$(d_1^{k_1},\dots,d_n^{k_n})$.
Furthermore, define an ordering on the type of generating subspaces by:
\smallskip\noindent
$(d_1^{k_1},\dots,d_n^{k_n}) < ({d'}_1^{{k'}_1},\dots,{d'}_{n'}^{{k'}_{n'}})$
if
\roster
\item $d_i^{k_i} = {d'}_i^{{k'}_i}$ for $i < i_0 \le \min(n,n')$
      and either $d_{i_0} < d'_{i_0}$ or $d_{i_0} = {d'}_{i_0}$ and
      $k_{i_0} < {k'}_{i_0}$ or
\item $n<n'$ and $d_i^{k_i} = {d'}_i^{{k'}_i}$ for $i=1,\dots,n$.
\endroster
\endremark

In the physical literature the type of $\w$-algebras is used
frequently:
\definition{Definition (Type of a $\w$-algebra)}
A finitely generated $\w$-algebra $V$ is said to be of type
$\w(d_1^{k_1},\dots,d_n^{k_n})$
if the minimum (with respect to the order above)
of the type of all generating subspaces of $V$
is given by $(d_1^{k_1},\dots,d_n^{k_n})$ .
\enddefinition

\remark{Remarks}
\roster
\item
Examples of $\w$-algebras can be constructed directly from the Virasoro
and Kac-Moody algebras. They are of type $\w(1^n)$,
respectively $\w(2)$ for the Virasoro algebra \cite{FZ}.
\item
Starting from a Kac-Moody algebra associated to a simple Lie algebra
${\Cal K}$ one can construct a 1-parameter family $\w\Cal K$ of
$\w$-algebras, the parameter being the central charge
(see e.g.\ \cite{BS}) (Note that this construction
is different from the one mentioned in (1)).
For all but a finite number of central charges these $\w$-algebras
are of type $\w(d_1,\dots,d_n)$ where  $n$ is the rank of $\Cal K$ and
the $d_i$ ($i=1,\dots,n$) are the orders of the Casimir
operators of $\Cal K$.
The remaining ones, called truncated, are of type
$\w(d_{i_1},\dots,d_{i_r})$  where the $d_{i_k}$ form a proper
subfamily of the $d_i$ above.
Note that the $\w$-algebras constructed from the Virasoro algebra
mentioned in (1) are exactly the  Casimir $\w$-algebras
associated to ${\Cal A}_1$.
The rational models of Casimir $\w$-algebras
(sometimes called minimal models) have been determined,
assuming certain conjectures, in \cite{FKW} (some corresponding data
can be also be found in Appendix 7.4).
\item
In the physical literature one refers to a finitely generated
$\w$-algebra by giving their type (although this does not specify the
$\w$-algebra uniquely in many cases).
\endroster
\endremark

One can try to construct finitely generated $\w$-algebras directly
from their axioms (see e.g\. \cite{KW,BFKNRV}).
In this direct construction of $\w$-algebras one starts with
a finite number of `simple' elements which are defined by
\definition{Definition (Simple elements of a $\w$-algebra)}
A quasi-primary element $\phi$ of a simple $\w$-algebra is called
simple if
$$(\phi, \psi_m\chi_{-1} \cdot 1) = 0 \qquad (m\le-1;\psi,\chi\in\ker(L_1)) $$
where $(\cdot,\cdot)$ is the invariant bilinear form whose
existence is guaranteed by one of the theorems given above.
\enddefinition

One of the main ingredients in the direct construction
of $\w$-algebras is the following commutator formula
(see e.g.\ \cite{Na,FRT}):
\proclaim{Theorem (Commutator formula for $\w$-algebras) }
Let $\phi$ and $\psi$ be two quasi-primary elements of a $\w$-algebra $V$
of dimension $d,d'\ge 1$, respectively. Then there exist
quasi-primary elements $\chi^{d''}\in V_{d''}$ ($0\le d'' < d+d'$) such that
$$ [\phi_{n},\psi_{m}] = \sum_{d''=0}^{d+d'-1} p(d,d',d'',m,n) \,
                                           \chi^{d''}_{d''+1-d-d'+m+n}
$$
where the $p(d,d',d'',m,n)$ are universal polynomials given by
$$ \align
   &p(d,d',d'',m,n) = \\
   &\sum_{r+s=d+d'-1-d''} (-1)^r \ r!\ s!\
     \binom{D-r}{s} \binom{D'-s}{r} \binom{D-m}{r} \binom{D'-n}{s}
   \endalign
$$
and $D=2(d-1)$ and $D' = 2(d'-1)$.
\endproclaim
\demo{Proof}
For the proof we need the simple
\proclaim{Lemma}
For a quasi-primary element $\psi\in V_d$ and integers
$0\le a\le b$ one has
$$ \left(L_{-1}^a\psi\right)_b = (-1)^{a+1} a! \binom{b}{a} \psi_{b-a} $$
and
$$ [L_1^a,L_{-1}^b] \psi = (a!)^2\binom{b}{a}
                           \binom{2d-1+b}{a}L_{-1}^{b-a}\psi.$$
\endproclaim
\demo{Proof} We leave the easy calculation to the reader.
\enddemo
With remark (1) after the definition of vertex algebras we have
$$ [\phi_m,\psi_n] = \sum_{i\ge 0} \binom{m}{i}  (\phi_i\psi)_{m+n-i}. $$
Since $V$ is quasi-primary generated we know that there exist
quasi-primary  elements $\chi^{d''}\in V_{d''}$ such that
$$ \phi_{0} \psi = \sum_{d''=0}^{d+d'-1} L_{-1}^{d+d'-1-d''} \, \chi^{d''}.$$
Assume w.l.o.g.\ that $d\ge d'$.
Using that $\phi$ is quasi-primary and applying the Lemma above we find
$$\align
   \phi_{i}\psi &= \frac{1}{\prod_{n=0}^i (2d-2-n)}
                   \left( \ad(L_1)^i \phi_{0} \right) \psi
                 = \left( i! \binom{2d-2}{i} \right)^{-1}
                         L_1^i \phi_{0}  \psi \\
                &= \left( i! \binom{2d-2}{i} \right)^{-1}
                   \sum_{d''} [L_{1}^i,L_{-1}^{d+d'-1-d''}] \chi^{d''} \\
                &= i! \binom{2d-2}{i}^{-1}
                   \sum_{d''} \binom{d+d'-1-d''}{i}
                              \binom{d+d'+d''-2}{i}
                        L_{-1}^{d+d'-1-d''-i} \chi^{d''}.
\endalign
$$
Since
$$ \align
    &\left( L_{-1}^{d+d'-1-d''-i} \chi^{d''}\right)_{m+n-i} = \\
   &(-1)^{d+d'-1-d''-i+1} (d+d'-1-d''-i)!
             \binom{m+n-i}{d+d'-1-d''-i}
        \chi^{d''}_{m+n+d'+1-d-d''}
\endalign
$$
we find
$$ [\phi_{n},\psi_{m}] = \sum_{d''=0}^{d+d'-1} p(d,d',d'',m,n) \,
                                           \chi^{d''}_{m+n+d'+1-d-d''}
$$
where
$$ \align
   &p(d,d',d'',m,n) =   (-1)^{d+d'-d''} (d+d'-1-d'')! \ \cdot \\
                   &\sum_{i=0}^{d+d'-1-d''}
                               \binom{m}{i}
                               \binom{2d-2}{i}^{-1}
                               \binom{d+d'+d''-2}{i}
                               \binom{-m-n+d+d'-d''-2}{d+d'-1-d''-i}
\endalign
$$
(here we have used $\binom{x}{r} = (-1)^r \binom{r-x-1}{r}$).
Finally, note that this polynomial equals (up to a constant factor
depending on $d,d'$ and $d''$)
$$ \sum_{r+s=d+d'-1-d''} (-1)^r \ r!\ s!\
    \binom{D-r}{s} \binom{D'-s}{r} \binom{D-m}{r} \binom{D'-n}{s}
$$
where $D=2(d-1)$ and $D' = 2(d'-1)$ as one can e.g.\ see by comparing
the zeros of the two polynomials. \qed
\enddemo
\remark{Remark}
Note that polynomials similar to the polynomials above occur in the theory of
modular forms (cf.\ \cite{BEH$^3$} and \cite{Za$2$}).
\endremark
\mn
We make the notion of `rational models' precise.
\definition{Definition (Rational model)}
A rational model (or rational model of a $\w$-algebra) is a rational
$\w$-algebra $V$ which satisfies Zhu's finiteness condition.
The {\it effective central charge} of a rational model is
defined by
$${\tilde c} = c - 24 \min \{ h_{M_i} \}$$
where $M_i$ runs through a complete set of inequivalent irreducible
representations of $V$.
\enddefinition
\remark{Remarks}
\roster
\item
In the literature rational models are frequently called
rational conformal field theories (RCFTs) and we will also do so.
\item
Examples of rational models are given by
certain vertex operator algebras constructed from Kac-Moody
algebras \cite{FZ} or the Virasoro algebra \cite{Wa}
(for more details see also below).
\item
One can show that the effective central charge of a rational
model with a minimal generating subspace of dimension $n$ lies
in the range \cite{EFH$^2$NV}
$$ 0 \le {\tilde c}  < n.$$
\item
Historically the term `rational models' was used in the physical
literature \cite{BPZ} for field theories in which the operator
product expansion of any two local quantum fields decomposes into
finitely many conformal families from a finte set.
\endroster
\endremark

The following theorem justifies the terminology `rational models':

\proclaim{Theorem (\cite{AM}) }
Assume that the representation of the modular group acting on
the space spanned by the conformal characters of a rational
model is unitary.
Then the central charge and the conformal dimensions of the
rational model are rational numbers.
\endproclaim

\subhead
2.2 Definition of fusion algebras
\endsubhead

Consider a rational model consisting of a $\w$-algebra $V$
and its (finitely many) inequivalent irreducible modules $M_i$
($i=0,\dots,n-1$). Here $M_0$ denotes the vacuum representation,
i.e\. the representation of $V$ acting on itself.
Recall, that for a module $M$ of $V$ there is the notion of the dual
(or adjoint or conjugate) module $M'$ and that one has
$(M')' \cong M$. Since $V$ is rational the
conjugation defines a permutation $\pi$ of order two of the
irreducible modules  $M'_i \cong M_{\pi(i)}$.

The structure constants $N_{i,j}^k$ of the `fusion algebra' associated
to $V$ are given by the dimension of the corresponding space of intertwining
operators of three modules (cf\. \S2.1).
{}From now on we will assume that the
fusion coefficients related to the rational models under consideration are
always finite.

One of the important properties of the $N_{i,j}^k$ which is well
known in the physical literature is the fact that the numbers $N_{i,j}^k$
can be viewed  as the structure constants of an associative
commutative algebra, the fusion algebra. In the terminology of vertex
operator algebras a corresponding statement is proven under certain
assumptions in a recent series of papers \cite{Hu}
(see also the remark below the definition of fusion coefficients in \S2.1).
In the abstract definition of fusion algebras the properties
of all known examples associated to RCFTs are collected.

\definition{Definition (Fusion algebra)}
A {\bf fusion algebra} $\Cal F$ is a finite dimensional algebra over $\Q$
with a distinguished basis $\Phi_0=\id,\dots,\Phi_{n-1}$
($n=\dim(\Cal F )$) satisfying the following axioms:
\roster
\item
 $\Cal F$ is associative and commutative.
\item
 The structure constants $N_{i,j}^k$  ($i,j,k=0,\dots,n-1$)
 with respect to the distinguished basis $\Phi_i$ are nonnegative integers.
\item
 There exists a permutation $\pi\in S_n$ of order two such that for
 the structure constants in (2)
 one has
 $$ N_{i,j}^0 = \delta_{i,\pi(j)},\qquad
    N_{\pi(i),\pi(j)}^{\pi(k)} = N_{i,j}^k,\qquad  i,j,k=0,\dots,n-1.$$
\endroster
\enddefinition
\remark{Remarks}
\roster
\item
An isomorphism $\phi$ of two fusion algebras ${\Cal F},{\Cal F}'$
is an isomorphism of unital algebras
which maps the distinguished basis to the distinguished basis, i.e.\
there exists a permutation $\sigma\in S_n$ such that
$\phi(\Phi_i) = \Phi'_{\sigma(i)}$ ($i=0,\dots, n-1$).
\item
The tensor product of two fusion algebras $\Cal F$ and $\Cal F'$ is again
a fusion algebra, its distinguished basis is given by
$\Phi_{i_1}\otimes\Phi'_{i_2}$
($i_1=0,\dots,\dim(F)-1,\ i_2=0,\dots,\dim(F')-1$).
\item
The permutation $\pi$ of order two is called charge conjugation.
Fusion algebras with trivial charge conjugation are called selfconjugate.
\item
Note that it is an open question whether two nonisomorphic fusion
algebras can be isomorphic as unital algebras.
\endroster
\endremark

It is known that fusion algebras arising from
RCFTs have additional properties believed to be generic.
One of these additional
properties is their relation to conformal characters.
Recall, that one can show for rational vertex operator algebras
satisfying Zhu's finiteness condition \cite{Zh} that the conformal characters
defined in \S2.1 become holomorphic functions in the upper complex half
plane by setting $q = e^{2\pi i \tau}$. Furthermore, for these RVOAs the
space spanned by the finitely many conformal characters  is invariant
under the action of the modular group $\SL = \SLZ$
(note that it is conjectured that Zhu's finiteness condition is not
a necessary assumption for rational VOAs).
It was conjectured in 1988 by E.\ Verlinde \cite{Ve} that
for any rational model there exists a
representation  $\rho:\SL\to\GL{n}$ of $\SL$ such that
$$\align
  \chi_i(A\tau) &= (\chi_i|A)(\tau)
               = \sum_{m=0}^{n-1} \rho(A)_{j,i} \chi_j(\tau)
  \qquad A \in \SL \\
  N_{i,j}^0 &=  \rho(S^2)_{i,j} \\
  N_{i,j}^k &= \sum_{m=0}^{n-1}
               \frac{\rho(S)_{i,m}\rho(S)_{j,m}\rho(S^{-1})_{m,k}}
                    {\rho(S)_{0,m}}. \\
\endalign$$
We will refer to this formula as `Verlinde's formula' in the
following. The above conjecture motivates the definition of modular
fusion algebras.

\definition{Definition (Modular fusion algebra)}
A {\bf modular fusion algebra} $({\Cal F},\rho)$ is a fusion algebra
$\Cal F$ together with a unitary representation $\rho:\SLZ\to\GL{n}$
satisfying the following additional axioms:
\roster
\item
 $\rho(S)$ is a symmetric and $\rho(T)$ is a diagonal matrix.
\item
 $N_{i,j}^0 = \rho(S^2)_{i,j}$,
\item
 $N_{i,j}^k = \sum_{m=0}^{n-1}
               \frac{\rho(S)_{i,m}\rho(S)_{j,m}\rho(S^{-1})_{m,k}}
                    {\rho(S)_{0,m}}$
\endroster
where $N_{i,j}^k$ ($i,j,k=0,\dots,n-1$) are the structure constants of $\Cal F$
with respect to the distinguished basis.
\enddefinition
\remark{Remarks}
\roster
\item
Note that property (3) already implies that $\Cal F$ is associative
and commutative.
\item
Two modular fusion algebras $({\Cal F},\rho)$ and $({\Cal F'},\rho')$ are
called isomorphic
if: 1) $\Cal F$ and $\Cal F'$ are isomorphic as fusion algebras,
    2) $\rho$ and $\rho'$ are equivalent,
    3) $\rho(T)_{i,j} = \rho'(T)_{\sigma(i) \sigma(j)}$ where
       $\sigma\in S_n$ is the permutation defined by the isomorphism
       of the fusion algebras.
\item
The tensor product of two modular fusion algebras
$({\Cal F},\rho),({\Cal F'},\rho')$ is defined by
$({\Cal F}\otimes {\Cal F'},\rho\otimes\rho')$ and is
again a  modular fusion algebra.
\item
A (modular) fusion algebra is called composite if it is isomorphic to
a tensor product of two nontrivial (modular) fusion algebras.
Here a (modular) fusion algebra is called trivial if it is one dimensional. A
noncomposite (modular) fusion algebra is also
called simple.
\item
Note that for a modular fusion algebra with trivial charge conjugation
($\rho(S^2) = \id$) the matrix $\rho(S)$ is real.
\item
For modular fusion algebras associated to rational models
the eigenvalues of $\rho(T)$ are given by
the conformal dimensions $h_i$ ($i=0,\dots,n-1$) of the irreducible
modules $M_i$ ($h_i$ is the smallest $L_0$ eigenvalue in the module
$M_i$) and the central charge $c$ of the theory:
$$\rho(T) = \diag(e^{2\pi i (h_0-c/24)},\dots,e^{2\pi i (h_{n-1}-c/24})).$$
\item
Quite often nonisomorphic modular fusion algebras are isomorphic as
fusion algebras.
\endroster
\endremark

In the later sections we will investigate which representations of
$\SL$ are related to modular fusion algebras.

\definition{Definition (Admissible representation of $\SLZ$)}
A representation of the modular group $\rho:\SLZ\to\GL{n}$  is called
{\bf conformally admissible} or simply {\bf admissible} if there
exists a fusion algebra $\Cal F$ such that $({\Cal F},\rho)$ is a
modular fusion algebra.
\enddefinition

It is known that modular fusion algebras associated to rational
models have many additional properties.
In particular, the central charge and the conformal dimensions
are rational \cite{Va,AM}.
Furthermore, there exist certain compatibility conditions between
the central charge $c$, the
conformal dimensions $h_i$ and the fusion coefficients $N_{i j}^k$
(the so-called Fuchs conditions)
(see e.g.\ \cite{MMS},\cite{CPR}\footnotemark):

\footnotetext{ Note that the formula connecting the central
 charge with the conformal dimension in \cite{CPR}
 contains a misprint.}

$$ \align
   & \frac{n(n-1)}{12}
     -\sum_{m=0}^{n-1} \left( h_i- \frac{c}{24} \right)
     \in \frac16 (\N \backslash \{1\}), \\
    &\sum_{m=0}^{n-1}
    \left(
    (h_i+h_j+h_k+h_l) N_{i,j}^m N_{k,m}^l
    -h_m (N_{i,j}^m N_{k,m}^l + N_{i,k}^m N_{j,m}^l + N_{i,l}^m N_{k,j}^m)
    \right) \\
   &-\frac12 \left( \sum_{m=0}^{n-1} N_{i,j}^m N_{k,m}^l \right)
           \left( 1- \sum_{m=0}^{n-1} N_{i,j}^m N_{k,m}^l \right) \in \N
\endalign$$

In many contexts the so-called quantum dimensions of the irreducible
representations of the symmetry algebra are of particular interest.
\definition{Definition (Quantum dimension)} Let $V$ be a rational model.
Then the real number
$$ \delta_{M_i} := \lim_{\tau\to i\infty}
                           \frac{\chi_i(\tau)}{\chi_0(\tau)}$$
associated to an irreducible representation $M_i$ of $V$ is called
the quantum dimension of $M_i$.
\enddefinition
Of course, the quantum dimensions are nonnegative.
If there exists a unique irreducible representation $M_\lambda$
with minimal conformal dimension $h_\lambda$ then the quantum dimensions
are given by
$$ \delta_{M_i}  = \frac{ \rho(S)_{i,\lambda}}{\rho(S)_{0,\lambda}}.$$

In the rest of this thesis we will extensively rely on the observation that
in all known examples of RCFTs the conformal characters are modular
functions on some congruence subgroup of $\SL$. Therefore, the
corresponding representation $\rho$ factors through a representation
of $\Gamma(N)$. Here we have used $\Gamma(N)$ for
the principal congruence subgroup of $\SL$ of level $N$
$$ \Gamma(N) = \{\ A\in\SL\ \vert\ A\equiv\id \bmod N\ \}. $$

\definition{Definition (Strongly-modular fusion algebra)}
A modular fusion algebra $({\Cal F},\rho)$ is called {\bf strongly-modular}
if the kernel of the representation $\rho$ contains a congruence
subgroup of $\SL$.
\enddefinition

In this case $\rho$ defines a representation of $\Spl{N}{}$  and is
called a level $N$ representation of $\SL$
(here and in the following we use $\Z_N$ for $\Z/N\Z$).
A level $N$ representation $\rho$ will be called even or odd
if $\rho(S^2) = \id$ or $\rho(S^2)=-\id$, respectively.
Furthermore, one can show that for strongly-modular fusion algebras
associated to rational models the representation
$\rho$ is defined over the field $K$ of $N$-th roots of unity, i.e.\
$\rho:\SL\to\GLK{n}$ if the corresponding conformal characters are modular
functions on some congruence subgroup \cite{ES$1$}.
Indeed, we expect that this is true for all RCFTs thus motivating the
following definition and conjecture.

\definition{Definition ($K$-rational representation of $\SLZ$)}
A level $N$ representation $\rho:\SLZ \to \GL{n}$ is called
$K$-rational if it is defined over the field $K$ of the $N$-th roots of
unity, i.e.\ $\rho:\SLZ \to \GLK{n}$.
\enddefinition

\definition{Conjecture}
All modular fusion algebras associated to rational models
are strongly-modular fusion algebras and
the corresponding representations of the modular group are $K$-rational.
\enddefinition

\subhead
2.3 Some simple properties of modular fusion algebras
\endsubhead

In this section we prove some simple lemmas about modular fusion
algebras which will be needed in the proofs of the main theorems
in \S3.

\proclaim{Lemma 1}
Let $({\Cal F},\rho)$ be a modular fusion algebra.
Assume that $\rho(T)$ has nondegenerate eigenvalues.
Then $\rho$ is irreducible.
\endproclaim
\demo{Proof}
 Assume that $\rho$ is reducible and $\rho(T)$ has nondegenerate
 eigenvalues. Then $\rho(S)$ has block diagonal form and
 therefore $\rho(S)_{0,m}=0$ for some $m$.
 This is a contradiction to property (3) in the
 definition of modular fusion algebras.
\enddemo

\definition{Definition ((Non-)degenerate modular fusion algebra)}
A modular fusion algebra $({\Cal F},\rho)$ is called {\bf degenerate} or
{\bf nondegenerate} if $\rho(T)$ has degenerate or nondegenerate
eigenvalues, respectively.
\enddefinition

\proclaim{Lemma 2}
Let $\rho,\rho':\SL\to\GL{n}$ be equivalent, irreducible, unitary
representations of the modular group.
Assume that $\rho(T) = \rho'(T)$ is a diagonal matrix with
nondegenerate eigenvalues.
Then there exists a unitary diagonal matrix $D$ such that
$\rho = D^{-1}\rho' D$.
\endproclaim
\demo{Proof}
Since $\rho$ and $\rho'$ are equivalent there exists a matrix $D'$ such that
$\rho = {D'}^{-1}\rho' D'$. Since $\rho(T) = \rho'(T)$ is a diagonal
matrix with nondegenerate eigenvalues $D'$ is diagonal.
Finally, the irreducibility of
$\rho$ implies by Schur's lemma that ${D'}^+ D' = \alpha\id$ for some positive
real number $\alpha$ so that $D=\frac1{\sqrt\alpha}D'$ satisfies the
desired  properties.
\enddemo

\proclaim{Lemma 3}
Let $({\Cal F},\rho)$ and $({\Cal F'},\rho')$ be two nondegenerate modular
fusion algebras.
Assume that $\rho$ is equivalent to $\rho'$ and $\rho(T) = \rho'(T)$.
Then $\Cal F$ and $\Cal F'$ are isomorphic as fusion algebras.
\endproclaim
\demo{Proof}
The lemma follows directly from the definition of (modular) fusion
algebras and Lemma 2.
\enddemo

\proclaim{Lemma 4}
Let $({\Cal F},\rho)$ be a modular fusion algebra.
Then $\rho$ is not isomorphic to a direct sum of one
dimensional representations.
\endproclaim
\demo{Proof}
If $\rho$ is the direct sum of one dimensional  representations
$\rho(S)$ is also a diagonal matrix. This implies that one cannot apply
Verlinde's formula giving a contradiction since we have assumed
that $({\Cal F},\rho)$ is a modular fusion algebra.
\enddemo

Since there are exactly 12 one dimensional representations
of $\SL$ one has the following trivial lemma.
\proclaim{Lemma 5}
\roster
\item
Let $\rho$ be a one dimensional representation of $\SL$.
Then $\rho$ is equivalent to one of the following representations
$$ \rho(S) = e^{2\pi i \frac{3n}4},\qquad \rho(T) = e^{2\pi i \frac{n}{12}},
   \qquad n=0,\dots,11.$$
\item
Let $({\Cal F},\rho)$ be a one dimensional modular fusion algebra.
Then $({\Cal F},\rho)$ is strongly-modular, $\Cal F$ is trivial and $\rho$ is
given by
$$ \rho(S) = (-1)^n,\qquad
   \rho(T) = e^{2\pi i \frac{n}6},\qquad n=0,\dots,5. $$
\endroster
\endproclaim

\proclaim{Lemma 6}
Let $({\Cal F},\rho)$ be a strongly-modular fusion algebra associated to a
rational model. Then $\rho$ is $K$-rational.
\endproclaim
\demo{Proof}
For a rational vertex operator algebra satisfying Zhu's finiteness condition
the characters are holomorphic functions on the upper complex half plane.
Since we have assumed that $({\Cal F},\rho)$ is strongly-modular, $\rho$ is a
level $N$ representation for some $N$. This implies that the characters
are modular functions on $\Gamma(N)$. Moreover, their Fourier coefficients
are positive integers so that one can apply the theorem on $K$-rationality
of ref. \cite{ES$1$} implying that $\rho$ is $K$-rational.
\enddemo

Although Lemma 6 will not be used in the following it provides us with a good
motivation for looking at $K$-rationality of level $N$
representations.

\vfill\eject
\head
3. On the classification of modular fusion algebras
\endhead

In this section we develop several tools, following
references \cite{E2,E3}, which enable us to classify
all strongly-modular fusion algebras of dimension less than or equal to four
(for a definition of strongly-modular fusion algebras see \S2.2).
Our approach is based on the known classification of the irreducible
representations of the groups $\Spl{p}{\lambda}$ \cite{NW}.

With the tools developed in this section we obtain another partial
classification, namely of those strongly-modular fusion algebras
of dimension less than 24 where the corresponding representation $\rho$
of the modular group is such that $\rho(T)$ has nondegenerate eigenvalues.
The nondegeneracy of the eigenvalues of $\rho(T)$ means that the difference
of any two conformal dimensions of a possibly underlying RCFT is not
an integer.
The restriction on the dimension is of purely technical nature so that
it should  be possible to obtain a complete classification of all
nondegenerate strongly-modular fusion algebras with the methods
described in this thesis by using systematically Galois theory.

This section is organized as follows:
In \S 3.1 state our main results on the classification of strongly-modular
fusion algebras. Section 3.2 contains some remarks about the realization
of strongly-modular fusion algebras in rational models.
In the next four subsections we give a short review of the
classification of the irreducible representations of $\Spl{p}{\lambda}$
which will be the main tool in the proof of the main theorems.
The proofs of our main theorems 1 and 2 are contained in the last three
subsections.
Finally, the three Appendices 7.1-7.3 at the end of the thesis contain
the explicit form of the modular fusion algebras occurring in our
classifications as well as the explicit form of the irreducible
level $p^\lambda$ representations of dimension less than or equal to four.

\subhead
3.1 Results on the classification of strongly-modular fusion algebras
\endsubhead

We summarize our results on the classification of low dimensional
strongly-modular fusion algebras in the following two main theorems
(note that the terminology used for the fusion algebras will be explained
in detail in \S3.2):
\proclaim{Main theorem 1 (Classification of strongly-modular fusion
                          alge\-bras of dimension $\le 4$)}
Let $({\Cal F},\rho)$ be a two, three or four dimensional
simple strongly-modular fusion algebra.
Then $\Cal F$ is isomorphic to one of the following fusion algebras:
$$ \Z_2,\ "(2,5)",\ \Z_3,\ "(2,7)",\ "(3,4)",\ \Z_4,
   \ \Z_2\otimes\Z_2,\ "(2,9)". $$
Furthermore, $({\Cal F},\rho)$ is isomorphic to the tensor product of
a one dimensional modular fusion algebra with one of the
modular fusion algebras in Table 7.2a or 7.2b (see Appendix).
\endproclaim
\mn
In the nondegenerate case we have the
\proclaim{Main theorem 2 (Classification of nondegenerate strongly-modular
                          fusion algebras of dimension $<24$)}
Let $({\Cal F},\rho)$ be a simple nondegenerate strongly-modular fusion
algebra of dimension less than 24.
Then $\Cal F$ is isomorphic to one of the following types of fusion algebras
$$ \Z_2,\ "(3,4)",\ "(2,9)",\ "(2,q)",\ B_9,\ B_{11},\ G_9,\ G_{17},\ E_{23}$$
where $q<47$ is an odd prime.
Moreover, $\Cal F$ is isomorphic to $\Q[x]/<P(x)>$ with
distinguished basis $p_j(x)$ ($j=0,\dots,n-1$).
Here $P$ and $p_j$ are the unique polynomials satisfying
$$  \align
    P(x) &= \det( {\Cal N}_1 - x ) \\
    p_0(x) &= 1, \qquad p_1(x) = x,\qquad
    p_j(x) = \sum_{k=0}^{n-1} ({\Cal N}_1)_{j,k} \ p_k(x).
\endalign
$$
where the $({\Cal N}_1)_{j,k} := N_{1,j}^k$ are the fusion matrices
given in Appendix 7.3.
Furthermore, $\rho$ is isomorphic to the tensor product of an even
one dimensional representation of $\SL$ with one of the
representations in Table 7.3 (see Appendix).
\endproclaim
In the next subsection we comment on the question which of the strongly-modular
fusion algebras described by the the main theorems 1 and 2 occur in known
RCFTs.

  \subhead
  3.2 Realization of strongly-modular fusion algebras in RCFTs
      and data of certain rational models
  \endsubhead

Let us first comment on the fusion algebras related to
the theorems in \S3.1.

The fusion algebras of type "$(2,q)$" occur in the Virasoro minimal models
with central charge $c = c(2,q)$.
Here the rational models of the Virasoro vertex operator algebra
for $c = c(p,q) = 1-6\frac{(p-q)^2}{pq}$ ($p,q>1,\ (p,q)=1$)
\cite{BPZ,Wa} are called Virasoro minimal models and
the corresponding fusion algebras are denoted by "$(p,q)$".
A list of conformal dimensions for these models can be found at the end of this
subsection.

The fusion algebra of type $\Z_n$ occurs in the so-called $\Z_n$-models
(see e.g.\ \cite{De}).
This fusion algebra are isomorphic to the group algebra of $\Z_n$ with the
distinguished basis given by the group elements. We will call the fusion
algebra given by the group algebra of $\Z_n$ in the following $\Z_n$
fusion algebra.

For all fusion algebras in the main theorem 2 apart from  $\text{B}_9$
there indeed exist RCFTs where the associated fusion
algebras are isomorphic to the ones in Table 7.3:
The fusion algebra in the first row occurs in the so-called $\Z_2$-model,
the ones in row 2, 3 and 4 in the
corresponding Virasoro minimal models (see above) and, finally,
the ones in row 6, 7, 8 and 9  occur as fusion
algebras of certain rational models, so-called minimal models
of Casimir $\w$-algebras (cf\. \S2.1 and Appendix 7.4), namely for
${\Cal WB}_2$ and $c=-\frac{444}{11}$,
${\Cal WG}_2$ and $c=-\frac{590}{9}$,
${\Cal WG}_2$ and $c=-\frac{1420}{17}$ and
${\Cal WE}_7$ and $c=-\frac{3164}{23}$ \cite{E$2$}
(central charges, conformal dimensions and characters of Casimir $\w$-algebras
are described in Appendix 7.4; the data for five of the particular
rational models
mentioned here is also collected at the end of the subsection in Table 3.2c).
The fusion algebras of type $\text{B}_9$ seems to
be related to ${\Cal WB}_2$ and $c=-24$. However, in
this case the model is not rational.

The fact that we do not know examples of RCFTs for all of the
{\bf modular} fusion algebras in our classification can be
understood as follows.
The classification of the strongly-modular fusion algebras implies
restrictions on the central charge and the
conformal dimensions of possibly underlying RCFTs.
In Table 3.2a we have collected the possible values of $c$ and the $h_i$
for the simple strongly-modular fusion algebras of dimension less than or equal
to four. Note, however, that these restrictions are not as strong as the ones
in \cite{Ki} for the two dimensional case or in \cite{CPR} for the
two and three dimensional case.
A natural way to obtain stronger restrictions
than the ones presented in Table 3.2a is to look whether there exist
vector valued modular functions transforming under the corresponding
representation of the modular group which have the correct pole order
at $i\infty$. This can be done using the methods which will be developed in
\S4  and indeed leads to much stronger restrictions on $c$
and the $h_i$ as we shall discuss elsewhere.
Of course, we expect that for any RCFT the corresponding characters
are modular functions so that these stronger restrictions have to be valid
explaining that our classification contains modular fusion algebras
for which we do not know any realization in RCFTs.
\mn
\centerline{Table 3.2a:  Central charges and conformal dimensions related to}
\centerline{$\qquad$ simple strongly-modular fusion algebras of dimension $\le
4$}
\smallskip\noindent
\centerline{
\vbox{ \offinterlineskip
\def\Tablespace{ height2pt&\omit&&\omit&&\omit&\cr }
\def\Tablerule{ \Tablespace
                \noalign{\hrule}
                \Tablespace      }
\hrule
\halign{&\vrule#&
  \strut\quad\hfil#\hfil\quad\cr
\Tablespace
& $\Cal F$    && $c \ (\bmod 4)$  && $h_i \ (\bmod\Z)$    &\cr \Tablerule
& $\Z_2$      && $1$          && $0,\frac14$  &\cr \Tablespace\Tablespace
& \omit       && $3$          && $0,\frac34$  &\cr \Tablerule
& $\Z_3$      && $2$          && $0,\frac13,\frac13$  or $0,\frac23,\frac23$
  &\cr \Tablerule
& $\Z_4$      &&  $1$ && $0,\frac18,\frac12,\frac18$ or
                         $0,\frac58,\frac12,\frac58$
  &\cr \Tablespace\Tablespace
& \omit       && $3$ && $0,\frac38,\frac12,\frac38$  or
                        $0,\frac78,\frac12,\frac78$
  &\cr \Tablerule
& $\Z_2\otimes\Z_2$  && $0$ && $0,0,0,\frac12$  or
                               $0,\frac12,\frac12,\frac12$
  &\cr \Tablerule
& "$(2,5)$"   && $\frac65$    && $0,\frac35$  &\cr \Tablespace\Tablespace
& \omit       && $\frac{14}5$ && $0,\frac25$  &\cr \Tablespace\Tablespace
& \omit       && $\frac25$    && $0,\frac15$  &\cr \Tablespace\Tablespace
& \omit       && $\frac{18}5$ && $0,\frac45$  &\cr \Tablerule
& "$(2,7)$"   && $\frac{16}7$ && $0,\frac47,\frac57$
  &\cr \Tablespace\Tablespace
& \omit       && $\frac{12}7$ && $0,\frac37,\frac27$
  &\cr \Tablespace\Tablespace
& \omit       && $\frac47$    && $0,\frac37,\frac17$
   &\cr \Tablespace\Tablespace
& \omit       && $\frac{24}7$ && $0,\frac47,\frac67$
   &\cr \Tablespace\Tablespace
& \omit       && $\frac87$    && $0,\frac67,\frac27$
  &\cr \Tablespace\Tablespace
& \omit       && $\frac{20}7$ && $0,\frac17,\frac57$
 &\cr \Tablerule
& "$(2,9)$"   && $\frac{10}3$ && $0,\frac13,\frac23,\frac{2n}9$
  &\cr \Tablespace\Tablespace
& \omit       && $\frac23$    && $0,\frac13,\frac23,\frac{n}9$
  &\cr \Tablespace\Tablespace
& \omit       && \omit        && $n=1,4,7$
  &\cr \Tablerule
& "$(3,4)$"   && $\frac{3n}2$ && $0,\frac12,\frac{3n}{16}$
  &\cr \Tablespace\Tablespace
& \omit       && $n=0,\dots,15$ && \omit
  &\cr \Tablespace
}
\hrule}
}
\mn
We finally give concrete lists of the central charges and conformal
dimensions of certain rational models
which will appear again in the main theorem 3 on uniqueness of conformal
characters in \S4 and by the main theorem 5 on theta formulas for
conformal characters in \S5.
\mn
\leftline{$\quad\underline{\text{\bf
   Central charges and conformal dimensions of certain rational models
  }}$.}
\mn

Note that some of the results summarized in this section are
not yet proved on a mathematically rigorous level.
However, taken as an input into the formalism developed in \S4
the central charges and sets of conformal dimensions given below
will lead to consistent representations of the modular group on
spaces of modular functions.
This section serves rather as a
motivation than as a background for the considerations in the
subsequent sections.

Firstly, we review some known rational models with effective
central charge less than 1.
The simplest $\w$-algebras are those which can be constructed from
the Virasoro algebra (as already mentioned in\S2.1).
The rational models among these are called
the Virasoro minimal models (see e.g.\ \cite{BPZ,RC,Wa}).
They can be parameterized by a set of two coprime integers
$p,q\ge 2$.
The rational model corresponding to such a
set $p,q$ has central charge
$$ c = c(p,q) = 1-6\frac{(p-q)^2}{pq} $$
and its conformal dimensions  are given by:
$$ h(p,q,r,s) = \frac{(rp-sq)^2-(p-q)^2}{4pq}
   \quad (1\le r<q,\ (2,r)=1, \ 1\le s<p),
$$
where we assume $q$ to be odd.

The Virasoro minimal models are special examples of the larger
class of rational models with $\tilde c < 1$ which emerges from
the $ADE$-classification  of modular invariant partition functions
\cite{CIZ,EFH$^2$NV}.
Their central charges and conformal dimensions are given in Table 3.2b:
The first column describes the type of modular invariant
partition function, the central charge is always $c= c(p,q)$
where $p$ and $q$ are the parameters of the respective row
under consideration. Moreover, $c(p,q)$ and $h(p,q,\cdot,\cdot)$
are as defined above.
Note that the listed models exist also for $p,q,m$ not
necessarily prime. The primality restrictions have been added for
technical reasons only which will become clear in the next section.
\mn
\centerline{Table 3.2b: Data of certain rational $\w$-algebras related
                     to the $ADE$-classification}
\smallskip\noindent
\centerline{
\vbox{ \offinterlineskip
\def\tablespace{ height2pt&\omit&&\omit&&\omit&\cr }
\def\tablerule{ \tablespace
                \noalign{\hrule}
                \tablespace      }
\hrule
\halign{&\vrule#&
  \strut\quad\hfil#\hfil\quad\cr
\tablespace
& type
  && type of $\w$-algebra
  && $H_{c(p,q)}\qquad (I_n := \{1,\dots,n\})$ &\cr
\tablerule
\tablerule
& $(A_{q-1},A_{p-1})$
  &&   $ \w(2)$
  &&  $\{ h(p,q,r,s) \ \vert \ r\in I_{q-1},
                             \ s\in I_{p-1},\ (2,r)=1\}$
&   \cr \tablespace
&\omit
  && $p>q$  odd primes
  &&\omit
&\cr \tablerule
&$(A_{q-1},D_{m+1})$
  &&   $\w(2,\frac{(m-1)(q-2)}{2})$
  &&  $\{ h(p,q,r,s) \ \vert \ r\in I_{(q-1)/2},
                             \ s\in I_m,\ (2,s)=1 \}$
&   \cr \tablespace
&\omit
  && $p=2m$
  && \omit
&\cr \tablespace
&\omit
  && $q$ and $m$ odd primes
  &&\omit
&\cr \tablerule
& $(A_{q-1},E_6)$
  &&   $\w(2,q-3)$
  &&  $\{ \min(h(p,q,r,1),h(p,q,r,\ 7)) \ \vert
          \ r\in I_{(q-1)/2} \}  \cup$
& \cr\tablespace
&\omit
  && $p=12,\ q\ge5$
  &&  $\{ \min(h(p,q,r,5),h(p,q,r,11)) \ \vert
          \ r\in I_{(q-1)/2} \}  \cup$
& \cr\tablespace
& \omit
  &&  $q$ prime
  && $ \{ h(p,q,r,4) \ \vert \ r\in I_{(q-1)/2} \}$
&   \cr  \tablerule
&$(A_{q-1},E_8)$
  &&   $\w(2,q-5)$
  &&  $\ \{ \min(h(p,q,r,1),h(p,q,r,11)) \ \vert
        \ r\in I_{(q-1)/2} \}  \cup$
& \cr\tablespace
&\omit
  && $p=30,\ q\ge7$
  && $\{ \min(h(p,q,r,7),h(p,q,r,13)) \ \vert
           \ r\in I_{(q-1)/2} \}\ $
&   \cr \tablespace
&\omit
  &&$q$ prime
  && \omit
&\cr \tablespace
}
\hrule}
}
\mn
The second list of rational models which we shall consider
are special cases of the so-called Casimir $\w$-algebras (cf\. \S2.1).

In Table 3.2c we list the central charges $c$, effective central
charge $\tilde c$ and the sets of conformal dimensions $H_c$  of
$5$ rational models  with $\tilde c > 1$.

The last three are minimal models of Casimir $\w$-algebras associated to
${\Cal B}_2,{\Cal G}_2$ and ${\Cal E}_7$.

The first two $\w$-algebras are `tensor products' of
the rational $\w$-algebra with $c=-22/5$ constructed
from the Virasoro algebra and the rational $\w$-algebras
with $c= 14/5$ or $c=26/5$ constructed from the Kac-Moody
algebras associated to  ${\Cal G}_2$ or ${\Cal F}_4$, respectively.
We denote them by $\w_{{\Cal G}_2}(2,1^{14})$ and
$\w_{{\Cal F}_4}(2,1^{26})$, respectively.
Here the construction of the $\w$-algebras in question is the
one mentioned in \S2.1 in the remark after the definition of the type of
$\w$-algebras in (2).

\mn
\centerline{Table 3.2c:  Data of the five rational models }
\smallskip\noindent
\centerline{
\vbox{ \offinterlineskip
\def\tablespace{ height2pt&\omit&&\omit&&\omit&&\omit&\cr }
\def\tablerule{ \tablespace
                \noalign{\hrule}
                \tablespace      }
\hrule
\halign{&\vrule#&
  \strut\quad\hfil#\hfil\quad\cr
\tablespace
& $\w$-algebra && $c$    && $\tilde{c}$  && $H_c$ &\cr
\tablerule
& $\w_{G_2}(2,1^{14})$  &&  $-{8\over5}$  && ${16\over5}$ &&
  $ {1\over5} \{0,-1,1,2 \} $
                    &\cr \tablerule
& $\w_{F_4}(2,1^{26})$  &&  ${4\over5}$  && ${28\over5}$ &&
  $ {1\over5} \{0,-1,2,3 \} $
                    &\cr \tablerule
& $\w(2,4)$  &&  $-{444\over11}$  && ${12\over11}$ &&
  $ -{1\over11}
               \{0,9,10,12,14,15,16,17,18,19  \} $
                    &\cr \tablerule
& $\w(2,6)$  &&  $-{1420\over17}$ && ${20\over17}$ &&
  $ -{1\over17}
               \{0,27,30,37,39,46,48,49,50,$&\cr
&\omit  && \omit    && \omit             &&
           \qquad $52,53,55,57,58,59,60  \} $
                    &\cr \tablerule
& $\w(2,8)$  &&  $-{3164\over23}$ && ${28\over23}$ &&
  $-{1\over23}
               \{ 0,54,67,81,91,94,98,103,111,$&\cr
&\omit      && \omit   &&\omit           && \quad
                $112,116,118,119,120,122,124, $ &\cr
&\omit       && \omit  &&\omit           &&
          \quad $125,129,130,131,132,133 \} $
                    &\cr \tablespace}
\hrule}
}
\mn
We give some comments on these $5$ rational models.
Using \cite{RC} and \cite{Ka} the central charges, conformal
characters and  dimensions of the two composite rational models
can be computed. For the rational models of type
$\w(2,d)$   lists of the associated conformal dimension
can be found in \cite{EFH$^2$NV}.

As it will be seen in the next section the
first five rational models in Table~3.2c do have a common feature:
The representations of $\SL$ afforded by their
conformal characters belong, up to multiplication by certain 1-dimensional
$\SL$-representations, to one and the same series $\rho_l$
(cf. \S4.4 and \S5.2 for details). So one could ask whether there exist
more rational models with this property.
A more detailed investigation of the fusion algebras associated
to such potentially existing models showed that this is not
the case \cite{E$2$} (cf. also the speculation in \cite{EfH$^2$NV}).

\subhead
3.3 Some theorems on level $N$ representations of $\SLZ$
\endsubhead

In this subsection we will consider level $N$ representations of $\SLZ$.
Firstly, we review the fact that all irreducible representations of
$\Spl{N}{}$ can be obtained by those of $\Spl{p}{\lambda}$
where $p$ is a prime and $\lambda$ is a positive integer.
Secondly, we discuss the construction of level $p^\lambda$
representations using Weil representations (in this part we follow
ref. \cite{NW}).
\vfil\eject
\proclaim{Lemma 7}
Let $\rho$ be a finite dimensional representation of $\Spl{N}{}$
where $N$ is a positive integer.
Then the representation $\rho$ is completely reducible.
Furthermore, each irreducible component $\omega$ of $\rho$
has a unique product decomposition
$$ \omega \cong \otimes_{j=1}^{n} \pi({p_j^{\lambda_j}}) $$
where $N=\prod_{j=1}^{n} p_j^{\lambda_j}$ is the prime factor
decomposition of $N$ and the $\pi({p_j^{\lambda_j}})$ are
irreducible representations of $\Spl{p_j}{\lambda_j}$.
\endproclaim
\demo{Proof}
Since $\Spl{N}{}$ is a finite group, $\rho$ is completely reducible.

For a proof of the second statement note that
$$\Spl{N}{} = \Spl{p_1}{\lambda_1}\times\dots\times\Spl{p_n}{\lambda_n}$$
where $N=\prod_{j=1}^{n} p_j^{\lambda_j}$ (see e.g.\ \cite{G}).
Obviously, the tensor product of irreducible representations
$\pi({p_j^{\lambda_j}})$ of $\Spl{p_j}{\lambda_j}$ is an
irreducible representation of $\Spl{N}{}$. Using now
Burnside's lemma we obtain the second statement.
\enddemo

In order to deal with the representations of the groups
$\Spl{p}{\lambda}$ we describe their structure by the following theorem.

\proclaim{Theorem (Structure of $\Spl{p}{\lambda}$
                   \cite{NW, Satz 1, p. 466})}
The finite group  $\Spl{p}{\lambda}$ is generated by the elements
$$ S= \left(\matrix 0&-1\\1&0\endmatrix\right),\quad
   T= \left(\matrix 1&1\\0&1\endmatrix\right),\quad
$$
and the relations
$$\align
   T^{p^\lambda} & = \id,\qquad
   S^2 = H(-1) \\
   H(a) H(a') &= H(a a'), \qquad
   H(a) T     = T^{a^2} H(a),\qquad
   S H(a)     = H(a^{-1}) S   \\
\endalign $$
where  $ H(a) := T^{-a} S T^{-a^{-1}} S^{-1} T^{-a} S^{-1}$
and  $a,a' \in \Z_{p^\lambda}^*$.
\endproclaim
\remark{Remark}
As elements of $\Spl{p}{\lambda}$ the  $H(a)$
($a\in \Z_{p^\lambda}^*$) are given by
$$ H(a) = \left(\matrix a&0\\0&a^{-1}\endmatrix\right).$$
\endremark

We will now describe the construction of representations of $\Spl{p}{\lambda}$
by means of Weil representations.

\definition{Definition (Quadratic form)}
Let $M$ be a finite $\Z_{p^\lambda}$ module. A quadratic form $Q$ of $M$
is a map $Q:M\to\ p^{-\lambda}\Z / \Z$ such that
\roster
\item
 $Q(-x) = Q(x)$ for all $x\in M$.
\item
 $B(x,y) := Q(x+y) - Q(x) -Q(y)$ defines a  $\Z_{p^\lambda}$-bilinear map
 from $M\times M$ to $  p^{-\lambda}\Z / \Z$.
\endroster
\enddefinition

\definition{Definition (Quadratic module)}
A finite $\Z_{p^\lambda}$ module $M$ together with a quadratic form $Q$
is called a quadratic module of $\Z_{p^\lambda}$.
\enddefinition

\definition{Definition (Weil representation)}
Let $(M,Q)$ be a quadratic module.
Define a right action of $\Spl{p}{\lambda}$ on the space of $\C$ valued
functions on $M$ by
$$ \align
  (f|T)(x)       &= e^{2\pi i Q(x)} \ f(x) \\
  (f|H(a))(x)    &= \alpha_Q(a) \alpha_Q(-1) \ f(x)
                    \qquad \forall a \in \Z_{p^\lambda}^* \\
  (f|S^{-1})(x)  &= \frac{\alpha_Q(-1)}{\vert M\vert ^{1/2}}
                       \sum_{y\in M}  e^{2\pi i B(x,y)} \ f(y) \\
\endalign
$$
where $ \vert M \vert$ denotes the order of $M$,
$$ \alpha_Q(a) = \frac1{\vert M\vert} \sum_{x\in M} e^{2\pi i a Q(x)} $$
and $f$ is any $\C$ valued function on $M$.

If this right action of $\Spl{p}{\lambda}$ defines a representation of
$\Spl{p}{\lambda}$ it is called the (proper) Weil representation associated
to the quadratic module $(M,Q)$ and denoted by $W(M,Q)$.
\enddefinition

Note that the above right action always defines a projective
representation of $\SL$. A necessary and sufficient condition for it
to define a proper representation is given by the following theorem.
\proclaim{Theorem (Proper Weil representation \cite{NW, Satz 2, p. 467}) }
The above right action of $\Spl{p}{\lambda}$ defines a representation of
$\Spl{p}{\lambda}$ if and only if
$$ \alpha_Q(a) \alpha_Q(a') = \alpha_Q(1) \alpha_Q(a a') \
   \qquad a,a'\in \Z_{p^\lambda}^*. $$
\endproclaim

In the following we will only deal with proper Weil representations
and, therefore, call them simply Weil representations.

\subhead
3.4 Weil representations associated to binary quadratic forms
\endsubhead

Although the classification of the irreducible representations of
the finite groups $\Spl{p}{\lambda}$ is contained in \cite{NW}
we will give a short review here. Our main motivation for this is the
fact that we will strongly rely on this classification in the proofs
of the main theorems 1 and 2 in \S3.7 and \S3.9.
(Furthermore, ref. \cite{NW} is not written in English but in German.)

In the this subsection we describe how to obtain
irreducible level $p^\lambda$ representations as
subrepresentations of Weil representations.
In subsections \S3.5 and \S3.6 we give complete lists of the
irreducible representations for the cases $p\not=2$ and $p=2$,
respectively.

In addition to the review we investigate in some cases
whether the irreducible representations are $K$-rational or not.

Most of the irreducible representations of $\Spl{p}{\lambda}$ can be obtained
as subrepresentations of Weil representations $W(M,Q)$ associated to
a module $M$ of rank one or two.
The following two theorems describe the Weil representations needed in the
later sections.

\proclaim{Theorem (Weil representations of $\Spl{p}{\lambda}$ ($p\not=2$)
                   \cite{NW, Lemma 1, Satz 3, p.\ 474}) }
Let $p\ne 2 $ be a prime. Then the following quadratic modules of
$\Z_{p^\lambda}$ define Weil representations:
$$ \align
  &(1)\quad M = \Z_{p^\lambda},\qquad \qquad  \ \ \,
            Q(x) = p^{-\lambda} r x^2
      \qquad \qquad \quad \quad\, (\lambda \ge 1)
      \qquad \left( R_{\lambda}(r) \right) \\
  &(2)\quad M = \Z_{p^\lambda}\oplus\Z_{p^\lambda},\quad \quad\,
            Q(x) = p^{-\lambda}x_1 x_2
        \qquad \qquad\quad\ \, \,(\lambda \ge 1)
        \qquad \left( D_{\lambda} \right) \\
  &(3)\quad M = \Z_{p^\lambda}\oplus\Z_{p^\lambda},\quad \quad\,
            Q(x) = p^{-\lambda}(x_1^2 -u x_2^2)
        \qquad \quad\, (\lambda \ge 1)
        \qquad \left( N_{\lambda} \right) \\
  &(4)\quad M = \Z_{p^\lambda}\oplus\Z_{p^{\lambda-\sigma}},\quad \,
            Q(x) = p^{-\lambda} r (x_1^2 -p^\sigma t x_2^2)
        \qquad (\lambda \ge 2)
        \qquad \left( R_\lambda^\sigma (r,t) \right)
\endalign $$
where $r,t$ run through $\{1,u \}$ with $(\frac u p) = -1$
($(\frac{\cdot}{\cdot})$ denotes the Legendre symbol),
where $\sigma = 1,\dots,\lambda-1$ and where the last column contains
the name of the corresponding Weil representation.
\endproclaim

\proclaim{Theorem (Weil representations of $\Spl{2}{\lambda}$
                   \cite{NW, Satz 4, p.\ 474}) }
Let $p=2$. Then the following quadratic modules of $\Z_{2^\lambda}$
define Weil representations:
$$ \align
  &(1)\quad M = \Z_{2^\lambda}\oplus\Z_{2^\lambda},\qquad \quad\ \
            Q(x) = 2^{-\lambda}x_1 x_2
            \qquad \qquad \quad \ \,(\lambda \ge 1)
            \qquad \left( D_{\lambda} \right)  \\
  &(2)\quad M = \Z_{2^\lambda}\oplus\Z_{2^\lambda},\qquad \quad\ \
            Q(x) = 2^{-\lambda}(x_1^2 + x_1 x_2 + x_2^2)
            \ (\lambda \ge 1)
            \qquad \left( N_{\lambda} \right) \\
  &(3)\quad M = \Z_{2^{\lambda-1}}\oplus\Z_{2^{\lambda-\sigma-1}},\quad
            Q(x) = 2^{-\lambda} r (x_1^2 +2^\sigma t x_2^2)
        \quad\ \ (\lambda \ge 2)
            \qquad \left( R_\lambda^\sigma (r,t) \right)
\endalign $$
where $\sigma = 0,\dots,\lambda-2$,
where $(r,t)$ run through a system of representatives of the classes
of pairs defined by $ (r_1,t_1) \cong (r_2,t_2)$
\sn
if $t_1 \equiv t_2 \bmod \min(8,2^{\lambda-\sigma})$
and
$$\cases
     r_2 \equiv r_1 \bmod 4 \quad \text{or} \quad
     r_2 \equiv r_1 t_1 \bmod 4   &\text{for} \quad \sigma=0\\
     r_2 \equiv r_1 \bmod 8 \quad \text{or} \quad
     r_2 \equiv r_1 + 2r_1 t_1 \bmod 8 &\text{for} \quad \sigma=1\\
     r_2 \equiv r_1 \bmod 4  &\text{for} \quad \sigma=2\\
     r_2 \equiv r_1 \bmod 8  &\text{for} \quad \sigma\ge3\\
\endcases$$
and where the last column contains the name of the corresponding
Weil representation.
\endproclaim

All irreducible representations of $\Spl{p}{\lambda}$ can be obtained
as subrepresentations of Weil representations $W(M,Q)$.
One possibility to extract subrepresentations of such representations
is to use characters of the automorphism group of the quadratic form
$Q$:
\proclaim{Theorem (Subrepresentation of a Weil representation
                   \cite{NW, p\. 480}) }
$ $
Let $W(M,Q)$ be a Weil representation described by one of the
theorems on Weil representations of $\Spl{p}{\lambda}$ above,
$\Cal U$ an abelian subgroup of $\text{Aut}(M,Q)$ and $\chi$ a
character of $\Cal U$.
Then the subspace
$$V(\chi) := \{ \ f:M\to \C\  \vert\  f(\epsilon x) = \chi(\epsilon) f(x),
               \quad x\in M,\ \epsilon \in {\Cal U} \ \}$$
of $\C^M$ is invariant under $\Spl{p}{\lambda}$. The corresponding
subrepresentation is denoted by $W(M,Q,\chi)$.
\endproclaim

\remark{Remarks}
\roster
\item
 The space $V(\chi)$ is spanned by $ V(\chi) = < f_x(\chi)>_{x\in M} $
where
$$f_x(\chi)(y) = \sum_{\epsilon\in {\Cal U}} \chi(\epsilon)
                                              \delta_{\epsilon x,y},
  \qquad  \delta_{x,y} = \cases 1 & \text{for}\ $x=y$ \\
                                0 & \text{otherwise.}
                        \endcases
$$
\item
The automorphism group of the quadratic forms in Theorem 4 contain a
conjugation $\kappa$: $\kappa(x_1,x_2) = (x_2,x_1)$ in case (1) and
 $\kappa(x_1,x_2) = (x_1,-x_2)$  in the cases (2) and (3).
In these cases the space
$$V(\chi)_{\pm} := \{ \ f\in V(\chi)\ \vert\  f(\kappa x) = \pm f(x),
                     \quad x\in M \ \}$$
is invariant under $\Spl{2}{\lambda}$.  The corresponding
subrepresentation is denoted by $W(M,Q,\chi)_\pm$.
\endroster
\endremark

{}From now on we will denote the trivial character $\chi \equiv 1$ by
$\chi_1$. Indeed, almost all irreducible representations of $\Spl{p}{\lambda}$
can be obtained as subrepresentations of the Weil representations described
by the main theorems on Weil representations of $\Spl{p}{\lambda}$
using `primitive' characters:

\definition{Definition (Primitive character of a Weil representation)
            \footnotemark}
$\qquad\quad$
Let $W(M,Q)$ be a Weil representation described by one of the two theorems
on Weil representations of $\Spl{p}{\lambda}$ above
and let ${\Cal U} = \text{Aut}(M,Q)$. A character $\chi$ of $\Cal U$ is called
primitive iff there exists an element $\epsilon\in {\Cal U}$ with
$\chi(\epsilon)\not=1$ such that each element of $pM$ is a fixed point of
$\epsilon$. The set of primitive characters of $\Cal U$ will be denoted by
$\P$.
\enddefinition
\footnotetext{ In the case of $M=\Z_{2^{\lambda-1}}\oplus\Z_{2}$
 ($\lambda \ge 5$) the definition of primitive characters is
 slightly different \cite{NW, p.\ 491}:
 Here $\Cal U \cong <-1> <\alpha>$ with
 $\alpha = \cases 1+4t+\sqrt{-8t} & \lambda=5\\
                  1-2^{\lambda-3}+\sqrt{-2^{\lambda-2}t} & \lambda>5
           \endcases\qquad\qquad\qquad$
 and $\chi$ is primitive if $\chi(\alpha) = -1$. }
\proclaim{Theorem (Isomorphy of Weil representations
                    \cite{NW, Hauptsatz 1, p. 492})}
$\ \ $
Let $W(M,Q)$ and $W(M',Q')$ be Weil representation described by one
of the two theorems on Weil representations above and  $\chi,\chi'$
primitive characters. Then one has
\roster
\item
 $W(M,Q,\chi)$ is an irreducible level $p^\lambda$ representation.
\item
 $W(M,Q,\chi)$ and $W(M',Q',\chi')$ are isomorphic if and only if
 the quadratic modules $(M,Q)$ and $(M',Q')$ are isomorphic and
 $\chi = \chi'$ or $\chi= \bar\chi'$.
\endroster
\endproclaim

The second main theorem of ref. \cite{NW} describes
the classification of the
irreducible representations of $\Spl{p}{\lambda}$.

\proclaim{ Theorem (Classification of irreducible representations of
                    $\Spl{p}{\lambda}$ \cite{NW, Hauptsatz 2, p.\ 493})}
The Weil representations described by the two theorems on Weil representations
of $\Spl{p}{\lambda}$ above contain all
irreducible representations of the groups $\Spl{p}{\lambda}$
(in general they are of the form $W(M,Q,\chi)$ for a primitive character
$\chi$) apart from 18 exceptional representations for $p=2$. These exceptional
representations can be obtained as tensor products of two representations
contained in some $W(M,Q)$ (described by the theorem on Weil
representations of $\Spl{2}{\lambda}$ above).
\endproclaim

\S3.5 and \S3.6 contain lists of all irreducible level $p^\lambda$
representations.

\subhead
3.5 The irreducible representations of $\SLZpl$ for $p\ne2$
\endsubhead

In the classification of the irreducible representations of
$\SLZpl$ for $p\not=2$ one has to distinguish the cases $\lambda=1$
and $\lambda>1$. Therefore, we treat them separately.

Following \cite{NW} we denote the trivial representation by $C_1$.

\proclaim{Theorem (Classification of irreducible representations of
                   $\SLZp$ ($p\not=2$) \cite{NW})}
A complete set of irreducible representations of $\SLZp$ for
a prime $p$ with $p\not= 2$ is given by the representations collected in
Table~3.5a. In Table 3.5a the $\chi$ run through the set of characters
of $\Cal U$ and $\chi_{-1}$ is the unique nontrivial character of $\Cal U$
taking values in $\pm1$. Furthermore, we denote by $\#$ (here
and in the following) the number of inequivalent representations.
\endproclaim
\mn
\centerline{Table 3.5a: Irreducible representations of $\SLZp$ for $p\not=2$}
\smallskip\noindent
\centerline{
\vbox{ \offinterlineskip
\def\Tablespace{ height2pt&\omit&&\omit&&\omit&&\omit&\cr }
\def\Tablerule{ \Tablespace
                \noalign{\hrule}
                \Tablespace      }
\hrule
\halign{&\vrule#&
  \strut\quad\hfil#\hfil\quad\cr
\Tablespace
& type of rep. &&  && dimension  && \# &\cr
\Tablerule
& $D_1(\chi)$        && $\chi\in\P$ && $p+1$
  && $\frac12(p-3)$ &\cr \Tablerule
& $N_1(\chi)$        && $\chi\in\P$ && $p-1$
  && $\frac12(p-1)$ &\cr \Tablerule
& $R_1(r,\chi_1)$    && $\bigl({r\over p} \bigr) = \pm 1 $
  && $\frac12(p+1)$   && $2$            &\cr \Tablerule
& $R_1(r,\chi_{-1})$ && $\bigl({r\over p} \bigr) = \pm 1$
  && $\frac12(p-1)$  && $2$            &\cr \Tablerule
& $N_1(\chi_1)$      && \omit
  && $p$            && $1$            &\cr \Tablespace
}
\hrule}
}
\mn

We will denote the 3 one dimensional level 3 representations
$C_1$, $R_1(1,\chi_{-1})$ and $R_1(2,\chi_{-1})$ by $B_1$, $B_2$ and
$B_3$, respectively.
\smallskip
The explicit form of these representations is well known
(see e.g.\ \cite{E$2$})
and one can address the question which of these representations are
$K$-rational (see also \S4). Note that, in view of the
results in \S2, this question is natural in the context of
admissible representations.
\proclaim{Lemma 8}
Let $p\not=2$ be a prime.
\roster
\item
For $p\equiv 1 \pmod 3$ there is exactly one and for
$p \not\equiv 1 \pmod 3$ there is no $K$-rational representation of
type $D_1(\chi)$.
\item
For $p\equiv 2 \pmod 3$ there is exactly one and for
$p \not\equiv 2 \pmod 3$ there is no $K$-rational
representation of type $N_1(\chi)$ ($\chi\in\P$).
\item
The representations of type $R_1(r,\chi_{\pm 1})$ and
$N_1(\chi_1)$ are $K$-rational.
\endroster
\endproclaim
\demo{Proof}
Using a character table for the above representations (see e.g.\ \cite{Do})
one easily finds that the characters of representations of type
$D_1(\chi)$ or $N_1(\chi)$ take values in the field of $p$-th roots
of unity only if $p\equiv 1 \pmod 3$ or $p\equiv 2 \pmod 3$
and if $\chi$ is a character of order 3.
Therefore, there is at most one $K$-rational representation of
type $D_1(\chi)$ or $N_1(\chi)$ for the corresponding values of $p$.
Using the explicit form of these representations  (see e.g.\ \cite{E$2$})
one finds that these two representations are indeed $K$-rational.
For the other two types of representations the $K$-rationality follows
directly from the fact that $\chi_{\pm 1}$ takes values in $\pm1$.
\enddemo

\proclaim{Theorem (Classification of irreducible representations of
                   $\Spl{p}{\lambda}$ ($p\not=2;\lambda>1$) \cite{NW})}
A complete set of irreducible representations of $\Spl{p}{\lambda}$
for $p\not=2$ prime and $\lambda>1$ is given by the representations
in Table 3.5b.
Where $\chi_{-1}$ is the unique nontrivial
character with values in $\pm1$ and $R_\lambda(r,\chi_{\pm1})_1$ is the unique
level $p^\lambda$ subrepresentation of $R_\lambda(r,\chi_{\pm1})$
which has dimension $\frac12(p^2-1)p^{\lambda-2}$.
\endproclaim

\centerline{Table 3.5b: Irreducible representations of $\SLZpl$ for
                     $p\not=2$ and $\lambda>1$ }
\smallskip\noindent
\centerline{
\vbox{ \offinterlineskip
\def\Tablespace{ height2pt&\omit&&\omit&&\omit&&\omit&\cr }
\def\Tablerule{ \Tablespace
                \noalign{\hrule}
                \Tablespace      }
\hrule
\halign{&\vrule#&
  \strut\quad\hfil#\hfil\quad\cr
\Tablespace
& type of rep. &&  && dimension  && \# &\cr
\Tablerule
& $D_\lambda(\chi)$    && $\chi\in \P$ && $(p+1)p^{\lambda-1}$
  && $\frac12(p-1)^2p^{\lambda-2}$                             &\cr \Tablerule
& $N_\lambda(\chi)$    && $\chi\in \P$ && $(p-1)p^{\lambda-1}$
  && $\frac12(p^2-1)p^{\lambda-2}$                             &\cr \Tablerule
& $R_\lambda^\sigma(r,t,\chi)$
  && $\bigl({r\over p}\bigr) = \pm 1,\ \bigl({t\over p}\bigr) = \pm 1$
  && $\frac12(p^2-1)p^{\lambda-2}$
  && $4 \sum_{\sigma=1}^{\lambda-1} (p-1)p^{\lambda-\sigma-1}$ &\cr \Tablerule
& $R_\lambda(r,\chi_{\pm1})_1$ && $\bigl({r\over p}\bigr) = \pm 1$
  && $\frac12(p^2-1)p^{\lambda-2}$ && $4$                      &\cr \Tablespace
}
\hrule}
}

\proclaim{Lemma 9}
Let $p\not=2$ be a prime and $\lambda>1$ an integer.
\roster
\item
The representations of type  $R_\lambda^\sigma(r,t,\chi)$
are $K$-rational for $p\not=2$ and $\lambda>1$.
\item
The representations of type $R_\lambda(r,\chi_{\pm1})_1$
are $K$-rational for $p\not=2$ and $\lambda>1$.
Furthermore, the image of $T$ under these representations
has nondegenerate eigenvalues only if  $p=3$ and $\lambda=2$.
\endroster
\endproclaim
\demo{Proof}
Since the automorphism group of the quadratic form of
$R_\lambda^\sigma(r,t,\chi)$ is given by \cite{NW, p. 495}
${\Cal U}  \cong \Z_2 \times \Z_{p^{\lambda-\sigma}}$
we obtain (1).
In the second case one obviously has ${\Cal U} \cong \Z_2$ so that
the $K$-rationality follows directly.
The statement concerning the eigenvalues of the image of $T$ for the
representations of type $R_\lambda(r,\chi_{\pm1})_1$ is proved in
Satz 4 of \cite{NW}.
\enddemo

\subhead
3.6 The irreducible representations of $\Spl{2}{\lambda}$
\endsubhead

The classification of the irreducible representations
of $\Spl{2}{\lambda}$ is complicated since there are a lot of
exceptional representations for $\lambda < 6$ \cite{NW}.
Since these representations have small dimensions
and we will be interested in such representations
in \S3.7 and \S3.9 we describe them in the rest of this subsection.
The Tables 3.6a-3.6f list complete sets of inequivalent irreducible
representations
of the groups
$\Spl{2}{\lambda}$ for the corresponding values of $\lambda$.

For $\lambda=1$ there are only two irreducible representations (see
Table 3.6a).
The representation $C_2$ is given by $C_2(S) = C_2(T) = -1$ and
both level 2 representations are $K$-rational.

For $\lambda=2$ there are seven irreducible representations (see Table 3.6b).
The representation $C_3$ is given by $C_3(S) = C_3(T) = - i$,
$C_4$ by $C_4(S) = C_4(T) = i$ and $R_2^0(1,3)_1$ is defined by
$ R_2^0(1,3) \cong R_2^0(1,3)_1\oplus C_1$.
All level 4 representations are $K$-rational.

For $\lambda=3$ there are 20 irreducible representations (see Table 3.6c).
Here $\hat\chi$ is one of the two characters of $\Cal U$ of order 4 and
the representation $R_3^0(1,3,\chi_1)_1$ is defined by
$R_3^0(1,3,\chi_1) \cong R_3^0(1,3,\chi_1)_1
                         \oplus N_1(\chi_1) \oplus C_2\oplus C_2.$

For $\lambda=4$ there are  46 irreducible representations (see Table 3.6d).
Here the representation $R_4^2(r,3,\chi_1)_1$ is given by the equality
$R_4^2(r,3,\chi_1) \cong R_4^2(r,3,\chi_1)_1 \oplus R_2^0(r,t)$.

\vfill\eject

\centerline{Table 3.6a: Irreducible representations of $\Spl{2}{}$ }
\centerline{
\vbox{ \offinterlineskip
\def\Tablespace{ height2pt&\omit&&\omit&&\omit&&\omit&\cr }
\def\Tablerule{ \Tablespace
                \noalign{\hrule}
                \Tablespace      }
\hrule
\halign{&\vrule#&
  \strut\quad\hfil#\hfil\quad\cr
\Tablespace
& type of rep. &&  && dim  && \#  &\cr
\Tablerule
& $C_2 = N_1(\chi)$ &&  $\chi \in \P$     && $1$  && $1$      &\cr \Tablerule
& $N_1(\chi_1)$     && \omit              && $2$  && $1$      &\cr \Tablespace
}
\hrule}
}

\centerline{Table 3.6b: Irreducible representations of $\Spl{2}{2}$}
\centerline{
\vbox{ \offinterlineskip
\def\Tablespace{ height2pt&\omit&&\omit&&\omit&&\omit&\cr }
\def\Tablerule{ \Tablespace
                \noalign{\hrule}
                \Tablespace      }
\hrule
\halign{&\vrule#&
  \strut\quad\hfil#\hfil\quad\cr
\Tablespace
& type of rep. &&  && dim && \# &\cr
\Tablerule
& $D_2(\chi)_+$ && $\chi \not\equiv 1$    && $3$  && $1$      &\cr \Tablerule
& $D_2(\chi)_-$ && $\chi \not\equiv 1$    && $3$  && $1$      &\cr \Tablerule
& $R_2^0(1,3)_1$ && \omit                 && $3$  && $1$      &\cr \Tablerule
& $C_2 \otimes R_2^0(1,3)_1$ &&\omit      && $3$  && $1$      &\cr \Tablerule
& $N_2(\chi)$ &&
  $\chi\in \P;\ \chi \not\equiv 1$         && $2$  && $1$      &\cr \Tablerule
& $C_3=R_2^0(3,1,\chi)$ &&
  $\chi\not\equiv 1$                      && $1$  && $1$      &\cr \Tablerule
& $C_4=R_2^0(1,1,\chi)$ &&
  $\chi\not\equiv 1$                      && $1$  && $1$      &\cr \Tablespace
}
\hrule}
}

\sn

\centerline{Table 3.6c: Irreducible representations of $\Spl{2}{3}$}
\centerline{
\vbox{ \offinterlineskip
\def\Tablespace{ height2pt&\omit&&\omit&&\omit&&\omit&\cr }
\def\Tablerule{ \Tablespace
                \noalign{\hrule}
                \Tablespace      }
\hrule
\halign{&\vrule#&
  \strut\quad\hfil#\hfil\quad\cr
\Tablespace
& type of rep. &&  && dim  && \# &\cr
\Tablerule
& $D_3(\chi)_{\pm}$ &&
  $\chi\in \P$                       && $6$  && $4$      &\cr \Tablerule
& $R_3^0(1,3,\chi_1)_1$ &&\omit      && $6$  && $1$      &\cr \Tablerule
& $C_3 \otimes R_3^0(1,3,\chi_1)_1$ &&
  \omit                              && $6$  && $1$      &\cr \Tablerule
& $N_3(\chi)$ &&
  $\chi\in\P;\ \chi^2 \not\equiv 1$  && $4$  && $2$      &\cr \Tablerule
& $N_3(\chi)_{\pm}$ &&
  $\chi\in\P;\ \chi^2 \equiv 1$      && $2$  && $4$      &\cr \Tablerule
& $R_3^0(r,t,\hat\chi)$ &&
  $r=1,3;\ t=1,5$                    && $3$  && $4$      &\cr \Tablerule
& $R_3^0(1,t,\chi)_{\pm}$ &&
  $\chi\not\equiv 1;\ t=3,7$         && $3$  && $4$      &\cr \Tablespace
}
\hrule}
}

\sn

\centerline{Table 3.6d: Irreducible representations of $\Spl{2}{4}$ }
\smallskip\noindent
\centerline{
\vbox{ \offinterlineskip
\def\Tablespace{ height2pt&\omit&&\omit&&\omit&&\omit&\cr }
\def\Tablerule{ \Tablespace
                \noalign{\hrule}
                \Tablespace      }
\hrule
\halign{&\vrule#&
  \strut\quad\hfil#\hfil\quad\cr
\Tablespace
& type of rep. &&  && dim  && \# &\cr
\Tablerule
& $D_4(\chi)$ && $\chi\in\P$                            && $24$  &&  $2$
&\cr \Tablerule
& $N_4(\chi)$ && $\chi\in\P$                            &&  $8$  &&  $6$
&\cr \Tablerule
& $R_4^0(r,t,\chi)$ &&
  $\chi\in\P;\ \chi\not\equiv 1;\ r=1,3;\ t=1,5$        &&  $6$  &&  $4$
&\cr \Tablerule
& $R_4^0(r,t,\chi)_\pm$ &&
  $\chi\in\P;\ \chi^2\equiv 1;\ r=1,3;\ t=1,5$            &&  $3$  && $16$
&\cr \Tablerule
& $R_4^0(1,t,\chi)_\pm$ && $\chi\in \P;\ t=3,7$         &&  $6$  &&  $8$
&\cr \Tablerule
& $R_4^2(r,t,\chi)$ &&
  $\chi\not\equiv 1;\ r,t \in \{1,3\}$                  &&  $6$  &&  $4$
&\cr \Tablerule
& $C_2 \otimes R_4^2(r,3,\chi)$ &&
  $\chi \not\equiv 1;\ r=1,3$                           &&  $6$  &&  $2$
&\cr \Tablerule
& $R_4^2(r,3,\chi_1)_1$ && $r=1,3$                      &&  $6$  &&  $2$
&\cr \Tablerule
& $N_3(\chi)_+\otimes R_4^0(1,7,\psi)_+$ &&
  $\chi\in\P;\ \chi^2\equiv1;\ \psi \not\equiv 1;$      && $12$  &&  $2$
&\cr \Tablespace
&\omit &&
   $\psi^2\equiv 1;\ \psi(-1) = 1$  && \omit && \omit
&\cr \Tablespace
}
\hrule}
}
\vfill\eject
For $\lambda=5$ there are 92 irreducible representations (see Table 3.6e).
Here for fixed $r=1,3$ the 2 irreducible representations of type
$R_5^2(\cdot,1,\chi)_1$ ($ \chi\not\in\P$) are given by the 2
two dimensional irreducible level 5 subrepresentations of
$R^2_5(r,1)$.

\centerline{Table 3.6e: Irreducible representations of $\Spl{2}{5}$ }
\smallskip\noindent
\centerline{
\vbox{ \offinterlineskip
\def\Tablespace{ height2pt&\omit&&\omit&&\omit&&\omit&\cr }
\def\Tablerule{ \Tablespace
                \noalign{\hrule}
                \Tablespace      }
\hrule
\halign{&\vrule#&
  \strut\quad\hfil#\hfil\quad\cr
\Tablespace
& type of rep. &&  && dim && \# &\cr
\Tablerule
& $D_5(\chi)$ && $\chi\in \P$            && $48$  &&  $4$      &\cr \Tablerule
& $N_5(\chi)$ && $\chi\in\P$             && $16$  && $12$      &\cr \Tablerule
& $R_5^0(r,t,\chi)$ &&
  $\chi\in\P;\ r=1,3;\ t=1,5$             && $12$  && $16$      &\cr \Tablerule
& $R_5^0(1,t,\chi)_\pm$ &&
  $\chi\in\P;\ t=3,7$                    && $24$  &&  $4$      &\cr \Tablerule
& $R_5^1(r,t,\chi)_\pm$ &&
  $\chi\in\P;\ r,t\in\{1,5\}\ \text{or}$ && $12$  && $16$      &\cr \Tablespace
& \omit && $r=1,3\ \text{and}\ t=3,7$ && \omit && \omit        &\cr \Tablerule
& $R_5^2(r,t,\chi)_\pm$ &&
  $\chi\in\P;\ r=1,3;\ t=1,3,5,7$        &&  $6$  && $32$      &\cr \Tablerule
& $R_5^2(r,1,\chi)_1$ &&
  $\chi\not\in\P;\ r=1,3$                && $12$  &&  $4$      &\cr \Tablerule
& $C_3 \otimes R_5^2(r,1,\chi)_1$ &&
  $\chi\not\in\P;\ r=1,3$                && $12$  &&  $4$      &\cr \Tablespace
}
\hrule}
}
\mn
For $\lambda>5$ there are the following irreducible representations (see
Table 3.6f). Here $\chi$ are always primitive characters and
$R_\lambda^{\lambda-3}(r,t,\chi_{\pm 1})_1$ is the unique irreducible
level $2^\lambda$ subrepresentation of
$R_\lambda^{\lambda-3}(r,t,\chi_{\pm 1})$ which has dimension
$3\cdot 2^{\lambda-4}$.
\mn
\centerline{Table 3.6f: Irreducible representations of $\Spl{2}{\lambda}$ for
                     $\lambda > 5$ }
\smallskip\noindent
\centerline{
\vbox{ \offinterlineskip
\def\Tablespace{ height2pt&\omit&&\omit&&\omit&&\omit&\cr }
\def\Tablerule{ \Tablespace
                \noalign{\hrule}
                \Tablespace      }
\hrule
\halign{&\vrule#&
  \strut\quad\hfil#\hfil\quad\cr
\Tablespace
& type of rep. \footnotemark &&  && dim  && \# &\cr
\Tablerule
& $D_\lambda(\chi)$ && \omit
  && $3 \cdot 2^{\lambda-1}$  &&  $2^{\lambda-3}$       &\cr \Tablerule
& $N_\lambda(\chi)$ && \omit
  && $2^{\lambda-1}$    && $3 \cdot 2^{\lambda-3}$      &\cr \Tablerule
& $R_\lambda^0(1,7,\chi)$ && $t=3,7$
  && $3 \cdot 2^{\lambda-2}$  && $2^{\lambda-3}$       &\cr \Tablerule
& $R_\lambda^\sigma(r,t,\chi)$ &&
  $\cases  r=1,3;\ t=1,5                     & \text{for}\ \sigma=0\\
           r,t\in \{1,5\}\ \text{or}\        & \\
           r=1,3\ \text{and}\ t=3,7         & \text{for}\ \sigma=1\\
           r=1,3;\ t=1,3,5,7                & \text{for}\ \sigma=2\\
   \endcases$
  && $3 \cdot 2^{\lambda-3}$  &&  $5 \cdot 2^{\lambda-2}$      &\cr \Tablerule
& $R_\lambda^\sigma(r,t,\chi)$ &&
  $\sigma=3,\dots,\lambda-3;\ r,t\in \{1,3,5,7\}$
  &&  $3 \cdot 2^{\lambda-4}$
  &&  $4\cdot \sum_{\sigma=3}^{\lambda-3} 2^{\lambda-\sigma}$
  &\cr \Tablerule
& $R_\lambda^{\lambda-2}(r,t,\chi)$ &&
  $r=1,3,5,7;\ t=1,3$
  && $3 \cdot 2^{\lambda-4}$  &&  $16$      &\cr \Tablerule
& $R_\lambda^{\lambda-3}(r,t,\chi_{\pm 1})_1$ &&
  $r=1,3,5,7;\ t= 1,3$
  && $3 \cdot 2^{\lambda-4}$  &&  $16$      &\cr \Tablespace
}
\hrule}
}
\footnotetext{For $\lambda=6$ one has to use representation of type
              $R_6^4(r,t,\chi_1)_1$ and
              $C_2\otimes R_6^4(r,t,\chi_1)_1$ ($r=1,3$)
              instead of those of type
              $R_\lambda^{\lambda-3}(r,t,\chi_{\pm 1})_1$.
              The representations $R_6^4(r,t,\chi_1)_1$ are the unique
              level 6 subrepresentations of  $R_6^4(r,t,\chi_1)$
              with dimension 12.
              }

\vfill\eject
  \subhead
  3.7 Proof of the classification of the strongly-modular fusion algebras
      of dimension less than or equal to four
  \endsubhead

\demo{Proof of the main theorem 1 for $\dim(\Cal F)=2$}

Let $({\Cal F},\rho)$ be a two dimensional strongly-modular fusion
algebra. Lemma 4 implies that $\rho$ is irreducible.
Therefore, we have to consider all irreducible two
dimensional representations of
$\SL$ which factor through a congruence subgroup.
By Lemma 7 we know that these representations can be obtained by
taking the tensor products of all irreducible two dimensional level
$p^\lambda$ representations with all one dimensional representations
of $\SL$.

There are exactly 11 inequivalent irreducible two dimensional
level $p^\lambda$ representations. Their explicit form is given in Appendix
7.1.
We are interested in the classification of the two dimensional
strongly-modular fusion algebras up to tensor products with
one dimensional fusion algebras. Therefore, we can restrict our
investigation to one of the two dimensional representations of level
$2$, $2^3$, $3$ and the two representations of level 5 (see
Appendix 7.1).
For the remaining 5 two dimensional representations
the eigenvalues of the image of $T$ are nondegenerate.
Hence, Lemma 2 implies that the corresponding matrix
representations are unique up to conjugation with
unitary diagonal matrices and permutation of the basis elements.
One can easily apply Verlinde's formula and check whether
the resulting coefficients $N_{i,j}^k$ have integer absolute
values for the two possible choices of
the basis element $\Phi_0$ corresponding to the vacuum
(conjugation with a unitary diagonal matrix does not change
the absolute value of $N_{i,j}^k$).
In particular for the level 2 representation $N_1(\chi_1)$ and the
level 3 representation $N_1(\chi)$ we obtain for both possible
choices of the distinguished basis elements $\Phi_0$ and $\Phi_1$
$$ \vert N_{1,1}^1 \vert =
   \cases \frac2{\sqrt{3}},& \qquad \text{for}\ N_1(\chi_1),\ p = 2 \\
          \frac1{\sqrt{2}},& \qquad \text{for}\ \ N_1(\chi),\ p = 3.
   \endcases$$
Since $\vert N_{1,1}^1\vert $ is not an integer we can
exclude these two representations.
For the level $2^3$ and $5$ representations one obtains integer
values for the $N_{i,j}^k$. Moreover, in all three cases both
possible choices of  the distinguished basis elements $\Phi_0$
and $\Phi_1$ lead to isomorphic fusion algebras.
We conclude that the representation of the modular group given by a
two dimensional strongly-modular fusion algebra is isomorphic to
the tensor product of a one dimensional representation  and
$N_3(\chi)_+$ ($p^\lambda = 2^3$) or $R_1(r,\chi_{-1})$
($r=1,2;p^\lambda = 5$).
Using that $\rho(S^2)$ should be a matrix consisting of
nonnegative integers one can determine the one dimensional
representation of $\SL$ up to an even one dimensional representation.
Therefore, $({\Cal F},\rho)$ is determined
up to tensor products with one dimensional modular fusion algebras.
The resulting representations and fusion algebras are collected
in Table 7.2a.
\qed\enddemo

\demo{Proof of the main theorem 1 for $\dim(\Cal F)=3$}

Let $({\Cal F},\rho)$ be a three dimensional strongly-modular
fusion algebra. By Lemma 7, $\rho$ is either irreducible
or isomorphic to a sum of a two dimensional and a one dimensional
irreducible representation.
We will now consider these two cases separately.
\sn
Firstly, assume that $\rho$ is irreducible. By Lemma 7, $\rho$ is
isomorphic to the tensor product of a one dimensional representation
and one of the three dimensional irreducible level $p^\lambda$
representations. There are exactly 33 inequivalent
irreducible 3 dimensional level $p^\lambda$ representations.
Their explicit form is given in Appendix 7.1.
We are interested in the classification up to tensor products
with one dimensional modular fusion algebras. Therefore, we
can restrict our investigation to a set of irreducible representations which
are not related via
tensor products with one dimensional representations.
This means that we have to consider one representation of
level $3$ and $2^2$, two representations of level $5$ and $7$
and, finally, four representations of level $2^4$  (see Appendix 7.1).

For these representations the eigenvalues of the image
of $T$ are nondegenerate so that we can proceed now as in the proof of
the main theorem 1.

Using Verlinde's formula for the representation
$N_1(1,\chi_1)$ ($p=3$) we obtain $\vert N_{1,1}^1 \vert = \frac12$
for all possible choices of the distinguished basis.
In the same way one finds for $R_1(r,\chi_1)$ ($r=1,2;p=5$) that
$$ \cases \vert N_{1,1}^2 \vert = \frac1{\sqrt{2}}
          & \text{for}\
             \rho(T) = \diag(1,e^{2\pi i \frac{r}5},
                             e^{2\pi i \frac{4r}5}) \\
          & \ \text{or}\
             \rho(T) = \diag(1,e^{2\pi i \frac{4r}5},
                             e^{2\pi i \frac{r}5}) \\
         \vert N_{1,1}^1 \vert = \frac1{\sqrt{2}}
          & \text{for}\
            \rho(T) =  \diag(e^{2\pi i \frac{r}5},1,
                             e^{2\pi i \frac{4r}5}) \\
          \vert N_{1,1}^1 \vert = \frac1{\sqrt{2}}
          & \text{for}\
            \rho(T) =  \diag(e^{2\pi i \frac{4r}5},1,
                             e^{2\pi i \frac{r}5}). \\
\endcases $$
Here the different cases correspond to the different
possible choices of the distinguished basis.
We conclude that $\rho$ cannot be isomorphic to a tensor product of
a one dimensional representation and $N_1(1,\chi_1)$ ($p=3$)
or $R_1(r,\chi_1)$ ($r=1,2;p=5$).

An analogous calculation shows that for the representations of type
$R_1(r,\chi_{-1})$ one has $\vert N_{i,j}^k\vert \in \N$
for all 3 possible choices of the distinguished basis.
For the remaining representations one also
has $\vert N_{i,j}^k\vert \in \N$ for the two possible
choices of the distinguished basis (here the matrix $\rho(S)$
contains a zero so that there are only two possible choices
of the distinguished basis).

Hence, $\rho$ is isomorphic to a tensor product of a one
dimensional representation with one of these 7 representations.
Using that for a modular fusion algebra $\rho(S^2)_{i,j}$
equals $N_{i,j}^0$ one can determine the possible one dimensional
representations. The corresponding strongly-modular fusion
algebras are contained in Table 10 in the second and third row.

\mn
Secondly, assume that $\rho$ decomposes into a direct sum of
two irreducible representations $\rho \cong \rho_1 \oplus \rho_2$
with $\dim(\rho_j) = j$. Then $\rho_2$ is isomorphic to the
tensor product of a one dimensional representation with one of
the two dimensional irreducible level $p^\lambda$ representations
contained in Table 7.1a.

Using Lemma 1 we conclude that $\rho(T)$ has degenerate
eigenvalues so that $\rho_2(T)$ must have an eigenvalue of the
form $e^{2\pi i \frac{n}{12}}$.
Hence, $\rho_2$ cannot be isomorphic to the tensor product of a one
dimensional representation and one of the two dimensional
irreducible level $5$ and $2^3$ representations in Table 7.1a.
Using once more that $\rho(T)$ has degenerate eigenvalues
we obtain that $\rho$ is isomorphic to the tensor product of a
one dimensional representation with either $N_1(\chi_1)\oplus C_j$
($j=1,2;p=2$) or $N_1(\chi)\oplus B_j$ ($j=2,3;p=3$).
In order find out whether these four representations are
admissible we have to look for distinguished bases.

Let us first consider the case
$\rho \cong C\otimes(N_1(\chi)\oplus B_j)$
($j=2,3;p=3$) where $C$ is a one dimensional representation.
Here $\rho(S^2)$ has two different eigenvalues since $N_1(\chi)$
is odd and the representations $B_j$ are even.
Since the vacuum is selfconjugate, i.e.\ $\rho(S^2)_{0 0} = 1$
the representation $C$ has to be odd. Without loss of generality
we choose $C=C_4$ for $j=2$ and $C=C_3$ for $j=3$.
Furthermore, the fact that $\rho(S^2)$ has two different eigenvalues
implies that we must have
$$ \rho(S^2) = \pmatrix 1 &0 &0 \\ 0 &0 &1 \\ 0 &1 &0 \endpmatrix.$$
Using these two conditions it follows that in a basis in which $\rho(S^2)$
has this form and $\rho(T)$ is diagonal we must have
$$  \rho(S) = \frac1{\sqrt{3}}
        \pmatrix \epsilon & \epsilon           & \epsilon \\
                 \epsilon & e^{2\pi i \frac13} & e^{2\pi i \frac23} \\
                 \epsilon & e^{2\pi i \frac23} & e^{2\pi i \frac13}
       \endpmatrix,\qquad \epsilon^2 = 1
$$
and
$$\rho(T) = \cases \diag(e^{2\pi i\frac5{12}},e^{2\pi i\frac1{12}},
                         e^{2\pi i\frac1{12}}) &\text{or}  \\
                   \diag(e^{2\pi i\frac7{12}},e^{2\pi i\frac{11}{12}},
                        e^{2\pi i\frac{11}{12}}) &\\
            \endcases
$$
up to conjugation with a unitary diagonal matrix
(the two possibilities for $\rho(T)$ correspond  to
the two possible choices of the distinguished basis).

Applying now Verlinde's formula leads to a modular fusion algebra
iff $\epsilon=1$ for both choices of the distinguished basis.
The corresponding fusion algebra, $\rho(S)$ and
$\rho(T)$ are listed in the third row of Table 7.2a.

Finally, consider the case
$\rho \cong C\otimes(N_1(\chi_1)\oplus C_j)$ ($j=1,2$).
Since $N_1(\chi_1)$ ($p=2$) and $C_j$ ($j=1,2$) are even $\rho$ has to be even,
too. Therefore, $C$ is even and w.l.o.g. we choose
$C=C_1$ for $j=1$ and $C=C_2$ for $j=2$.
Since $\rho$ is even one must have $\rho(S^2)= \id$ and,
therefore, $\rho(S)$ is real (c.f.\ the second remark in \S2.2).
Plugging this in we find (up to permutation of the basis elements)
that
$$\rho(S) = \frac12 \pmatrix 1          & -\sqrt{3}a & \sqrt{3}b \\
                             -\sqrt{3}a & 2-3a^2     & 3ab       \\
                              \sqrt{3}b & 3ab        & 3a^2-1
                    \endpmatrix, \qquad
  \rho(T) = (-1)^j \diag(1,-1,-1)
$$
where $a,b \in \R$ and $a^2+b^2 =1$.
Using Verlinde's formula we obtain as conditions for $\rho$ to
be admissible
$$ \cases  \frac{(1-3a^2)(3a^2-2)}{\sqrt{3}a} \in \N
           & \text{for}\ \rho(T) = (-1)^j \diag(1,-1,-1) \\
           \frac{1}{\sqrt{3} a (3a^2-2)},
           \frac{3a^2-1}{\sqrt{3} a (3a^2-2)} \in \N
           & \text{for}\ \rho(T) = (-1)^j \diag(-1,-1,1). \\
  \endcases
$$
The first case implies that $a^2 = \frac13$ or $a^2 = \frac23$
and the second one $a^2 = \frac13$, respectively. Inserting these values of
$a$ in the explicit form of $\rho(S)$ above we indeed obtain
modular fusion algebras if we choose the signs of $a$ and $b$
correctly. The resulting modular fusion algebras are contained
in the fifth row of Table 7.1a. As fusion algebras they are
of type "$(3,4)$", also called Ising fusion algebra.

This completes the proof in the three dimensional case.
\qed\enddemo

\demo{Proof of the main Theorem 1 for $\dim(\Cal F)=4$}

Let $({\Cal F},\rho)$ be a strongly-modular fusion algebra.
Then, by Lemma 4, we have the following possibilities for
$\rho$:
\roster
\item
$\rho$ is irreducible,
\item
$\rho\cong \rho_1\oplus\rho_2$ with $\dim(\rho_1) =3$, $\dim(\rho_2) = 1$,
\item
$\rho\cong \rho_1\oplus\rho_2$ with $\dim(\rho_1) = \dim(\rho_2) = 2$,
\item
$\rho\cong \rho_1\oplus\rho_2\oplus\rho_3$ with $\dim(\rho_1) =2$,
$\dim(\rho_2) = \dim(\rho_3) = 1$
\endroster
where $\rho_i$ ($i=1,2,3$) are irreducible representations.
\mn
\leftline{ $ \underline{ \text{(1) $\rho$ is irreducible}}$}
Assume that $\rho$ is irreducible. Then $\rho$
is either isomorphic to the tensor product of 2 two
dimensional representations of coprime levels  or it is
isomorphic to the tensor product of a one dimensional representation
with a four dimensional irreducible level $p^\lambda$ representation.
In the first case we obviously  have that $\rho$ is only admissible
iff both two dimensional representations are admissible (look at
Table 7.1a). In this case the corresponding modular fusion algebra
is a tensor product of two modular fusion algebras contained in Table 7.2a.
Let us now consider the other case, namely that
$\rho \cong C\otimes \rho_1$ where $C$ is a one dimensional
representation and $\rho_1$ is a four dimensional irreducible level
$p^\lambda$ representation.
In this case $\rho_1$ is given by one of the 9 representations
in Table 7.1c. Note that for all of these representations the
eigenvalues of the image of $T$ are nondegenerate so that we
can use the argumentation used in the proof of the main theorem 1
for $\dim({\Cal F})=2$.

For the representation $N_1(\chi)$ ($\chi^3\not\equiv 1;p=5$) we
find by Verlinde's formula
$$ \vert N_{1,1 }^1 \vert  = \sqrt{3},\qquad \text{for}\ \
   \rho(T) = \diag( e^{2\pi i \frac{n}5},
                    e^{2\pi i \frac{3n}5},
                    e^{2\pi i \frac{2n}5},
                    e^{2\pi i \frac{4n}5})
   \qquad (n=1,\dots,4)
$$
where again the different possibilities for $\rho(T)$ correspond
to the different possible distinguished basis.
This shows that $\rho_1$ cannot be isomorphic to this representation.

Since the representation $N_1(\chi)$ ($\chi^3\equiv 1;p=5$) is
isomorphic to the tensor product of the two different level 5
representations in Table 7.1a it is clear that this representation is
admissible. Since the image of $T$ under this representation has nondegenerate
eigenvalues the corresponding modular fusion algebras
are isomorphic to the tensor product of 2 two dimensional modular
fusion algebras (as fusion algebras they are of type "$(2,5)$").

Consider now the representations $R_1(r,\chi_1)$ ($r=1,2;p=7$).
Here Verlinde's formula  implies that
$$ \vert N_{1,1}^1 \vert = \frac1{\sqrt{2}}\quad\text{for}\quad
   \rho(T) = \diag(e^{2\pi i \frac{n}7},1,\cdot,\cdot)\quad (n=1,\dots 6) $$
and
$$ \vert N_{1,1}^2 \vert  = \frac1{\sqrt{2}}\quad\text{for}\quad
   \rho(T) = \cases
     \diag(1,e^{2\pi i \frac27},e^{2\pi i \frac47},e^{2\pi i \frac17}) &
        \text{or} \\
     \diag(1,e^{2\pi i \frac57},e^{2\pi i \frac37},e^{2\pi i \frac67}).
    \endcases
$$
As above this removes these representations from the list of candidates
leading to modular fusion algebras.

For the representation  $N_3(\chi)$ ($\chi^3\not\equiv1; p=2^3$) one has
$$ \vert N_{1,1}^1 \vert = \sqrt{\frac43} \quad \text{for}\quad
   \rho(T) = \diag(e^{2\pi i \frac{2n+1}8},e^{2\pi i \frac{2n+5}8},\cdot,\cdot)
   \quad (n=1,\dots,4)
$$
so that this representation is also excluded.

Consider now the representations $R^1_2(r,1,\chi)$
($r=1,2; \chi^3\not\equiv 1; p=3^2$).
Here one has
$$ \vert N_{1,1}^1 \vert = \frac1{\sqrt3} \quad\text{for}\quad
   \rho(T) = \diag(e^{2\pi i \frac{r n^2}9},e^{2\pi i\frac{r}3},\cdot,\cdot)
   \quad (n=1,2,3).
$$
The basis element in the representation space corresponding to the
$\rho(T)$ eigenvalue of order three cannot correspond to $\Phi_0$
since in the corresponding row of $\rho(S)$ contains a zero.

Finally, the only remaining four dimensional irreducible level $p^\lambda$
representations that might lead to modular fusion algebras are those of type
$R_2^1(r,1,\chi)$ ($r=1,2; \chi^3\equiv 1;p^\lambda=3^2$).
Indeed, these representations lead to modular fusion algebras.
To be more precise one has to consider the tensor product of an odd
one dimensional representation with them because the $R_2^1(r,1,\chi)$
($\chi^3\equiv 1$) are odd themselves.
The corresponding fusion algebras are of type "$(2,9)$"
and the explicit form is given in Table 7.2b.
The different modular fusion algebras result from the two
different representations and the fact that the distinguished
basis can be chosen in different ways.

\mn
\leftline{$\underline{\text{ $\rho\cong \rho_1\oplus\rho_2$ with
              $\dim(\rho_1) =3$, $\dim(\rho_2) = 1$}}$}
Assume that $\rho$ is isomorphic to the direct sum of a
one dimensional and an irreducible three dimensional representation.
Then one has $\rho \cong C\otimes ( \rho_1 \oplus D )$ where
$C$ and $D$ are one dimensional representations and $\rho_1$ is one
of the three dimensional irreducible level $p^\lambda$ representations
in Table 7.1b. By Lemma 1 we know that $\rho(T)$ has degenerate
eigenvalues. Therefore, $\rho_1$ is of type $N_1(\chi_1)$ ($p=3$),
$R_1(r,\chi_1)$ ($r=1,2;p=5$), $D_2(\chi)_+$ ($p^\lambda=2^2$) or
$R_3^0(1,3)_{\pm}$ ($p^\lambda=2^3$).

Consider first the representation $N_1(\chi_1)$ ($p=3$).
In this case we can have $D=B_j$ ($j=1,2,3$).
Since $B_j$ and  $N_1(\chi_1)$ are even we can choose without loss of
generality $C=C_1$.
Using Verlinde's formula we find that
$$ \vert N_{1,1}^1 \vert = \frac12 \quad\text{for}\quad
   \rho(T) = \diag( e^{2\pi i \frac{j+1}3}, e^{2\pi i \frac{j+2}3},
                    e^{2\pi i \frac{j}3}, e^{2\pi i \frac{j}3})
$$
giving a contradiction for these choices of the distinguished basis.
For $\rho(T) = \diag( e^{2\pi i \frac{j}3},
                      e^{2\pi i \frac{j}3},
                      e^{2\pi i \frac{j+1}3},
                      e^{2\pi i \frac{j+2}3})$
the line of reasoning is a little bit more involved.
Here $N_{i,j}^0 = \rho(S^2)_{i,j} = \delta_{i,j}$ implies that
$\rho(S)$ is given by
$$ \rho(S) = \frac13\pmatrix 4b^2-1 &    4ab &  2a  &  2a \\
                                4ab & 3-4b^2 & -2b  & -2b \\
                                 2a &    -2b &  -1  &   2 \\
                                 2a &    -2b &   2  &   -1
                    \endpmatrix
$$
up to conjugation with an orthogonal diagonal matrix,
with $a,b \in \R$ and $a^2+b^2=1$.
With the explicit form of $\rho(S)$ we find as conditions for
$\rho$ to be admissible
$$  N_{1,1}^1 = \frac1{2a(3-4a^2)} \in \Z,\qquad
    N_{1,1}^2 = \frac{2a^2-1}{2a(3-4a^2)} \in \Z.
$$
However, the only solutions  that satisfy these two conditions
are those $a$ which equal $\frac1{2m}$ for an integer $m$ and
satisfy  $m^3 \equiv 0 \bmod 3m^2-1 $. It follows that
$ m \equiv 0 \bmod 3m^2-1 $ which gives a contradiction.
Therefore, the representations $N_1(\chi_1)\oplus B_j$ ($p=3$)
do not lead to modular fusion algebras.

Next we consider the representations $R_1(r,\chi_1)$ ($r=1,2;p=5$).
In this case the one dimensional representation $D$ has to be the
trivial one. Since these two representations are even we can
choose without loss of generality $C=C_1$, too. Using that $N_{i,j}^0 =
\delta_{i,j}$
we find that the matrix which describes the basis in the two
dimensional eigenspace corresponding to the eigenvalue
1 of $\rho(T)$ is orthogonal. Furthermore, by looking at suitable
$N_{i,j}^k$ we find that there are only two possibilities for
this matrix. In the corresponding basis we indeed find modular
fusion algebra given by the tensor product of two modular fusion
algebras of type "$(2,5)$".
That $\rho$ is admissible can also be interfered from the equality
$R_1(r,\chi_1)\oplus C \cong R_1(r,\chi_{-1}) \otimes
 R_1(r,\chi_{-1})$ ($r=1,2;p=5$).

Finally, we have to consider $D_2(\chi)_+$ ($p^\lambda=2^2$)
and $R_3^0(1,3,\chi)_{\pm}$ ($p^\lambda = 2^3$).
The corresponding possibilities for $\rho$ are
$ C_3 \otimes D_2(\chi)_+ \oplus C_j $ ($j=1,3,4$),
$ C_4 \otimes R_3^0(1,3,\chi)_+ \oplus C_3 $ or
$ C_3 \otimes R_3^0(1,3,\chi)_- \oplus C_4 $.
For the case  $\rho \cong  C_3 \otimes D_2(\chi)_+ \oplus C_1$
we obtain a modular fusion algebra given by the tensor product
of two $\Z_2$ fusion algebras. This can also be seen by looking
at the identity
$$C_3 \otimes D_2(\chi)_+ \oplus C_1 \cong
  D_2(\chi)_+ \otimes D_2(\chi)_+.$$
For $ C_4 \otimes R_3^0(1,3,\chi)_+ \oplus C_3 $ or
$ C_3 \otimes R_3^0(1,3,\chi)_- \oplus C_4 $ we obtain
$\Z_4$ type fusion algebras (see Table 7.2b).
The other two representations
($ C_3 \otimes D_2(\chi)_+ \oplus C_j $ ($j=3,4$))
are not admissible as one can easily check by applying
Verlinde's formula.

\mn
\leftline{$\underline{\text{$\rho\cong \rho_1\oplus\rho_2$ with
              $\dim(\rho_1) = \dim(\rho_2) = 2$}}$}
Assume that $\rho$ decomposes into a direct sum of 2
two dimensional irreducible representations.
In this case we have
$\rho = C \otimes (\rho_1 \oplus D \otimes \rho_2)$ where
$C$ and $D$ are  one dimensional representations and $\rho_1,\rho_2$
are some level $p^\lambda$ representations contained in Table 7.1a.
Since $\rho$ is reducible we know that $\rho(T)$ has
degenerate eigenvalues. This together with the parity of
the representations in Table 7.1a implies that $\rho$ equals
(up to a tensor product with an even one dimensional representation)
one of the following representations:
$$ \align
  & N_1(\chi_1) \oplus N_1(\chi_1)  \\
  &C_3 \otimes (N_1(\chi)   \oplus B_i\otimes N_1(\chi)) \quad
    (i=1,2) \\
  &C_4 \otimes (R_1(r,\chi_{-1})\oplus R_1(r,\chi_{-1})) \quad
    (r=1,2)\\
  &C_4 \otimes (N_3(\chi)_+ \oplus N_3(\chi)_+). \\
\endalign $$
In all cases we have that $\rho(S)$ is conjugate to a matrix
of block diagonal form. More precisely, this matrix consists
of two identical two by two matrices. A simple calculation
shows now that conjugation of $\rho(S)$ with a matrix which
leaves $\rho(T)$ diagonal leads to a matrix which has at least one
zero element in every row.
This is a contradiction since we have assumed that
$\rho$ is admissible and one can apply Verlinde's formula.

\mn
\leftline{$\underline{\text{ $\rho\cong \rho_1\oplus\rho_2\oplus\rho_3$
              with $\dim(\rho_1) =2$,
              $\dim(\rho_2) = \dim(\rho_3) = 1$}}$}
Assume that $\rho$ decomposes into a direct sum of an
irreducible two dimensional and 2 one dimensional representations.
Then, again by Lemma 1, $\rho(T)$ has degenerate eigenvalues
and a simple parity argument shows that the only possibilities
for $\rho$ are (up to a tensor product with an even one
dimensional representation):
$$ N_1(\chi_1) \oplus C_1 \oplus C_1 \qquad \text{or} \qquad
   N_1(\chi_1) \oplus C_1 \oplus C_2 $$
where $N_1(\chi_1)$ is the level 2 representation in Table A1.
We have to consider these two cases separately.

Firstly, let $\rho$ be conjugate to  $N_1(\chi_1) \oplus C_1 \oplus C_1$.
Then the requirements that $\rho(S)$ has to be symmetric and real
and that $\rho(T)$ has to be diagonal imply that
(up to permutation of the basis elements and conjugation with an
orthogonal diagonal matrix):
$$ \rho(S) = -\frac12
   \pmatrix
             -1&  \sqrt{3} a &  \sqrt{3} b &  \sqrt{3} c \\
    \sqrt{3} a & 3 a^2 - 2   &  3 a b      &  3 a c      \\
    \sqrt{3} b &  3 a b      &  3 b^2 - 2  &  3 b c      \\
    \sqrt{3} c &  3 a c      &  3 b c      &  3 c^2 - 2  \\
   \endpmatrix
$$
where $a,b,c\in \R$ with $a^2+b^2+c^2=1$ and
$\rho(T) = \diag(-1,1,1,1)$.

Fixing the distinguished basis such that $\Phi_0$ corresponds to the
eigenvector of $\rho(T)$ with eigenvalue $-1$  we obtain
$$ \align
   N_{1 1}^1 &= \frac{(2-3a^2)(1-3a^2)}{\sqrt{3}a},\ \
   N_{2 2}^2  = \frac{(2-3b^2)(1-3b^2)}{\sqrt{3}b},\ \
   N_{3 3}^3  = \frac{(2-3c^2)(1-3c^2)}{\sqrt{3}c} \\
   N_{1 1}^2 &= \sqrt{3} (3a^2-1)b,\qquad
   N_{1 1}^3  = \sqrt{3} (3a^2-1)c \\
   N_{2 2}^1 &= \sqrt{3} (3b^2-1)a,\qquad
   N_{2 2}^3  = \sqrt{3} (3b^2-1)c.
\endalign $$
This implies that $a^2 = b^2 = c^2 = \frac13$. The resulting
structure constants indeed define a fusion algebra, namely
the tensor product of two fusion algebras of type $\Z_2$.
As a modular fusion algebra this fusion algebra is
{\bf simple}, i.e.\ it is not a tensor product of two
nontrivial modular fusion algebras. The resulting modular
fusion algebra is contained in Table 7.2b.

For the other choice of the distinguished basis where
$\Phi_0$ corresponds to an eigenvector $\rho(T)$ with
eigenvalue 1 we find
$$\align
 N_{3 3}^1 &= \frac{(3a^2-1)b}{a(3a^2-2)},\qquad
 N_{3 3}^2 = \frac{(3a^2-1)c}{a(3a^2-2)}, \\
 N_{3 3}^3 &= \frac{3a^2-1}{\sqrt{3}a(3a^2-2)},\qquad
 N_{2 2}^3 = \frac{1-3b^2}{\sqrt{3}a(3a^2-2)}
\endalign
$$
where the basis was chosen such that $\rho(T) = \diag(1,1,1,-1)$.
Let now $n := (N_{3 3}^1)^2 + (N_{3 3}^2)^2$ and
$m :=  (N_{3 3}^3)^2$.
It is now easy to verify that $n$ and $m$ satisfy the equation
$$ m^3 + (1-5 n) m^2 +(4 n^2+7 n) m + 4 n^2 - 3 n^3 = 0.$$
By Lemma 10 in \S3.8 below the only nonnegative integer
solution of this equation with $m$ being a square is given by $n=m=0$.
Therefore, we find as the only possible solution $a^2=b^2=c^2=\frac13$.
The resulting structure constants define a fusion algebra
isomorphic to the tensor product of two $\Z_2$ fusion algebras.
However, analogous to the case of the other distinguished basis
discussed above this modular fusion algebra is {\bf simple}
and contained in Table 7.2b.

Secondly, assume that $\rho$ is conjugate to
$N_1(\chi_1) \oplus C_1 \oplus C_2$.
Requiring that $\rho(S)$ is a symmetric real matrix and
that $\rho(T)$ is diagonal implies
(up to a permutation of the basis elements and conjugation
with an orthogonal diagonal matrix)
$$ \rho(S) = \frac12
\pmatrix
       3b^2-1 &        -3 a b & -\sqrt{3}  a c & \sqrt{3} a d \\
       -3 a b &        3a^2-1 & -\sqrt{3} b c  & \sqrt{3} b d \\
-\sqrt{3} a c & -\sqrt{3} b c &        3c^2-2  & -3 c d \\
 \sqrt{3} a d &  \sqrt{3} b d & -3 c d         & 3d^2-2\\
\endpmatrix
$$
where $a,b,c,d \in \R$ and $a^2+b^2 =1, c^2 + d^2 = 1$ and
$\rho(T) = \diag(1,1,-1,-1)$.
Using Verlinde's formula we obtain for the choice of the distinguished
basis in which $\Phi_0$ corresponds to the eigenvector of
$\rho(T)$ with eigenvalue 1
$$ ( N_{1 1}^1 )^2 =  \frac{(3a-1)^2(6a-5)^2}{9a^2(1-a^2)(3a-2)^2},\ \
   ( N_{1 1}^2 )^2 =  \frac{c^2}{3a^2(3a^2-2)},\ \
   ( N_{1 1}^3 )^2 =  \frac{d^2}{3a^2(3a^2-2)}.
$$
For the other choice of the distinguished basis
($\Phi_0$ corresponding to eigenvalue $-1$) one finds the
same expressions with $a$ and $c$ exchanged.

Let $n := ( N_{1 1}^2 )^2 + ( N_{1 1}^3 )^2$ and let
$m := ( N_{1 1}^1 )^2$. It is easy to verify that the
following equation for $n$ and $m$ holds true
$$ \align
   &(1-3n) m^3 +(12-37n+31n^2) m^2 + (48-152n+155n^2-53n^3) m \\
   &\quad    + 64 -208n +249n^2 -130n^3 + 25n^4 = 0.
\endalign
$$
By Lemma 10 in \S3.8 below the only nonnegative integer
solution of this equation with $m$ being a square is given by  $m=0,n=1$.
This is a contradiction to the explicit form of $n$ and $m$
in terms of $a$ above.  Hence the representation
$N_1(\chi_1)\oplus C_1 \oplus C_2$ is not admissible.

This proves the main theorem 1.
\qed\enddemo

\vfil\eject
\subhead
3.8 Proof of a Lemma on diophantic equations
\endsubhead

\proclaim{Lemma 10 \footnotemark{}}
Let $n$ be a nonnegative integer, $m$ a square of an integer
and $n,m$ solutions of
\roster
\item
$ m^3 + (1-5n) m^2 +(4n+7n^2) m + 4n^2-3n^3 = 0$
or
\item
$(1-3n) m^3 +(12-37 n+31 n^2) m^2 + $
$ (48-152 n+155 n^2-53 n^3) m $
$+ 64 -208 n +249 n^2 -130 n^3 + 25 n^4 = 0$
\endroster
Then either $n=m=0$ for $(1)$ or  $m=0,n=1$ for $(2)$.
\endproclaim
\footnotetext{ I would like to thank D.\ Zagier
for discussion on this lemma \cite{Za}}
\demo{Proof}
Firstly, consider the equation (1). It can be written in the form
$$ (3n-m)(m-n)^2 = (m+2n)^2.$$
If $n=m$ then $m=n=0$. Otherwise, set
$t = \frac{m+2n}{m-n}$ implying
$$ m = \frac{(t+2)t^2}{2t-5},\qquad n = \frac{(t-1)t^2}{2t-5}. $$
If $m$ and $n$ are integral then also $t$ has to be integral
(any prime factor of the denominator of $t$ would divide the
denominator of $m$ and $n$).
Then $N=2t-5$ divides $(t-1)t^2 = \frac18 (N+5)^2(N+3)$ so that
$N$ divides $3\cdot 5^2$. None of the resulting 12 possibilities
leads to a nonnegative integer solution of $n,m$ where
$m\not=n$ and $m$ is a square.

Secondly, consider the equation (2).
Set $k = m-n+4$, then (2) is equivalent to
$$  k^3 + 2 k^2 n - 3 k^3 n + 125 n^2 - 92 k n^2 + 22 k^2 n^2 -
    11 n^3 = 0.$$
If $k=0$ then $n=0$ and $m=-4$ is not a square. Otherwise, (2)
is equivalent to
$$ (-3 t+22 t^2) k^2 + (1+2t-92 t^2-11 t^3) k + 125 t^2 = 0,
    \quad k\not=0$$
where $t=\frac{n}{k}$. This equation has discriminant
$  (1 + 18 t + t^2) (1 - 7 t + 11 t^2)^2 $ and this must be a
square. Setting
$\frac{p}{q} := (1- t - (1 + 18 t + t^2)^{1/2})/(10t) \in \Q$
(with coprime $p,q$ and $q>0$) we get
$$ t =  \frac{ q (p + q)}{ p (5p + q)}. $$
Hence, using the quadratic equation in $k$ we finally have
$$ m = \frac{ (2p+q)^2 (p-q)^2}{p^2 (2q-p)^2 (p+q)},\qquad
   n = \frac{q^3}{p^2 (2q-p)}.
$$
The parameterization of $n$ implies that $p=\pm1$ and, furthermore,
that $q^3 \equiv 0 \bmod (2q-p)$. Therefore, we have
$p^3\equiv 0 \bmod (2q-p)$ so that $2q-p=\pm1$.
{}From the resulting four possibilities only $p=q=1$
satisfies the desired properties and leads to $m=0, n=1$.
\qed\enddemo
\remark{Remark}
Note that the proof of Lemma 10 relies essentially on the fact that
the curves defined by the two above equations are rational.
\endremark

 \subhead
  3.9 Proof of the classification of the nondegenerate strongly-modular
      fusion algebras of dimension less than 24
  \endsubhead

\demo{Proof of the main theorem 2}

Let $({\Cal F},\rho)$ be a simple nondegenerate strongly-modular fusion algebra
of dimension less than 24. Lemma 1 implies that $\rho$ is irreducible.
Furthermore, since $({\Cal F},\rho)$ is strongly-modular we have
to consider all irreducible representations of $\Spl{N}{}$ of dimension
less than 24. Since $({\Cal F},\rho)$ is simple and nondegenerate
simple Lemma 7 shows that we can restrict our investigation to
irreducible representations of $\Spl{p}{\lambda}$.
Once again, since $({\Cal F},\rho)$ is nondegenerate we can follow
the line of reasoning in the proof of the main theorem 1 for the two
dimensional case.

Therefore, we can directly apply Verlinde's formula to any such matrix
representation $\hat \rho$ and look  whether the resulting coefficients
$N_{i,j}^k$ have integer absolute values for the different choices
of the basis element corresponding to $\Phi_0$.
If the resulting numbers $N_{i,j}^k$ do not have integer absolute
values we can conclude that there exists no nondegenerate
strongly-modular fusion algebra $({\Cal F},\rho)$ where $\rho$ is
conjugate to the tensor product of a one dimensional representation
of $\SL$ and $\hat \rho$.
We have investigated this for all irreducible representations
of $\Spl{p}{\lambda}$ of dimension less than 24 by constructing them
explicitly\footnotemark{}.
\footnotetext{Here we have used the computer algebra system
PARI-GP \cite{GP}.}

The proof of the theorem will consist of three separate cases:
We consider representations of  $\Spl{p}{}$ and $\Spl{p}{\lambda}$
and $\Spl{2}{\lambda}$ separately.
\mn
Firstly, let $\rho$ be isomorphic to a tensor product of a one dimensional
representation and an irreducible representation $\hat \rho$ of $\Spl{p}{}$
($p\not=2$).
Note that this case was already discussed in \cite{E$2$}.

For the representations of type $D_1(\chi)$ the matrix $\rho(T)$ has
degenerate eigenvalues so that we can leave out this type of representation.

For the representations of type $N_1(\chi)$ we find
modular fusion algebras only for $p=5,11,17$ and $23$ and $\chi^3 \equiv 1$.
For $p=5$ the modular fusion algebra is not simple but equals
the tensor product of two modular fusion algebras where the
corresponding fusion algebras are of type "$(2,5)$"
(cf.\ also the proof of the main theorem 3).
The modular fusion algebras corresponding to $p=11,17,23$ are
contained in the last three rows of Table 7.3.
As was already mentioned in \cite{E$2$} these four representations
are probably the only admissible ones of type $N_1(\chi)$.
However, we do not have a proof of this statement but numerical checks
show that there is no other admissible representation of this type for
$p<167$~\cite{E$2$}.

The representations of type $R_1(r,\chi_1)$ and $N_1(\chi_1)$ do not lead
to modular fusion algebras \cite{E$2$}.

For all $\hat \rho$ of type $R_1(r,\chi_{-1})$ we obtain modular fusion
algebras. Here $\rho \cong  (C_4)^{\frac{p+1}2} \otimes R_1(r,\chi_{-1})$
is admissible for all odd primes $p$.
The corresponding modular fusion algebras are of type "$(2,p)$".
They are contained in the third row of Table~7.3.
\mn
Secondly, let  $\rho$ be isomorphic to a tensor product of a one dimensional
representation and a irreducible representation $\hat \rho$ of
$\Spl{p}{\lambda}$ ($p\not=2,\ \lambda>1$).

For the representations of type $D_\lambda(\chi)$ the matrix
$\rho(T)$ has degenerate eigenvalues excluding these representations
from our investigation.

The only representations of type $N_\lambda(\chi)$ which have dimension less
than 24 are those corresponding to $(p=3; \lambda=2,3)$ and $(p=5;\lambda=2)$.
A calculation shows that exactly one of these
representations leads to a modular fusion algebra.
This is the representation with $(p=3;\lambda=2)$ and $\chi^3\equiv1$.
The corresponding strongly-modular fusion algebra is contained in Table~7.3.

Only those representations of type $R_\lambda^\sigma(r,t,\chi)$ and
$R_\lambda(r,\chi_{\pm1})_1$ with $(p=3;\lambda=2,3)$ or $(p=5;\lambda=2)$
have dimension less than 24.
The representations $R_2^1(r,1,\chi)$ $(p^\lambda=3^2; \chi^3\equiv 1)$
lead to nondegenerate modular fusion algebras (cf.\ the proof of the
main theorem 3).
{}From the other representations only those with
$p^\lambda=3^3;r=1,2;\chi^3\equiv1$ lead to modular
fusion algebras (see Table 7.3).

\mn
Thirdly, consider the irreducible representations of $\Spl{2}{\lambda}$.
All irreducible representations of dimension
less than or equal to 4 have been considered in the main theorems 1 to 3.
The corresponding admissible representations with
nondegenerate eigenvalues of $\rho(T)$ are contained in Table 7.3.

For $\lambda=1,2$  all irreducible representations have dimension less
than
or equal to~3.

For $\lambda=3$ we have to consider the representations of
type $R_3^0(1,3,\chi_1)_1$ and $D_3(\chi)_{\pm}$.
The former representation  does not lead to a modular fusion algebra but
the representations $D_3(\chi)_{\pm}$ lead to modular fusion algebras of type
$\Z_2\otimes \text{"}(3,4)\text{"}$. The corresponding  modular fusion
algebras are composite and therefore not contained in Table 7.3.

For $\lambda=4$ only the irreducible representations of type
$R_4^0(r,t,\chi)_{\pm}$, $R_4^2(r,3,\chi_1)_1$ and $R_4^2(r,t,\chi)$
lead to modular fusion algebras. The first one leads to a fusion algebra
of type "$(3,4)$" (see main theorem 2). The other two representations lead
to composite modular fusion algebras. These fusion algebras are of type
$\Z_2\otimes \text{"}(3,4)\text{"}$ and are not contained in Table 7.3.

For $\lambda=5,6$ there are no irreducible representations of dimension less
than 24 leading to modular fusion algebras
(some of them correspond to `fermionic fusion algebras'
of $N=1$-Super-Virasoro minimal models which we do not discuss here).
\qed
\enddemo

\vfill\eject
\head
4. Uniqueness of conformal characters
\endhead

In this section we show that given the central charge
and the finite set of conformal dimensions of certain rational models
the conformal characters are already uniquely determined.
More precisely, we shall state a few  general
and simple axioms which are satisfied by the conformal characters of all
known rational models of $\w$-algebras. These axioms state essentially
not more than the $\SLZ$-invariance of the space of functions spanned by
the conformal characters, the rationality of their Fourier coefficients
and an upper bound for the order of their poles. The only data of
the underlying rational model occurring in these axioms are the central
charge and the conformal dimensions, which give the upper bound for the
pole orders and a certain restriction on the $\SLZ$-invariance.
We then prove that, for various sets of central charges and conformal
dimensions, there is at most one set of modular  functions which satisfies
these axioms (cf\. the main theorem 3 in \S4.1).

In this section we restrict our attention to rational models of $\w$-algebras
where the associated representation $\rho$
turns out to be irreducible. This restriction is mainly
of technical nature: It simplifies the identification of $\rho$.
However, we believe that our main theorem on uniqueness of conformal
characters can be generalized, i.e\. that it can be extended to
rational models with composite $\rho$, possibly with a
slightly more restrictive set of axioms.

We have organized \S4 as follows:
In \S4.1 we state and comment on our main result: The theorem on uniqueness
of conformal characters. The sections \S4.2 and \S4.3, where we develop
the necessary tools needed for the proof of the main theorem 3,
may be of independent interest for those studying representations $\rho$
arising from conformal characters. Finally, in \S4.4 we prove the main
theorem 3.

  \subhead
  4.1 Results on the uniqueness of conformal characters of certain
      rational models
  \endsubhead

\proclaim{Main theorem 3 (Uniqueness of conformal characters)}
Let $c$ be any of the central charges of Table 3.2b or 3.2c, let $H_c$
denote the set of corresponding conformal dimensions, and let
$H$ be a subset of $H_c$ containing $0$. Assume that there exist
nonzero functions $\xi_{c,h}$ ($h\in H$), holomorphic on the upper half
plane, which satisfy the following conditions:
\roster
\item
The functions $\xi_{c,h}$ are modular functions for some
congruence subgroup of $\SL=\SLZ$.
\item
The
space of functions spanned by the $\xi_{c,h}$ ($h\in\ H$)
is invariant under $\SL$ with respect to the action
$(A,\xi)\mapsto\xi(A\tau)$.
\item
For each $h\in H$ one has $\xi_{c,h}=\LO(q^{-\tilde c/24})$
as $\Im(\tau)$  tends to infinity, where $\tilde c = c -24 \min H$.
\item
For each $h\in H$ the function $q^{-{(h-\frac c{24})}}\xi_{c,h}$
is periodic with period 1.
\item
The Fourier coefficients of the $\xi_{c,h}$ are rational numbers.
\endroster
Then $H=H_c$, and, for each $h\in H$, the function $\xi_{c,h}$
is unique up to multiplication by a scalar.
\endproclaim
\vfill\eject
\remark{Remarks}
\roster
\item
The theorem only ensures the uniqueness of the functions
$\xi_{c,h}$ not their existence. However, they do exist.
For Table 3.2b the existence of the corresponding functions is a well-known
fact \cite{CIZ,EFH$^2$NV}: explicit formulas for them can be given
in terms of the Riemann-Jacobi theta series
$$\sum\Sb x\in\Z\\x\equiv\lambda\bmod 2k\endSb\exp(2\pi i\tau x^2/4k).$$
The existence of the functions $\xi_{c,h}$ related to Table 3.2c will be
proved in \S5.
\item
The conformal characters $\chi_{M}$ of a rational model
with $H$ as set of conformal dimensions satisfy the properties listed
under (2) -- (5) by the very definition of rational models and Zhu's
theorem  if we set  $\xi_{c,h}=\chi_{M}$ ($h=$ conformal dimension of
$M$). Property (1) is not part of this definition, and it is not clear
whether it is implied by the axioms for rational models. However,
there are indications that it always holds true (cf\. the discussion below).
\item
If we assume for a rational model corresponding to a row in Table 3.2b
or Table~3.2c that its conformal characters satisfy (1) we can conclude
from our theorem that the corresponding set $H_c$ is exactly
the set of its conformal dimensions and that the properly normalized
functions  $\xi_{c,h}$  ($h\in H_c$) are its conformal characters.
\item
For the proof of the theorem for the five  models of Table 3.2c the
assumption $0\in H$ is not needed, and it can possibly be dropped in
all cases. However, we did not pursue this any further: From the
physical point of view the assumption $0\in H$ is natural since $h=0$
corresponds to the vacuum representation of the underlying $\w$-algebra,
i.e.\ the representation  given by the action of the algebra on itself.
\endroster
\endremark

For the first two cases of Table 3.2c the requirement that the $\xi_{c,h}$
are modular functions on some congruence subgroup is not necessary. Here
we have the
\proclaim{Supplement to the main theorem}
For $c=-\frac85$ and $c=\frac 45$ and with $H_c$ as in Table 3.2c the
equality $H=H_c$ and the uniqueness of the $\xi_{c,h}$ ($h\in H$)
are already implied by properties (2) to (5).
\endproclaim

For the other cases we do not know whether the statement about
the uniqueness of $H$ and the $\xi_{c,h}$ remains true if one also takes
into  account non-modular functions or non-congruence subgroups.

However, as already mentioned, it seems to be reasonable to expect that
the conformal characters associated to rational models satisfy (1).
Support for this is given by the following:

There is no example of a conformal character of any rational model
which is not a modular function on a  congruence subgroup.

As mentioned above the functions $\xi_{c,h}$, whose uniqueness is
ensured by the main theorem, exist. As it turns out they can be
normalized so that their Fourier coefficients are always nonnegative
integers (for the case of Table 3.2c cf\. \S5). This gives
further evidence that they are identical with the conformal characters
of the corresponding $\w$-algebra models whence the latter
therefore satisfy (1).

According to the main theorem 3, for each $H_c$ of Table 3.2b and 3.2c
the $\SL$-module spanned by the $\xi_{c,h}$ is uniquely determined.
In particular the $S$-matrix (i.e\. the matrix representing the action
of $S$ with respect to the basis given by the $\xi_{c,h}$  with the
normalization indicated in the preceding remark) is unique. Closed
formulas for the $S$-matrices  corresponding to the first four rows
of Table 3.2c will be given in \S5.2 (cf\. \cite{ES$2$}).
They can be compared with the
$S$-matrix of the corresponding $\w(2,4)$ rational model with
$c=-\frac{444}{11}$ as numerically computed in \cite{E$1$} using
so-called direct calculations in the $\w$-algebra. Both $S$-matrices
coincide within  the numerical precision.

The last three rational models listed in Table 3.2c are minimal models
of Casimir $\w$-algebras for which formulas for the corresponding
conformal characters have been obtained in \cite{FKW} under the assumption
of a certain conjecture.  Once more, the conformal characters obtained
in this way are modular functions on congruence subgroups
(cf\. Appendix 7.4 and the
discussion in \S5.3).

In the other subsections of \S4 we prove our main theorem. To this end we will
develop some general tools dealing with modular representations, i.e\. with
representations of $\SL=\SLZ$ on spaces of modular functions or forms.
These methods are introduced in the
next two subsections. In \S4.4 we conclude with the proof of the main
theorem 3.

  \subhead
  4.2 A dimension formula for vector valued modular forms
  \endsubhead

In this section we state dimension formulas for spaces of vector valued modular
forms on $\SLZ$.
These formulas are one of
the main tools in
the proof of the main theorem. It is quite natural in the context of conformal
characters, or more generally in the context of modular representations, to ask
for such formulas: The vector $\chi$ whose entries are the conformal characters
of a rational model, multiplied by a suitable power of $\eta$, is exactly what
we shall call a vector valued modular form, and is as such an element of a
finite dimensional space. (The latter holds true at least in the case where the
characters are invariant under a subgroup of finite index in $\SL$; see the
assumptions in the theorem below).

Multiplying $\chi$ by an odd power of $\eta$ yields a vector valued modular
form of half-integral weight. However, because of the ambiguity of the square
root of $c\tau+d$ ($c,d$ being the lower entries of a matrix in $\SL$) we now
do not deal with a vector valued modular form on $\SLZ$ but rather on a certain
double cover $\CSL=\CSLZ$ of this group.

We now make these notions precise.

The double cover $\CSL$ is defined as follows: the group elements are
the pairs $(A,w)$, where $A$ is a matrix in $\SL$ and $w$ is a holomorphic
function on $\PH$ satisfying $w^2(\tau)=c\tau+d$ with $c,d$  the
lower row of $A$. The multiplication of two such pairs is defined by
$$(A,w(\tau))\cdot(A',w'(\tau))=(AA',w(A'\tau)\cdot w'(\tau)).$$

For any $k\in\Z$ we have an action of $\CSL$ on functions $f$ on $\PH$
given by
$$(f|_k(A,w))(\tau)=f(A\tau)\,w(\tau)^{-2k}.$$
Note that for integral $k$ this action factors to an action of $\SL$,
which is nothing else than the usual `$|_k$'-action
of $\SL$ given by $(f|_kA)(\tau)=f(A\tau)(c\tau+d)^{-k}$.

For a subgroup $\Delta$ of $\SL$ we will denote by $D\Delta \subset \CSL$
the preimage of $\Delta$ with respect to the natural projection $\CSL\to\SL$
mapping elements to their first component.

Special subgroups of $D\Gamma$ which we have to consider below are the groups
$$ \Gamma(4m)^{\sharp} = \{ (A,j(A,\tau)) \vert A\in \Gamma(4m) \}.$$
Here, for $A\in\Gamma(4m)$, we use
$$ j(A,\tau) = \vartheta(A\tau)/\vartheta(\tau)$$
where $\vartheta(\tau) = \sum_{n\in\Z} q^{n^2}$.
It is well-known that indeed $j(A,\tau) = \epsilon(A) \sqrt{c\tau+d}$
where $c,d$ are the lower row of $A$ and $\epsilon(A) = \pm 1$.
Explicit formulas for $\epsilon(A)$ can be found in the literature, e.g\.
\cite{Sk}.

We can now define the notion of a vector
valued modular form on $\SL$ or $\CSL$.
\definition{Definition}
For any representation $\rho\colon\CSL\rightarrow \GL{n}$ and any
number $k\in\frac12\Z$ denote by $M_k(\rho)$ the space of all holomorphic
maps $F\colon\PH\to\C^n$ which satisfy $F|_k\alpha=\rho(\alpha)F$ for all
$\alpha\in\CSL$, and which are bounded in any region $\Im(\tau)\ge r>0$.
Denote by $S_k(\rho)$ the subspace of all forms $F(\tau)$ in $M_k(\rho)$
which tend to 0 as $\Im(\tau)$ tends to infinity.
\enddefinition

If $\rho$ is a representation of $\SL$ and $k$ is integral we use $M_k(\rho)$
for $M_k(\rho\circ\pi)$, where $\pi$ is the projection of $\CSL$ onto the first
component. Clearly, in this case the transformation law for the functions $F$
of
$M_k(\rho)$ is equivalent to $F|_kA=\rho(A)F$ for all
$A\in\SL$. In general, if $k$ is integral, the group $\CSL$ may be replaced
by $\SL$ in all of the following considerations.

Finally, for a subgroup $\Delta$ of $\CSL$ or $\SL$ we use $M_k(\Delta)$ for
the space of modular forms of weight $k$ on $\Delta$ in the usual sense. In the
case
$\Delta\subset\SL$ the weight $k$ has of course to be integral.
The reader may not mix the two kinds of spaces $M_k(\rho)$ and $M_k(\Delta)$;
it will always be clear from the context whether $\rho$ and $\Delta$
refer to a representation or a group.

Clearly, if the image of $\rho$ is finite, i.e\. if the kernel
of $\rho$ is of finite index in $\CSL$ then the components of
an $F$ in $M_k(\rho)$ are modular forms of weight $k$ on this
 kernel. In particular, the space $M_k(\rho)$ is then finite
dimensional. Formulas for the dimension of these spaces
can be obtained as follows: Let $V$ be the complex vector space
 of row vectors of length $n=\dim\rho$, equipped with the $\CSL$-right
 action $(z,\alpha)\mapsto z \rho(\alpha)$, where ${()}^t$ means transposition.
The space $M_k(\rho)$ can then be identified with the space
$\Hom_{\CSL}(V,M_k(\Delta))$ of $\CSL$-homomorphisms from $V$
to $M_k(\Delta)$, where $\Delta=\ker\rho$, via the correspondence
$$M_k(\rho)\ni F\mapsto\text{ the map which associates }z\in V
\text{ to }z  \cdot F\in M_k(\Delta).$$
By orthogonality of group characters the dimension of
$\Hom_{\CSL}(V,M_k(\Delta))$ can be expressed in terms of the traces
 of the endomorphisms defined by the action of elements of $\CSL$
on $M_k(\Delta)$. These traces in turn can be explicitly computed
using the Eichler-Selberg trace formula. This way one can derive
the following theorem (cf\. \cite{Sk, pp\. 100} for a complete proof):

\eject
\proclaim{Theorem (Dimension formula \cite{Sk})}
Let  $\rho:\CSLZ\rightarrow\operatorname{GL}(n,\C)$ be a representation with
finite image and such that $\rho((\epsilon^2\id,\epsilon))=\epsilon^{-2k}\id$
for all fourth roots of unity $\epsilon$. and let $k\in\frac12\Z$. Then the
dimension
of $M_k(\rho)$ is given by the following formula
$$\align
\dim M_k(\rho)-\dim S_{2-k}(\overline\rho)=&\frac{k-1}{12}\cdot n
+\frac14\Re\big(\e^{\pi ik/2}\tr\rho((S,\sqrt{\tau}))\big)\\
&+\frac2{3\sqrt3}\Re\big(\e^{\pi i(2k+1)/6}\tr\rho((ST,\sqrt{\tau+1}))\big)\\
&+\frac12a(\rho)
-\sum_{j=1}^n\B_1(\lambda_j).
\endalign$$
Here the $\lambda_j$ ($1\le j\le n$) are complex numbers such that
$\e^{2\pi i\lambda_j}$ runs through the eigenvalues of  $\rho(T)$,
we use $a(\rho)$ for the number of $j$ such that $\e^{2\pi i\lambda_j}=1$,
 and we use $\B_1(x)=x'-1/2$ if $x\in x'+\Z$ with $0<x'<1$, and $\B_1(x)=0$ for
$x$ integral.
Moreover, for $\tau\in\PH$, we use $\sqrt{\tau}$ and $\sqrt{\tau+1}$ for those
square roots which have
positive real parts.
\endproclaim
\remark{Remark}
For $k\ge 2$ the theorem gives an explicit formula for $\dim M_k(\rho)$ since
in this case $\dim(S_{2-k}(\rho))=0$ (the components of a vector valued modular
form are ordinary modular forms on $\ker\rho$, and there exist no nonzero
modular forms of negative weight and no cusp forms of weight 0).
\endremark

For $k=1/2$, $3/2$ and $\ker(\rho)\supset\Gamma(4m)^{\sharp}$ it is still
possible to give an explicit formula for $M_k(\rho)$ \cite{Sk}.
However, we do not need those dimension formulas in full generality
but need only the following corollary:
\proclaim{Supplement to the dimension formula \cite{Sk}}
Let $\rho:\CSLZ\rightarrow\operatorname{GL}(n,\C)$ be an
irreducible representation with $\Gamma(4m)^{\sharp} \subset \ker(\rho)$
for some integer $m$.  Then one has
$ \dim(M_{1/2}(\rho))= 0,1.$
Furthermore, if $\dim(M_{1/2}(\rho))=1$ then
the eigenvalues of $\rho(T)$ are of the form
$e^{2\pi i \frac{l^2}{4m}}$ with integers $l$.
\endproclaim
\remark{Remark}
A complete list of all those representations $\rho$ for which
$\dim(M_k(\rho))=1$ can be found in \cite{Sk}.
\endremark

A proof of this supplement can be found in \cite{Sk}.
It uses a theorem of Serre-Stark describing explicitly the modular forms
of weight $1/2$ on congruence subgroups.

  \subhead
  4.3 Three basic lemmas on representations of $\SLZ$
  \endsubhead

In this section we will prove some lemmas which are useful for identifying
a given representation $\rho$ of $\SL$ if one has certain information about
$\rho$, which can e.g\. -for representations related to rational models-
be easily computed from the central charge and the conformal dimensions.

Assume that the conformal characters of a rational model
are modular functions on some a priori unknown congruence subgroup.
Then the first step for determining the representation $\rho$, given by the
action of $\SL$ on the conformal characters,
consists in finding a positive integer $N$ such that
$\rho$ factors through $\Gamma(N)$.
The next theorem tells us that the optimal choice of $N$ is given by the
order of $\rho(T)$.

\eject

\proclaim{Theorem (Factorization criterion)}
Let $\rho\colon\SL\to\GL{n}$ be a representation, and let $N>0$ be an
integer. Assume that $\rho(T^N)=1$, and, if $N>5$, that the kernel  of
$\rho$ is a congruence subgroup. Then $\rho$ factors through a
representation of $\SL/\Gamma(N)$.
\endproclaim
\demo{Proof}
The kernel $\Gamma'$ of $\rho$ contains the normal hull in $\SL$ of the
subgroup generated by $T^N$. Call this normal hull $\Delta(N)$. By a
result of \cite{Wo} (but actually going back to Fricke-Klein) one has
$\Delta(N)=\Gamma(N)$ for $N\le 5$. If $N>5$ then by assumption we
have $\Gamma'\supset\Gamma(N')$ for some integer $N'$. Thus $\Gamma'$
contains $\Delta(N)\Gamma(NN')$, which, once more by \cite{Wo},
 equals $\Gamma(N)$.
\enddemo

By the last theorem the determination of the representation $\rho$ associated
to a rational model with modular functions as conformal characters
is reduced to the investigation of the finite list of irreducible
representations of $\SL/\Gamma(N) \approx \MG{N}$ with some easily computable
$N$.
The following theorem, or rather its subsequent corollary, allows to reduce
this list dramatically.

\proclaim{Theorem ($K$-Rationality of modular representations)}
Let $k$ and $N>0$ be integers, let $K=\Q(\e^{2\pi i/N})$. Then the
$K$-vector space $M_k^K(\Gamma(N))$ of all modular forms on $\Gamma(N)$
of weight $k$ whose Fourier developments with respect to $\e^{2\pi i\tau/N}$
have coefficients in $K$ is invariant under the action $(f,A)\mapsto f|_k A$
of $\SL$.
\endproclaim
\demo{Proof}
Let $j(\tau)$ denote the usual $j$-function, which has Fourier coefficients
 in $\Z$ and satisfies $j(A\tau)=j(\tau)$ for all $A\in\SL$. Assume that $k$
is even. Then the map $f\mapsto f/{j'}^{k/2}$ defines an injection of the
$K$-vector space $M_k^K(\Gamma(N))$ into the field of all modular functions
on $\Gamma(N)$
whose Fourier expansions have coefficients in $K$. It clearly suffices to
show that the latter field is invariant under $\SL$. A proof for this can
be found in \cite{Sh, p.\ 140, Prop\. 6.9 (1), equ\. (6.1.3)}.
The case $k$ odd can be reduced to the case $k$ even by
considering the squares of the modular forms in $M_k^K(\Gamma(N))$.
\enddemo

\proclaim{Corollary}
Let $\rho:\SL\rightarrow\operatorname{GL}(n,\C)$ be
a representation whose kernel contains $\Gamma(N)$ for some positive
integer $N$, and let $K=\Q(\e^{2\pi i/N})$. If, for some integer $k$,
there exists a nonzero element in $M_k(\rho)$ whose Fourier development
has Fourier coefficients in $K^n$, then $\rho(\SL)\subset\GLK{n}$.
\endproclaim
\demo{Proof}
If $F\in M_k(\rho)$ has Fourier coefficients in $K^n$ then $F|_kA$, by the
preceding theorem, has Fourier coefficients in $K^n$ too. Here $A$ is any
element in $\SL$. From $F|_kA=\rho(A)F$ we deduce that $\rho(A)$ has entries
in $K$.
\enddemo
\remark{Remark}
If one assumes that a vector valued modular form is related to the conformal
characters of a rational model which are modular functions of some congruence
subgroup then obviously all the Fourier coefficients are rational so that
the corollary applies.
\endremark

  \subhead
  4.4 Proof of the theorem on uniqueness of conformal characters
  \endsubhead

We will now prove our main theorem stated in \S4.1.
Pick one of the central charges $c$ in Table 3.2b or Table 3.2c.
Assume that for some $H\subset H_c$ containing $0$ there exist functions
$\xi_{c,h}$ ($h\in H$) which satisfy the properties (1) to (5) of the main
theorem. Let $\xi$ denote the vector whose components are the functions
$\xi_{c,h}$ ordered with increasing $h$. Note that the $h$-values are pairwise
different modulo 1. By (4) the $\xi_{c,h}$ are thus linearly independent.
Hence,
we have a well-defined $|H|$-dimensional representation $\rho$ of the modular
group if we set $\xi( A\tau)=\rho(A)\xi(\tau)$ for $A\in\SL$. Finally, recall
that the Dedekind eta function $\eta$ is a modular form of weight $1/2$ for
$\CSL$,
 more precisely, that there exists a one-dimensional representation $\theta$ of
$\CSL$ on the group of 24-th roots of unity such that $\eta \in
M_{\frac12}(\theta)$.

For any half integer $k\in \frac12\Z$ such that
$$k\ge \tilde c/2$$
we have
$F:=\eta^{2k}\xi\in M_k(\rho\otimes\theta^{2k})$, as is
as an immediate consequence of property (3) and the assumption that the
$\xi_{c,h}$ are
holomorphic in the upper half plane. Let $k$ be the smallest possible half
integer
satisfying this inequality. The actual value is given in Table 4.4 below.

We shall show that by property (1) to (5) the representation $\rho$ is uniquely
determined (up to equivalence). Its precise description can be read off from
the last column of Table 3, respectively (notations will be explained below).
In particular, $\rho$  has dimension equal to the cardinality of $H_c$, and
hence
we conclude $H=H_c$. The $h$-values are pairwise incongruent modulo 1, i.e.\
$\rho(T)$ has pairwise different eigenvalues. Since  $\rho(T)$ is a diagonal
matrix the representation $\rho$ is thus unique
up to conjugation by diagonal matrices.

Finally, the kernel of $\rho$ is a congruence subgroup by property (1).
In particular, $\rho\otimes\theta^{2k}$ has a finite image. Thus we can
apply the dimension formulas stated in \S4.2.
(For verifying the second assumption for the dimension formula note that $\rho$
is even and that $\theta((\epsilon^2\id,\epsilon))=\eta|_{\frac12}
(\epsilon^2\id,\epsilon)(\tau)/\eta(\tau)=\epsilon^{-1}$ for all
$\epsilon^4=1$.)
It will turn out that
$M_k(\rho\otimes\theta^{2k})$ is one-dimensional.
Thus, if there actually exist functions $\xi_{c,h}$ satisfying (1) to (5) then
$M_k(\rho\otimes\theta^{2k})=\C\cdot \xi\eta^{2k}$. Since $\rho$ is unique up
to conjugation by diagonal matrices we conclude that $\xi$ is unique up to
multiplication by such matrices, and this proves the theorem. We now give the
details.

\subhead
Determination of the representation $\rho$
\endsubhead
We first determine the equivalence class of the representation $\rho$.

For an integer $k'$ let $l(k')$ be the lowest common
denominator of the numbers $h-c/24+k'/12$ ($h\in H_c$), i.e\. let
$$l(k')=12d/\gcd(12d,\dots,12n_j+k'd,\dots),$$
where the $n_j/d$ denote the rational numbers $h-c/24$
($h\in H_c$) with integers $n_j$, $d$.
Clearly, the order
of $(\rho\otimes\theta^{2k'})(T)$ divides
$l(k')$. Let $k'$ the smallest nonnegative integer such that
$l=l(k')$ is minimal, and set $\rhot=\rho\otimes\theta^{2k'}$. The values of
$k'$ and $l$ are given in Table 4.4.

Note that $k'$ integral implies that $\rhot $ can be regarded as a
representation of $\SL$ (rather than $\CSLZ$). By property (1) its
kernel is a congruence subgroup. Thus we can apply the factorization
criterion of \S4.3 to conclude that this kernel contains $\Gamma(l)$.
Note that here the assumption (1), namely that the $\xi_{c,h}$ are
invariant under a congruence subgroup is crucial if $l>5$. For $l\le 5$,
this assumption is not necessary, which explains the supplement to the main
theorem.

We shall say that a representation of $\SL$ has level $N$ if its kernel
contains $\Gamma(N)$ (here $N$ is not assumed to be minimal).
Since any representation of level $N$ factors to a
representation of
$$\SL/\Gamma(N)\approx~\MG N,$$
it has a  unique decomposition as sum of irreducible level $N$
representations. Furthermore, there are only finitely many irreducible
level $N$ representation, and each such representation $\pi$ has a unique
product decomposition
$$\pi=\prod\Sb p^\lambda\Vert l\endSb\pi_{p^\lambda}$$
with irreducible level $p^\lambda$ representations  $\pi_{p^\lambda}$.
Here the product is to be taken over all prime powers dividing $N$ and such
that $\gcd(p^\lambda,N/p^\lambda)=1$.
Finally, $\pi_{p^\lambda}(T)$ has order dividing $p^\lambda$, i.e\. its
eigenvalues are $p^\lambda$-th roots of unity. Since any $N$-th root of
unity $\zeta$ has a unique decomposition as product of the $p^\lambda$-th
roots of unity $\zeta^{\frac N{p^\lambda}x_p}$ with
$\frac N{p^\lambda}x_p\equiv 1\bmod p^\lambda$, we conclude:

\proclaim{Lemma}
Let $\zeta_j$ ($1\le j\le n=\dim\pi$) be the eigenvalues of $\pi(T)$. Then,
for each $p^\lambda\Vert N$, the eigenvalues $\not =1$ of $\pi_{p^\lambda}(T)$
(counting multiplicities) are exactly those among the numbers
$\zeta_j^{\frac N{p^\lambda}x_p}$ ($1\le j\le n$) which are not equal
to 1.
\endproclaim
\bigskip
\centerline{Table 4.4: Representations of $\SL$ and weights related
                     to certain rational models}
\smallskip\noindent
\centerline{
\vbox{\offinterlineskip
\def\tablespace{ height2pt&\omit&&\omit&&\omit&&\omit&&\omit&&\omit&\cr }
\def\tablerule{ \tablespace
                \noalign{\hrule}
                \tablespace      }
\hrule
\halign{&\vrule#&\strut\kern2pt\hfil$#$\hfil\kern2pt\cr
\tablespace
& \w\text{-algebra}&& c && k && k' && l &&\rhot=\rho\otimes\theta^{2k'} &\cr
\tablerule
&\w(2)&&1-6\frac{(p-q)^2}{pq}&&\frac12&&2&&8pq
  &&R_1^p(q,\chi_{-1})\otimes R_1^q(p,\chi_{-1})\otimes D_8^{pq}&\cr
\tablerule
&\w(2,\frac{(m-1)(q-2)}{2})&&1-3\frac{(2m-q)^2}{mq}
  &&\frac12&&\frac{1-3mq}2\bmod{12}&&mq
  &&R_1^q(2m,\chi_{-1})\otimes R_1^m(2q,\chi_1)&\cr
\tablerule
&\w(2,q-3)&&1-\frac{(12-q)^2}{2q}&&\frac12&&-1-q\bmod 3&&16q
  &&R_1^q(3,\chi_{-1})\otimes D_{16}^q&\cr
\tablerule
&\w(2,q-5)&&1-\frac{(30-q)^2}{5q}&&\frac12&&\frac{1-5q}2 \bmod{12}&&5q
  &&R_1^q(30,\chi_{-1})\otimes R_1^5(q,\chi_{-1})&\cr
\tablerule
&\w_{G_2}(2,1^{14})&&-{8\over5}&&2&&4&&5&&\rho_5&\cr
\tablerule
&\w_{F_4}(2,1^{26})&&{4\over5}&&3&&10&&5&&\rho_5&\cr
\tablerule
&\w(2,4)&&-{444\over11}&&1&&6&&11&&\rho_{11}&\cr
\tablerule
&\w(2,6)&&-{1420\over17}&&1&&2&&17&&\rho_{17}&\cr
\tablerule
&\w(2,8)&& -{3164\over23}&&1&&10&&23&&\rho_{23}&\cr
\tablespace
}
\hrule}
}
Recall that in Table 4.4 the integers $p,q$ and $m$ are primes with
$q\not=p,m$.
\bigskip

\subhead
The representation $\rho$ in line 1 to 4 of Table 4.4
\endsubhead
First, we consider the rational models corresponding to the first 4 rows of
Table~4.4. By assumption $h=0$ is in $H$, i.e\. $\mu=\exp(2\pi i
(-c/24+k'/12))$
is an eigenvalue of $\rhot (T)$. Let $\pi$ be that irreducible level $l$
representation in the sum decomposition of $\rhot $ such that $\pi(T)$ has
the eigenvalue $\mu$. Since $\pi$ is irreducible it has a decomposition
as product of irreducible representations $\pi_{p^\lambda}$ as above. Since
$\mu$
is a primitive $l$-th root of unity the lemma implies that the
$\pi_{p^\lambda}$
are nontrivial.
\medskip
The minimal dimension of a nontrivial irreducible level $p^\lambda$
representation
is $2$, 3 or $(p-1)/2$ accordingly if $p^\lambda$ equals 8, 16 or is an odd
prime $p$ (cf\. \S3 or \cite{NW, p.\ 521ff}). Hence we have the inequalities
$$\dim\pi\ge
\cases
(p-1)(q-1)/2&\text{for row 1}\\
(m-1)(q-1)/4&\text{for row 2}\\
3(q-1)/2&\text{for row 3}\\
q-1&\text{for row 4}
\endcases.
$$
For row 1, 3 and 4 the right hand side equals the cardinality of $H_c$
respectively. In these cases we thus conclude that $\rhot =\pi$ is irreducible,
that it is equal to a product of  nontrivial level $p^\lambda$ representations
with minimal dimensions, and, in particular, that $H=H_c$.

For row 2 the right hand side is smaller than the cardinality of $H_c$.
However, here we can sharpen the above inequality:
First we note that the level $p$ representations of dimension $(p-1)/2$
have parity $(-1)^{(p+1)/2}$, whence the product of the corresponding
level $m$ and $q$ representations has parity $(-1)^{(mq-1)/2}$. On the
other hand any irreducible subrepresentation has the same parity $\rhot $,
 i.e\. the parity $(-1)^{k'}=(-1)^{(mq+1)/2}$.  Hence $\pi$ cannot equal
a product of two nontrivial level $m$ and $q$ representations of minimal
dimension. The dimension of the second smallest nontrivial irreducible
level $p$ representations is $(p+1)/2$. Under each of these representations
$T$ affords eigenvalue 1. Since $T$ under $\rhot $ affords no $m$-th root of
unity as eigenvalue, we conclude that $\pi$ cannot be equal to a product of a
$(q+1)/2$ dimensional level $q$ and a $(m-1)/2$ dimensional level $m$
representation. Thus,
$$\dim \pi\ge (m+1)(q-1)/4.$$
The right hand side equals $|H_c|$, and we conclude as above that $H=H_c$,
that $\rho$ is irreducible, and that $\rhot $ equals a product of an
irreducible  $(q-1)/2$ dimensional level $q$ and an
irreducible $(m+1)/2$ dimensional level $m$ representation.

To identify $\rho$ it thus remains to examine the nontrivial level
$p^\lambda$ representations with small dimensions (cf\. \S3 or
\cite{NW, p. 521ff}).

Let $p^\lambda=p$ be an odd prime. There exist exactly two irreducible
level $p$ representations with dimension $(p-1)/2$. The image of $T$
under these representations has the eigenvalues
$\exp(2\pi i \varepsilon x^2/p)$ ($1\le x\le (p-1)/2$) where  for one of
them $\varepsilon$ is a quadratic residue modulo $p$, and a quadratic
non-residue for the other one. Call these representations
accordingly $R_1^p(\varepsilon,\chi_{-1})$. Similarly there exist exactly 2
irreducible
level $p$ representations with dimension $(p+1)/2$, denoted by
$R_1^p(\varepsilon,\chi_1)$
(with $\varepsilon$ being a quadratic residue or non-residue modulo $p$). The
eigenvalues of $R_1^p(\varepsilon,\chi_1)$ are $\exp(2\pi i\varepsilon x^2/p)$
($0\le x\le (p-1)/2$).

Let  $p^\lambda=8$. There exist exactly 4 irreducible two dimensional level 8
representations which we denote by $D_8^x$ ($x$ being an integer modulo 4).
The eigenvalues of the image of $T$ under the representation $D_8^x$  are
$\exp(2\pi i(1+2x)/8)$ and $\exp(2\pi i(7+2x)/8)$.

Let $p^\lambda=16$. There are 16 irreducible three dimensional level 16
representations. These   can be distinguished by their eigenvalues
of the image of $T$. In particular, there are four of these representations,
denoted by $D_{16}^x$ ($x \bmod 4$), where the image of $T$ has the eigenvalues
$\exp(2\pi i (2x+3)/8),\exp(2\pi i (3x-6)/16),\exp(2\pi i(3x+2)/16)$.

Summarizing we find
$\rhot =R_1^p(n_q,\chi_{-1})\otimes R_1^q(n_p,\chi_{-1})\otimes D_8^{n_8}$,
$=R_1^q(n_q,\chi_{-1})\otimes R_1^m(n_m,\chi_1)$,
$=R_1^p(n_q,\chi_{-1})\otimes D_{16}^{n_{16}}$ or
$=R_1^p(n_q,\chi_{-1})\otimes R_1^5(n_5,\chi_{-1})$, respectively, with
suitable numbers
$n_p,\dots$. The latter can be easily determined using the Lemma and the
description
of $H_c$ in Table 1. The resulting values are given in Table 4.4.

\subhead
The representation $\rho$ in line 5 to 9 of Table 4.4
\endsubhead
We now consider the rational models corresponding to row 5 to 9 of Table 3.
Here the level of $\rhot$ is a prime $l$, the dimension of $\rho$ is $\le l-1$,
and the eigenvalues of $\rho(T)$ are pairwise different primitive $l$-th roots
of unity.

We show that $\rhot$ is irreducible with dimension $l-1$. Assume that
$\rhot$ is reducible or has dimension $<(l-1)$. The only irreducible
level $l$ representations with dimension $<(l-1)$ for which the image
of $T$ does not
afford eigenvalue 1 are $R_1^l(\varepsilon,\chi_{-1})$. Thus there are only two
possibilities: (a) $\rhot=R_1^l(\varepsilon,\chi_{-1})$ or (b)
$\rhot=R_1^l(\varepsilon,\chi_{-1})\otimes R_1^l(\varepsilon',\chi_{-1})$.
For $l=5,17$ the representations $R_1^l(\varepsilon,\chi_{-1})$ have parity
$-1$, whereas $\rhot$ has parity $+1$, a contradiction. For $l=11,23$
we note that $\xi\eta^2$ is an element of $M_1(\rhot\otimes\theta^{2-2k'})$.
We shall show in moment that the dimension of
$M_1(R_1^l(\varepsilon,\chi_{-1})\otimes\theta^{2-2k'})$ is 0, which gives the
desired contradiction (to recognize the contradiction in case (b) note
that the `functor' $\rho\mapsto M_k(\rho)$ respects direct sums).

Since the dimension formula gives explicit dimensions only for $k\not=1$ we
cannot apply it directly for calculating the dimension of
$M=M_1(R_1^l(\varepsilon,\chi_{-1})\otimes\theta^{2-2k'})$. For $l=11$ we note
that
$\eta^2 M$ is a subspace of
$M_2(R_1^l(\varepsilon,\chi_{-1})\otimes\theta^{4-2k'})$.
 To the latter we can apply the dimension formula, and find
(using $\tr R_1^l(\varepsilon,\chi_{-1})(S)=0$,
$\tr R_1^l(\varepsilon,\chi_{-1})(ST)=-1$)
that its dimension is 0.
For $l=23$ and $\varepsilon=1$ we consider
$M_{3/2}(R_1^l(1,\chi_{-1})\otimes\theta^{3-2k'})$
which contains $\eta M$. We find that its dimension equals
$$\dim S_{1/2}(R_1^l(-1,\chi_{-1})\otimes\theta^{-(3-2k')})\le
 \dim M_{1/2}(R_1^l(-1,\chi_{-1})\otimes\theta^{-(3-2k')})\,,$$
which equals 0 by the
supplement in \S4.2 (for applying the supplement note that
$(R_1^l(-1,\chi_{-1})\otimes\theta^{-(3-2k')}$ has a kernel containing
$\Gamma(23\cdot 24)^\sharp$ and represents $T$ with eigenvalues
$\exp(2\pi i (-24x^2+17\cdot23)/23\cdot24)$ ).
Finally, by the dimension formula we find
$$\dim M_1(R_1^l(-1,\chi_{-1})\otimes\theta^{2-2k'})=
\dim S_1(R_1^l(1,\chi_{-1})\otimes\theta^{-(2-2k')}),
$$
and the right hand side equals 0 since
$\dim S_{3/2}(R_1^l(1,\chi_{-1})\otimes\theta^{-(1-2k')})=0$ by the supplement.

Thus, $\rhot$ is irreducible of dimension $l-1$, which implies in particular
$H=H_c$. There exist exactly $(l-1)/2$ irreducible level $l$ representations
of dimension $l-1$ (cf\. Table 3.5a). We now use property (5) of the main
theorem, which implies that the Fourier coefficients of $\xi\cdot\eta^{2k'}$
are rational. Hence, by the corollary in \S4.3 we find that $\rhot$ takes
values in $\GLK{l-1}$ with $K$ being the field of $l$-th roots of unity.
There is exactly one irreducible level $l$ representations of dimension
$l-1$ whose character takes values in $K$ (Lemma 8 in \S3.5 or
\cite{Do, p\. 228}); denote it  by $\rho_l$. Then $\rhot=\rho_l$.
\eject
\subhead
Computation of dimensions
\endsubhead
It remains to show $d=\dim M_k(\rhot\otimes\theta^{2k-2k'})\le1$.
For the first 4 rows of Table 4.4 this follows from the supplement in \S4.2
and the irreducibility of $\rho$ (in fact it can be shown that $d=1$
\cite{Sk}). For row 5 and 6 we find $d=1$ by the dimension formula and
using $\tr\rho_l(S)=0$, $\tr\rho_l(ST)=1$ (valid for arbitrary primes $l$). For
the remaining cases (where $k=1$) we multiply $M_1(\rhot\otimes\theta^{2-2k'})$
by $\eta$ for obtaining  $d'=\dim M_{3/2}(\rhot\otimes\theta^{3-2k'})$ as upper
bound. Again, using the dimension formula and its supplement we find $d'=1$.

This concludes the proof of the main theorem 3.\qed

\vfill\eject
\head
5. Construction of conformal characters
\endhead

In \S4 we formulated a list of five axioms which are satisfied for all known
sets of conformal characters of rational models of $\w$-algebras. The only
data from an underlying rational model which occurs in these axioms is
its central charge and its conformal dimensions.
We showed that, for several rational models, these axioms uniquely
determine the conformal characters belonging to a given  central charge
and set of conformal dimensions.

Thus, once the central charge and conformal dimensions of a rational model
are known, the computation of its conformal characters can be viewed as
a problem which is completely independent from the theory
of $\w$-algebras, i.e\. for this computation one is left with a construction
problem, namely, the problem of finding, by whatever means, a set of
functions fulfilling the indicated list of axioms.

The purpose of this section is to describe such a mean which can solve in many
cases this construction problem. In particular, we shall apply our method to
the case of five special rational models related to Table 3.2c.
The reason for the choice of these
models is that the representation theory of the $\SLZ$-representation on their
conformal characters can be treated homogeneously in some generality, and that
the conformal characters of one of these models
(of type $\w(2,8)$ with central charge $c=-\frac{3164}{23}$)
could not be computed explicitly by the so far known methods.

This section is organized as follows:
In section 5.1 we describe a general procedure for
the construction of vector valued modular forms transforming under
a given matrix representation of $\SLZ$  (main theorem 4 on realization
by theta series). As already mentioned,
this procedure is useful in general
for finding explicit and easily computable formulas for conformal characters.
In \S5.2 we apply this general setup to the case of
the five special rational models, and we derive explicit formulas for their
conformal characters (main theorem 5 on theta formulas for conformal
characters).
Finally, in \S5.3 we compare our results with those formulas for the conformal
characters of the five models which can be obtained (assuming certain
conjectures) from the representation theory of Casimir $\w$-algebras.

  \subhead
  5.1  The general construction:
       Realization of modular representations by theta series
  \endsubhead

In this section we show how one can, under certain hypothesis, construct
systematically vector
valued modular forms in $M_k(\rho)$ for a given matrix representation
$\rho$ of $\Gamma$ and given weight $k$.

The first step is a realization of $\rho$ as subrepresentation of a Weil
representation.

Recall from section 3.4 that one has the following theorem.
\proclaim{Theorem \cite{NW}}
Each irreducible right-representation of $\Gamma$ whose kernel contains a
principal congruence subgroup is isomorphic to
a subrepresentation of a suitable Weil representation.
\endproclaim

We call two quadratic modules $(M,\QF)$ and $(M',\QF')$ isomorphic if there
exists an isomorphism (of abelian groups) $\pi\colon M\to M'$ such that
$\QF'\circ\pi=\QF$, and we denote such an isomorphism by
$$\pi\colon (M,\QF)\aproxeq (M',\QF').$$ It is easy to show that
isomorphic quadratic modules yield isomorphic Weil representations: an
isomorphism of
(projective or proper) $\SL$-representations is given by the map
$$\pi^\ast\colon\C^{M'}\to\C^M,\qquad
f\mapsto \pi^\ast f=f\circ\pi.$$

As the next step for constructing elements of spaces $M_k(\rho)$
we connect Weil representations and theta series by lifting quadratic modules
to lattices and quadratic forms on them.

More precisely, let $(M,\QF)$ be a quadratic module. Assume that $L$ is a
complete
lattice in some rational finite-dimensional vector
space $V$ and $Q$ a positive definite non-degenerate quadratic form on $V$
which takes on integral values on $L$, and such that there exists an
isomorphism
of quadratic
modules
$$\pi\colon(L^\sharp/L,\widetilde Q)\aproxeq (M,\QF).$$
Here we use
$L^\sharp$ for the dual lattice of $L$ with respect to $Q$, i.e\.
$L^\sharp$ is the set of all $y\in V$ such that
$B(L,y)\subset\Z$ with $B(x,y)=Q(x+y)-Q(x)-Q(y),$
and we use $\widetilde Q$ for the induced quadratic form
$$\widetilde Q:L^\sharp/L\to\Q/\Z\,,\qquad
x+L\mapsto Q(x) +\Z.$$
We shall call such a pair $(L,Q)$ a lift of the quadratic module
$(M,\QF)$.

Let $p$ a homogeneous spherical polynomial on $V$ with respect to $Q$ of
degree $\nu$,
i.e\. if we choose a basis $b_j$ of $V$, then $p\left(\sum b_j\xi_j\right)$
becomes a complex homogeneous polynomial in the
variables $\xi_j$ of degree $\nu$ satisfying
$$\nabla G^{-1}\nabla'\,p\big(\sum_j b_j\xi_j\big) = 0,$$ where
$\nabla=(\frac{\partial}{\partial\xi_1},\dots)$ and $G=(B(b_j,b_k))_{j,k}$ is
the
Gram matrix of $B$.

Finally, for $f\in\C^M$, set
$$\theta_f=\sum_{x\in L^\sharp}(\pi^\ast f)(x)\,p(x)\,q^{Q(x)}.$$
Here we view $\pi^\ast f$ as function on
$L^\sharp$ which is periodic with period lattice $L$.

We assume that $V$ has even dimension $2r$. Then the Weil representation
$\omega=\omega_{(M,\QF)}$ is proper as shown in ref. \cite{ES$2$}.
One has

\proclaim{Theorem (Representation by theta series)}
The map  $\C^M\ni
f\mapsto \theta_f$ has the property
$\theta_f|_{r+\nu}A=\theta_{f|\omega(A)}$ for all $A\in\SL$,
i.e\. it defines a homomorphism of $\SL$-modules.
\endproclaim

This is, in various different formulations, a well-known theorem. For the
reader's
convenience we shall sketch the proof in the Appendix at the end of this
subsection.

Let now $\rho\colon\SL\to\GL{n}$ be a congruence
matrix representation, and assume that we have determined a
quadratic module $(M,\QF)$
such that the associated Weil representation is proper and contains a
subrepresentation which is isomorphic to the (right-)representation
$\C^n\times\SL\ni(z,A)\mapsto z\rho(A)'$, where the prime denotes
transposition.
The existence of such an $(M,\QF)$ is guaranteed by the first theorem.
Thus, there exists a $\SL$-invariant subspace of $\C^M$ with basis $f_j$
such that
$$\Phi|\omega(A)=\rho(A)\Phi\qquad(A\in\SL),$$ where
$\Phi$ denotes the column vector build from the $f_j$.
Assume furthermore that there exists a lift $(L,Q)$ of $(M,\QF)$, i.e\.
an isomorphism
$$\pi\colon(L^\sharp/L,\widetilde Q)\aproxeq (M,\QF)$$
with a lattice $L$ of even rank $2r$.
Let $p$ be a homogeneous spherical polynomial w.r.t\. $Q$ of degree $\nu$.
{}From the last theorem it is then clear that we have the
following

\proclaim{Main theorem 4 (Realization by theta series)} The function
$$\theta=\sum_{x\in
L^\sharp}\Phi(\pi(x))\,p(x)\,q^{Q(x)}$$ is an element of $M_{r+\nu}(\rho)$.
\endproclaim

\subhead
$\underline{\text{Appendix}}$
\endsubhead
We proof the theorem on representation by theta series.
It is consequence of the following
\proclaim{Lemma (Basic transformation formula)}
Let $L$ be a lattice in a rational vector space $V$ of dimension $2r$,
let $Q$ be a
positive definite quadratic form on $V$ which takes on integral values on $L$,
let
$L^\sharp$ and $B$ be defined as above, let $w\in V\otimes\C$ with
$Q(w)=0$, let $\nu$ a non-negative integer, and let $z\in V$. Then one has
$$\align \tau^{-r-\nu}\sum\Sb x\in L\endSb &[B(w,x+z)]^\nu\,\e(-Q(x+z)/\tau)\\
&=\frac{i^{-r}}{\sqrt{[L^\sharp:L]}} \sum\Sb y\in L^\sharp\endSb
[B(w,y)]^\nu
\e(\tau Q(y)^t-B(y,z)),
\endalign$$
where $\tau$ is a variable in the complex upper half plane.
\endproclaim
The lemma is a well-known consequence of the Poisson summation
formula; for a proof cf\. \cite{Scho, p\. 206}. (For verifying that our formula
is
equivalent to the one given loc\. cit\. identify $L$ with $\Z^{2r}$
by choosing a $\Z$-basis $b_j$ of $L$, and note that then
$L^\sharp=G^{-1}\Z^{2r}$ and
$\det(G)=[L^\sharp:L]$ where $G=(B(b_j,b_k))$ is the Gram matrix of
$L$.
Moreover, the transformation formula loc\. cit\. is only stated for $\tau=it$
($t$ real); the general formula follows by analytic continuation.)

\demo{Proof of the theorem on representation by theta series}
Since any homogeneous spherical polynomial of degree $\nu$ can be written as
linear combination of the special ones $B(x,w)^\nu$ (where
$w\in\C$, $Q(w)=0$) we can assume that $p$ is of this special form.
Since $S$ and
$T$ generate $\SL$ it suffices to prove the asserted formula for these
elements.
For $A=T$ the formula is obvious. For proving the case $A=S$ let in the basic
transformation formula $z$ be an element of $L^\sharp$, multiply by $f(z)$ and
sum over a set of representatives $z$ for $L^\sharp/L$. Using $$\sum_{x\in
L^\sharp/L}\e(Q(x))=i^r\,\sqrt{[L^\sharp:L]}$$ (Milgram's theorem, e.g\.
\cite{MH, p\. 127}) we realize the claimed formula.  \qed\enddemo

  \subhead
  5.2 An example (I):
     Theta series associated to quaternion algebras and the conformal
     characters of the five special models
  \endsubhead

We shall use the notation introduced in \S4 and construct the conformal
characters related to the rational models in Table 3.2c.
To this end we follow the procedure outlined in the foregoing section
to construct elements of $M_k(\rho_l)$ where $l$ denotes an odd prime
$l\equiv -1\bmod 3$ and $\rho_l$ is the matrix representation introduced in
\S4.4, i.e.\  $\rho_l$ is the (up to equivalence) unique irreducible
representation whose kernel contains $\Gamma(l)$, and it takes its values in
$\operatorname{GL}(l-1,\Q(\e^{2\pi i/l}))$.
We shall give an explicit description of $\rho_l$ below.

{}From property (3) in the main theorem 3 on uniqueness of conformal characters
we have $\eta^{2k}\xi_c=\LO(q^\delta)$ for $q\to 0$, where
$\delta = -\tilde c + k/12$, and, in particular,  that $\eta^{2k}\xi_c$ is an
element of $M_k(\rho_l)$ (here $\xi_c$ is the vector whose components are the
functions $\xi_{c,h}$ ordered with increasing $h$).
The dimensions of these spaces can be computed using
the dimension formula in \S4.2. The resulting dimensions and the values of
$\delta$ are listed in Table 5.2.

Let $M^{(\delta)}_k(\rho_l)$ be the subspace of all $F\in M_k(\rho_l)$
satisfying $F=\LO(q^\delta)$. In \S4.4 it was shown that this subspace
is one-dimensional, which, by obvious arguments, implies that $\xi_c$ is
unique up to multiplication by diagonal matrices.
(Actually, it was shown
that $M_h(\rho_l\otimes \theta^{2h-2k})$
is one-dimensional, where $\theta^2(A)=(\eta^2|_1A)/\eta^2$.
However, this latter
space is obviously isomorphic to $M^{(\delta)}_k(\rho_l)$ via multiplication by
$\eta^{2k-2h}$.)

\mn
\centerline{ Table 5.2: Certain data related to five rational models}
\smallskip
\centerline{%
\vbox{\offinterlineskip
\def\tablespace{height2pt&\omit&&\omit&&\omit&&\omit&&\omit&&\omit&\cr}
\def\tablerule{\tablespace\noalign{\hrule}\tablespace}
\hrule
\halign{&\vrule#&\strut$\quad\hfil#\hfil\quad$\cr
\tablespace
&\w\text{-algebra}&&c&&l&&k&&\delta&&\dim M_k(\rho_l)&\cr
\tablerule
&\w_{G_2}(2,1^{14})&&-{8\over5}&&5&&4&&\frac15&&1&\cr
\tablerule
&\w_{F_4}(2,1^{26})&&{4\over5}&&5&&10&&\frac35&&3&\cr
\tablerule
&\w(2,4)&&-{444\over11}&&11&&6&&\frac5{11}&&5&\cr
\tablerule
&\w(2,6)&&-{1420\over17}&&17&&2&&\frac2{17}&&2&\cr
\tablerule
&\w(2,8)&&-{3164\over23}&&23&&10&&\frac{18}{23}&&17&\cr
\tablespace
}
\hrule}}
\mn

We first describe how to obtain $\rho_l$ from a proper Weil representation.
Let $\omega$ be the Weil representation associated to the quadratic module
$(\FF,\norm(x)/l)$. Here $\FF$
is the field with $l^2$ elements, and $\norm(x)=x\cdot\overline x$ with
$x\mapsto \overline x=x^l$  denoting the non-trivial automorphism of $\FF$.
Note that
$\tr(x\overline y)/l$ where
$\tr(x)=x+\overline x$ is the bilinear form associated to $n(x)/l$.
The Weil representation $\omega$ associated is thus a (right-)representation of
$\SL$ on the space
of functions $f\colon \FF\to\C$, and it is given by
$$f|\omega(T)(x)=\e(\norm(x)/l)\,f(x),\qquad
f|\omega(S)(x)=\frac{-1}l\sum\Sb y\in\FF\endSb\e(-\tr(\overline xy)/l)\,f(y).$$
Here we used
$$\sum_{x\in\FF}\e(\norm(x)/l)=-l,$$ as follows for instance from
Milgram's theorem and the considerations below where we shall obtain
$\FF=L^\sharp/L$
with a lattice of rank 4. Note that this identity implies in particular that
$\omega$ is a proper representation (cf\. the discussion in section 3.4).

Let $\chi$ be one of the two characters of order 3
of the multiplicative group of nonzero elements in $\FF$, and let $G$ be the
subgroup of elements with $\norm(x)=1$.
Note that the existence of $\chi$ follows from the assumption $l\equiv -1\bmod
3$.
Let $X(\chi)$ be the subspace
of all $\phi\in X$ which satisfy $\phi(gx)=\chi(g)\phi(x)$ for all $g\in G$. It
is
easily checked that $X(\chi)$ is a $\SL$-submodule of $X$. In fact,
 it is even an irreducible one \cite{NW, Satz 2}.
 As basis for $X(\chi)$ we may pick the functions $\chi_r$ ($1\le r\le l-1$)
which are defined by
$\chi_r(x)=\chi(x)$ if $\norm(x)=r$ and $\chi_r(x)=0$ otherwise. Let
$\Phi_\chi$ be the
complex column vector valued function on $\FF$ whose
$r$-th component equals $\chi_r$.
We then have $\Phi_\chi|A=\rho(A)\Phi_\chi$ with a unique
matrix representation $\rho\colon\SL\rightarrow
\GL{l-1}$.
It is an easy exercise to verify the
identities
$$\rho(T)=\diag (\e^{2\pi i 1/l},\cdots,\e^{2\pi i (l-1)/l}),\quad
\rho(S)=\left(\lambda(rs)\right)_{1\le r,s\le l-1},$$
where we use
$$\lambda(r)=\frac{-1}{l}\sum\Sb x\in\FF\\ \norm(x)=r\endSb
\chi(x)\e(\tr(x)/l).$$
(In the identity $\norm(x)=r$ the $r$ has to be viewed as an element of $\FF$.)

Note that $\lambda(r)$ does not depend on the choice of $\chi$,
 as is easily deduced by replacing in its defining sum $x$ by
$\overline x$ and by using $\chi(\overline x)=\overline\chi(x)$ and
$\tr(\overline x)=\tr(x)$.
The independence of the choice of $\chi$ implies
that $\lambda(r)$, for any $r$, is contained in the field of $l$-th roots of
unities
(actually, $\lambda(r)$ is even real as follows from the easily proved facts
that $\rho(S)$
is unitary, symmetric and satisfies $\rho(S)^2=1$.)
Thus $\rho$ satisfies the properties listed at the begin of the section,
and hence is equivalent to $\rho_l$. Indeed, by permuting the components of
the vector valued function $\xi_c$ occurring in
the definition of $\rho_l$ and by multiplying by a suitable diagonal matrix
we can even assume that $\rho=\rho_l$.

We now set $\Phi=\Phi_\chi+\Phi_{\overline\chi}$. The independence of
the matrices $\rho(A)$ ($A\in\SL$) of the choice of $\chi$ then implies
that the subspace spanned by the components of $\Phi$ is invariant under $\SL$,
and that
$\Phi|\omega(A)=\rho_l(A)\Phi$ for all $A\in\SL$. It is easily verified that
for all $x\in\FF$ one has
$$\Phi(x)\in\{0,-1,2\}^{l-1},\
(r\text{-th entry of }\Phi(x))\bmod l=\cases
\tr(x^{(l^2-1)/3})&\text{if }\norm(x)=r,\\
0&\text{otherwise},
\endcases
$$
where $r$ runs from 1 to $l-1$.

Next we describe lifts of $(\FF,\norm(x)/l)$.
Let $V$ be the quaternion algebra over $\Q$ ramified at $l$ and $\infty$. If we
set
$K=\Q(\sqrt{-l})$ then $V$ can be described as
$V=K+Ku$, where $u^2=-1/3$ and $\alpha u=u\overline\alpha$
for all $\alpha \in K$.
The map $c=\alpha+\beta u\mapsto \overline c:=\alpha-u\overline\beta$ defines
an
anti-involution of $V$. The reduced norm $\norm(c)$ and reduced trace
$\tr(c)$ of a $c\in V$ are given by
$$\norm(c)=c\overline
c=|\alpha|^2+\frac13|\beta|^2,\qquad \tr(c)=c+\overline
c=\alpha+\overline\alpha.$$

Let $\o$ be the ring of integers in $K$. Note that the rational
prime
$3$ splits in $K$ since $l\equiv -1\bmod 3$. i.e\.
$3=\p\overline\p$ with a prime ideal $\p$ in $K$. (Indeed, one can
take $\p=3\o+(1+\sqrt{-l})\o$.) We set
$$\O=\o+\p v,\qquad v=\cases
u&\text{for }l\equiv3\bmod4\\
\frac{1+u}2&\text{for }l\equiv1\bmod4
\endcases.
$$
It can be easily checked that $\O$ is an order in $V$ (i.e\. a subring
which, viewed as
$\Z$-module, is free of rank 4). In fact, $\O$ is even a maximal order since
the
determinant of the Gram matrix $(\tr(e_j\overline e_k))$, for any $\Z$-basis
$e_j$ of $\O$, equals
$l^2$ (cf\.  \cite{Vi, Chap\. III, Corollaire 5.3}.).

We now have
\proclaim{Lemma}
(1) The dual lattice of $\sqrt{-l}\,\O$ w.r.t\. the quadratic form $\norm(c)/l$
is
$\O$. The quotient ring $\O/\sqrt{-l}\O$ is the field with $l^2$
elements, and the anti-involution $c\mapsto\overline c$ on $\O$
induces the Frobenius automorphism $x\mapsto x^l$ on $\O/\sqrt{-l}\O$.

(2) Let $I\subset\O$ be an $\O$-left ideal, and let $n=\norm(I)$ be the reduced
norm
of $I$ (i.e\. the g.c.d\. of the integers $\norm(x)$ where $x$ runs through
$I$).
Then the dual lattice of
$\sqrt{-l}I$ with respect to $\norm(c)/ln$ is $I$. There exists a $c_0\in I$
such that
$n(c_0)/n\equiv 1\bmod l$, and for any such $c_0$ the map $c\mapsto c_0\cdot c$
defines an isomorphism
of quadratic modules
$(\O/\sqrt{-l}\O,\norm(x)/l)\aproxeq (I/\sqrt{-l}I,\norm(x)/ln)$.
\endproclaim

Here, for convenience, we use the same symbols $\overline n(x)/nl$ for the
quadratic form on $I$
 as well as for the quadratic form induced by it on $I/\sqrt{-l}I$.
The Lemma follows easily from standard facts in the theory of quaternion
algebras; for
the reader's convenience we sketch the proof in the Appendix to this section.

The Lemma provides us with lifts $(I,n(x)/nl)$ of $(\FF,\norm(x)/l)$, and we
now can write down
explicitly elements of $M_k(\rho_l)$.

To this end let
$\Phi\colon\FF=\O/\sqrt{-l}\O\to\{-1,0,2\}^{l-1}$ be defined as above.
Let $I$ be an $\O$-left ideal,
choose $c_0$ as in the Lemma, and let
$$\pi:I\to\O/\sqrt{-l}\O,\qquad\pi(c)=\lambda+\sqrt{-l}\O\quad
\text{with }c\equiv\lambda c_0\bmod\sqrt{-l}\O.$$

Finally, let $p$ be a  homogeneous spherical
polynomial function on $V$.
If we write polynomial functions on $V$ as polynomials $p$ in $\alpha$,
$\overline\alpha$, $\beta$, $\overline \beta$, then it is spherical of degree
$\nu$
(with respect to any nonzero multiple of $n(c)$) if and only if $p$ is
homogeneous
of degree $\nu$ and satisfies
$$
\big(\frac{\partial^2}{\partial\alpha\partial\overline\alpha}
+3\frac{\partial^2}{\partial\beta\partial\overline\beta}\big) p=0.
$$

Set
$$\theta(\tau;I,p)=\sum_{c\in I}\Phi(\pi(c))\,p(c)\,q^{n(c)/n(I)l}.$$
We suppress the dependence of this function on $c_0$ since a different choice
results only in
multiplying $\theta(\tau;I,p)$ by a scalar.
By the theorem on realization by theta series we then have
$$\theta(\tau;I,p)\in M_{2+\deg(p)}(\rho_l).$$

It is easy to compute these functions using a computer. In fact, by a computer
calculation we found

\proclaim{Theorem}
Let $l$, $k$ be as in Table 5.2. Then the space $M_k(\rho_l)$
is spanned by the series $\theta(\tau;I,p)$, where $I=\O$ for $l\not=17$, and
$I=\O,\O\p$ for $l=17$,
and where $p$ runs through the homogeneous polynomial functions on the
quaternion algebra
$V$ of degree $k-2$ which are spherical with respect to the quadratic form
$\norm(c)$.
\endproclaim

It is an open question whether the spaces $M_k(\rho_l)$, for arbitrary $k$ or
primes $l$
($\equiv -1\bmod l$), are always spanned by theta series of the form
$\theta(\tau;I,p)$, or,
more generally, which spaces $M_k(\rho)$ of vector valued modular forms at all
can be generated by theta series.

As explained above we are especially interested in the one-dimensional subspace
$M_k^{(\delta)}(\rho_l)$ of functions in $M_k(\rho_l)$ which
are $\LO(q^\delta)$ with $\delta$ as in Table 5.2. Here we have

\proclaim{Main theorem 5 (Theta formulas for conformal characters)}

(1) Let $c$, $l$, $k$ and $\delta$ be as in Table 5.2, and let $I=\O$ for
$l\not=17$
and $I=\O\p$ for $l=17$.
Then there exists a homogeneous spherical polynomial function $p$ of degree
$k-2$ such that the
$\theta(\tau;I,p)$ is nonzero and satisfies $\theta(\tau;I,p)=\LO(q^\delta)$.

(2) Moreover, for any $p$ with this property, there exists a nonzero constant
$\kappa$
 such that the components of
the Fourier coefficients of $\kappa\theta(\tau;I,p)$ are rational integers.
In particular, the components of $\kappa\eta(\tau)^{-2k}\theta(\tau;I,p)$
satisfy the properties (1) to (5) in the main theorem 3 on the uniqueness
of conformal characters in \S4.1.
\endproclaim

\demo{Proof}
(1) The existence of a $p$ with Fourier development starting at
$q^\delta$ follows from the preceding theorem and the fact that the
subspace $M^{(\delta)}_k(\rho_l)$ contains nonzero elements.
For the latter cf\. the discussion at the beginning of \S5.2;
of course, it can also be checked by a straight forward calculation
using the $\theta(\tau;I,p)$ that $M^{(\delta)}_k(\rho_l)$ is
one-dimensional.

(2) This last fact also shows that for proving the second
statement of the theorem, it suffices to prove that, for at least one $p$
satisfying the condition
$\theta(\tau;I,p)=\LO(q^\delta)$, the function $\theta(\tau;I,p)$ has rational
Fourier coefficients.

For proving this let $P_\nu(F)$ be the set of spherical homogeneous
functions $p$ on $V$ of degree $\nu$ which are defined over the subfield
$F\subset\C$. By the latter we mean that the coefficients of $p(c)$,
when written as polynomial in the coefficients of $c$
with respect to a fixed basis of $V$, are in $F$. Note that this property
does not depend
on the choice of the $\Q$-basis of $V$. Since $P_\nu(F)$ is the kernel of a
differential
operator which has constant rational coefficients, when written with respect to
any
$\Q$-basis of $V$, it is clear that $P_\nu(\C)=P_\nu(\Q)\otimes\C$, i.e. we can
find a
basis of $P_\nu(\C)$ which is contained in $P_\nu(\Q)$. But then we deduce,
using the
preceding theorem, that $M_k(\rho_l)$ has a basis $\theta_j$ ($1\le j\le d$)
whose
Fourier coefficients $a_{\theta_j}(r)$ ($r=1,2,\dots$) are elements of
$\Q^{l-1}$.
For deducing this note that $\theta(\tau;I,p)$ for $p\in P_\nu(\Q)$ has
rational Fourier
coefficients since $\Phi(x)$ is rational. The elements of
$M^{(\delta)}_k(\rho_l)$ are
now the linear combinations $\sum_j c_j\theta_j$ such that $\sum_j c_j
a_{\theta_j}(r)=0$
for all $1\le r< l\delta$. Since the latter system of linear equations is
defined over
$\Q$ and has a nonzero solution by part (1) we conclude the existence of
rational nonzero
solution, i.e\. the existence of a linear combination of the $\theta_j$ with
Fourier coefficients in $\Q$.
\qed\enddemo

If we pick a $p$ as described in the theorem, and if we denote by $\xi_{c,h}$
the
$r$-th component of
$\eta^{-2k}\theta(\tau;I,p)$, where $\frac rl-\frac k{12}\equiv h-\frac
c{24}\bmod\Z$
then it is clear that these functions satisfy properties (1) to (5) in the main
theorem 3 in \S 4.1
(after multiplied by a constant, if necessary). Hence, by the uniqueness result
proven in section 4.1, they are up to a constant the conformal
characters of the $\w$-algebras introduced in the same section. In fact, the
$\xi_{c,h}$
($l\not=5$) have interesting product expansions, which we shall discuss
elsewhere;
from these product expansions it can immediately read off that
they can be normalized such that their Fourier coefficients are even
non-negative
integers, as its should be for conformal characters.

\subhead
$\underline{\text{Appendix}}$
\endsubhead
\demo{Proof of the Lemma}
(1) Let $l\equiv-1\bmod4$. For $c=\alpha+u\beta\in V$ we have
$$\tr(c\overline\O\sqrt{-l})/l
=\tr(\alpha\,\o/\sqrt{-l})+\tr(\beta\,\overline\p/3\sqrt{-l}).$$
Thus the left hand side is in $\Z$
if and only if each of the two terms on the right are in
$\Z$. The latter is easily checked to be equivalent to
$\alpha\in\o$ and $\beta\in\p$, i.e\. to $c\in\O$. The case
$l\equiv1\bmod4$ can be treated similarly, and is left to the reader.

It is clear that
$\O/\sqrt{-l}\O$ is a ring of characteristic $l$ with $l^2$ elements.
Hence it is isomorphic to a ring extension of $\F(l)=\Z/l\Z$ with $l^2$
elements.
Moreover, it contains a root of $X^2+3$, namely $3u+\sqrt{-l}\O$. Since
$-3$ is not a quadratic residue modulo $l$ the polynomial $X^2+3$ is
irreducible over $\F(l)$, hence $\O/\sqrt{-l}\O$ is a field. The
anti-involution $c\mapsto\overline c$ induces an automorphism of the
field $\O/\sqrt{-l}\O$ which is nontrivial since it maps $u$ to $-u$, and
which hence is the Frobenius automorphism.

(2) If $I$ is an $\O$-left ideal then $I^*=\overline I\cdot\O^*/\norm(I)$,
where, for any left ideal $I$, we use
$$I^*=\{c\in V\,|\, \tr(Ic)\in\Z\}.$$
(We were not able to find a reference for this basic formula: it can easily be
proved
using adelic methods. However, we shall need it only for $I=\O$ or $I=\O\p$
(cf\. the
two theorems of the preceding section), and here it can be easily verified by
direct
computation. We omit the details for the general case.) Thus we find
$\tr(\sqrt{-l}I\overline c)/\norm(I)l\in\Z$
if and only if $\overline c\in\overline I\cdot\O^*\sqrt{-l}$. Using
$\O^*=\O/\sqrt{-l}$,
as follows from part (1), we find that the latter statement is indeed
equivalent to $c\in I$.

Left-multiplication in the quaternion algebra induces on $I/\sqrt{-l}I$
a structure of a one-dimensional $\O/\sqrt{-l}\O$-vector space.
Let $c_0+\sqrt{-l}I$ be a basis element. Clearly $\norm(c_0)/\norm(I)$ is not
divisible by $l$
since otherwise $\norm(c)/\norm(I)$ would be divisible by $l$ for any $c\in I$
contradicting
the definition of $\norm(I)$ as g.c.d\. of all $\norm(c)$ ($c\in I$). Thus we
can choose
a $\lambda\in\O$ with $\norm(\lambda)\norm(c_0)/\norm(I)\bmod l$. Replacing
$c_0$
by $\lambda c_0$ it is then clear that $c\mapsto c c_0$ induces the
isomorphism claimed to exist.
\qed\enddemo

  \subhead
  5.3 An example (II): Comparison to formulas derivable from the
      representation theory of
      Kac-Moody and Casimir $\w$-algebras
  \endsubhead

In this section we compare our explicit formulas for
the conformal characters with the ones
obtained from the representation theory of Casimir $\w$-algebras \cite{FKW},
the Virasoro algebra \cite{RC} and Kac-Moody algebras \cite{Ka}.

The last three rational models related to Table 5.2 are minimal models
of so-called Casimir $\w$-algebras.
For this kind of algebras the minimal models have been determined
(assuming a certain conjecture) in \cite{FKW}.
The representation theory of the two composite rational models
($\w_{{\Cal G}_2}(2,1^{14})$ and $\w_{{\Cal F}_4}(2,1^{26})$)
is well-known \cite{RC,Ka}.

In order to give the explicit formulas for the conformal characters of the
minimal models of the Casimir $\w$-algebras, the Virasoro algebra and
Kac-Moody algebras we have to fix some notation first.

Let $\Cal K$ be a simple complex Lie algebra of rank $l$ and dimension $n$,
$h \ ({h^\vee})$ its (dual) Coxeter number,
$\rho \ ({\rho^\vee})$ the sum of its (dual) fundamental
weights, $W$ the Weyl group and $\Lambda$ (${\Lambda^\vee}$)
the (dual) weight lattice of  $\Cal K$. For $\lambda \in \Lambda$
denote by $\pi_{\lambda}$ the highest weight representation
with highest weight $\lambda$.

Firstly, consider the case of the three rational models of the
Casimir $\w$-algebras.
Formulas for the central charge, the conformal dimensions and the conformal
characters of rational models of Casimir $\w$-algebras have been derived
assuming a certain conjecture \cite{FKW, p.\ 320} and are collected in Appendix
8.4.

Using these formulas for ${\Cal B}_2$ with $c(p,q) = c(11,6) = -{444\over11}$
and for ${\Cal G}_2$ with $c(p,q) = c(17,12) = -{1420\over17}$ for $\w(2,4)$
and $\w(2,6)$, respectively, one obtains the conformal characters given in
the last section (as can be checked by simply comparing a sufficient number
of  Fourier coefficients).
The last rational model, of type $\w(2,8)$, is a rational model of
${\Cal WE}_7$ with $c(p,q) = c(18,23)$. However, in this case
the above formula for the corresponding conformal characters contains a
sum over a rank 7 lattice (the dual weight lattice) and a
sum over the Weyl group of ${\Cal E}_7$ which has order $2.903.040$.
Therefore, this formula is of no practical use for explicit calculations in
this case. However, our formula in the foregoing section involves only a sum
over a rank 4 lattice which is easy to implement on a computer.

Secondly, consider the rational models
$\w_{{\Cal G}_2}(2,1^{14})$ and $\w_{{\Cal F}_4}(2,1^{26})$.
These rational models are `tensor products' of the Virasoro minimal
model with $c=-{22\over5}$ and the rational model associated to the level 1
Kac-Moody algebra  of ${\Cal G}_2$ or ${\Cal F}_4$, respectively.
The two conformal characters of the Virasoro minimal model with central charge
$c=-{22\over5}$  are given by \cite{RC}
$$  \chi_{0}^{Vir}(q) =
    q^{\frac{11}{60}}\prod_{n \equiv \pm 2 \bmod 5}(1-q^n)^{-1},\qquad
    \chi_{-1/5}^{Vir}(q) =
    q^{-\frac1{60}}  \prod_{n \equiv \pm 1 \bmod 5}(1-q^n)^{-1}.
$$

The characters of rational models associated to the level 1 Kac-Moody
algebras are well known from the Kac-Weyl formula \cite{Ka, p\. 173}.
The rational model associated to the level
$k$ Kac-Moody algebra of $\Cal K$ has the following central charge
and conformal dimensions
$$ c^{\Cal K}(k) = {12k\over {h^\vee}({h^\vee}+k)}\rho^2,
   \qquad h^{\Cal K}_{\lambda} =  {(\rho+\lambda)^2-\rho^2 \over
                             2({h^\vee}+k) }
   \quad( (\lambda, \psi) \le k)
   $$
where $\psi$ is the highest root of $\Cal K$.
The corresponding characters read
$$ \chi^{{\Cal K},\lambda}(q) =
   \eta(q)^{-n} q^{{n-c^{\Cal K}(k)\over24}}\sum_{t \in {\Lambda^\vee}}
                     \dim(\pi_{\rho+\lambda+({h^\vee}+k)t})
                     q^{ (\rho+\lambda+({h^\vee}+k)t)^2-\rho^2
                         \over 2({h^\vee}+k) }.
$$
The two conformal characters associated to the level 1 Kac-Moody algebras
of ${\Cal G}_2$ and ${\Cal F}_4$ are given by:
$$ \chi_{0}^{\Cal K} = \chi^{{\Cal K},0} \quad
   \chi_{h}^{\Cal K} = \chi^{{\Cal K},\lambda_1}
$$
where $h=h^{\Cal K}_{\lambda_1}=2/5$ or $3/5$ and  $\lambda_1$ is
the fundamental weight of ${\Cal G}_2$ or ${\Cal F}_4$
with $\dim(\pi_{\lambda})=7$ or $26$, respectively.

Using these formulas one obtains exactly the four conformal characters of the
models $\w_{{\Cal G}_2}(2,1^{14})$ and $\w_{{\Cal F}_4}(2,1^{26})$:
$$\chi_{0}     = \chi_{0}^{Vir}     \cdot \chi_{0}^{\Cal K}, \quad
  \chi_{-1/5}  = \chi_{-1/5}^{Vir}  \cdot \chi_{0}^{\Cal K}, \quad
  \chi_{h}     = \chi_{0}^{Vir}     \cdot \chi_{h}^{\Cal K}, \quad
  \chi_{h-1/5} = \chi_{-1/5}^{Vir}  \cdot \chi_{h}^{\Cal K}
$$
with ${\Cal K} = {\Cal G}_2,{\Cal F}_4 $ and $h=2/5,3/5$ respectively.
The product formulas for the Virasoro characters and the formula for
the conformal characters associated to the Kac-Moody algebras show that
the Fourier coefficients of the two rational models are positive integers.
Indeed, as one can show by comparing a sufficient number of Fourier
coefficients, these conformal characters are equal to the ones computed
in the last section.

\vfill\eject
\head
6. Conclusion and outlook
\endhead

Finally, we summarize and comment on the main results in this thesis.

Firstly, we have classified all strongly-modular fusion algebras of
dimension less than or equal to four and all nondegenerate strongly-modular
fusion algebras of dimension less than 24.
In order to obtain our results we have used the classification of the
irreducible representations of the groups $\Spl{p}{\lambda}$.
Not all modular fusion algebras in our classification show up in known RCFTs.
However, all corresponding fusion algebras
are realized in known RCFTs apart from the fusion algebra
of type $\text{B}_9$. This fusion algebra can formally be related to
the Casimir $\w$-algebra ${\Cal WB}_2$ at $c=-24$ and seems to be an analogue
of the fusion algebra formally associated to the Virasoro
algebra with central charge $c = c(3,9)$.

Unfortunately, the methods used in this thesis seem to be not
sufficient for obtaining a complete classification of
strongly-modular fusion algebras. For those strongly-modular fusion algebras
which are degenerate the corresponding representation of the modular
group is in general reducible and therefore there are a lot
of possible choices of the distinguished basis in the representation
space. In the proof of the  main theorem 1 on the classification of the
strongly-modular fusion
algebras of dimension less than or equal to four we have shown how one can
deal with this problem in the case of two, three and four dimensional fusion
algebras.  However, we do not know a general method to overcome this
problem for arbitrary dimension.

We would like to stress that the main assumption for obtaining
our classifications, namely that fusion algebras are induced by
representations of $\Spl{N}{}$, is valid for all known examples
of rational conformal field theories.
Nevertheless, the question whether all fusion algebras associated to
RCFTs are strongly-modular is not yet answered
(cf. the conjecture at the end of section 2.2).

\mn
Secondly, we have shown that the
conformal characters of certain rational models are uniquely
determined by the central charge and the set of conformal dimensions
of the model.

This result has several implications. It shows that the simple
constraints imposed on modular functions by the five axioms stated in the main
theorem 3 are surprisingly restrictive. Apart from giving an aesthetical
satisfaction
this observation gives further evidence that conformal characters are modular
functions of a rather special nature, which may deserve further studies,
even independent from the theory of $\w$-algebras.

Furthermore, it implies that, in the case of the rational models
considered in \S4, the conformal characters a priori do not encode  more
information about the underlying rational model than
the central charge and the conformal dimensions.
This is in perfect accordance with the more general belief that
these data already determine completely the rational models of
$\w$-algebras which do not contain currents (currents are nonzero
elements of dimension 1). In general one expects that a unique
characterization of rational models can be obtained if one takes into
account certain additional quantum numbers which can be defined in
terms of the zero modes of the currents.

Finally, our result has a useful practical consequence for the
computation of conformal characters.
Apart from several well-understood rational models where one has simple
closed formulas for the conformal characters, it is in general difficult
to compute them directly. Any attempt to obtain the first few Fourier
coefficients by the so-called direct calculations in the $\w$-algebra,
the so far only known method in the case where no closed formulas are
available, requires considerable computer power.  Our result indicates a
way to avoid the direct calculations:
Once the central charge and conformal dimensions are determined, the
computation of the conformal characters can be viewed as a problem which
belongs solely to the theory of modular forms, i.e\. a problem whose
solution affords no further data of the rational model in question.

Of course, one of the important open questions is whether a uniqueness result
like the main theorem 3 holds for more or even for all rational models.
For rational models with effective central charge less than 26 there is at
least some hope that the central charge and the set of conformal dimensions
already determine the conformal characters:
Looking at the dimension formula in \S4.2 we see that
`main' contribution to the dimension of the space of vector valued
modular forms of weight $k$ transforming under a representation $\rho$
is given by $\frac{k-1}{12} \dim(\rho)$. For rational models with
$\tilde c<26$ this contribution is less than the number of
conformal dimensions of the rational model.
Therefore, one might hope that the $\dim(\rho)$ conditions on the pole
orders of the conformal characters at $i\infty$ imposed by fixing the central
charge and the conformal dimensions already determine the conformal
characters uniquely. Note, however, that these $\dim(\rho)$ conditions will
in general not be independent and that one also has to take into account the
`correction' terms in the dimension formula..
To obtain more general results one would like to have a dimension
formula for vector valued modular forms having a prescribed vanishing
order at $i\infty$.
One might speculate that such a formula should be related to the
Atiyah-Singer index theorem since the dimension formula in \S4.2
is related to the Riemann-Roch theorem.

\mn
Thirdly, we have shown  that one can reconstruct the
conformal characters of certain rational models merely from the knowledge
of the central
charge and the set of conformal dimensions of the model by using theta series,
and, in particular, how one obtains in this way explicit closed formulas
for the conformal characters of certain nontrivial rational models which could
not be computed using known methods.

The main unsolved question concerning the construction procedure
described in section 5 is whether all spaces of vector valued modular forms
transforming under a congruence representation are generated by theta series.

\mn
Finally, I would like to stress that the methods and results developed
in this thesis have lead to a better understanding of the structure of RCFTs.
However, the classification program of rational conformal field theories is
still a fascinating open problem and deserves future effort.
\mn
I would like to end with the following quote \cite{M}:
\mn
{\it `` ... Our general approach follows the philosophy of nahmism
(called ``nahmsense'' by its detractors) in which one begins with
modular forms and then proceeds to try to deduce some interesting
physics from them. We will show that some of the interesting forms
do indeed arise in physical theories. ...''}

\vfill\eject

\head
 Acknowledgments
\endhead

It is a pleasure to thank my Ph.D. supervisor Professor Werner Nahm for
his constant support, constructive criticism and the warm and friendly
working atmosphere.

Furthermore, I would like to thank Professor N.-P. Skoruppa and
Professor D. Zagier for their advice and many interesting discussions.

I am grateful to all members of Werner Nahm's research group for lots
of stimulating discussions.

In particular, I would like to thank
A. Honecker, R. H\"ubel,
H. Arfaei, R. Blumenhagen, L. F\'eher, M. Flohr, J. Kellendonk, A. Kliem,
S. Mallwitz, N. Mohammedi, A.~Recknagel, M. R\"osgen,
M. Terhoeven, R.Varnhagen, K. de Vos and  A. Wi{\ss}kirchen.

Finally, I would like to thank Professor F. Hirzebruch,  and the
Max-Planck-Institut in Bonn Beuel for financial support.

\vfill \eject
\head
7. Appendix
\endhead

  \subhead
  7.1 The irreducible level $p^\lambda$ representations
      of dimension $\le 4$
  \endsubhead

Using the results in \S3 one obtains as a complete list of two dimensional
irreducible level $p^\lambda$ representations
$$ \align
 &p^\lambda=2^1, \qquad N_1(\chi_1) \\
 &p^\lambda=3^1,\qquad N_1(\chi_1) \otimes B_i \\
 &p^\lambda=5^1,\qquad R_1(1,\chi_{-1}),\ R_1(2,\chi_{-1}) \\
 &p^\lambda=2^2, \qquad N_1(\chi_1)\otimes C_3 \\
 &p^\lambda=2^3, \qquad N_3(\chi)_+\otimes C_j \\
 & \text{where}\ i=1,2,3; \ j=1,\dots,4.
\endalign $$
The explicit form of the representations which are not related by tensor
products with $B_i$ or $C_j$ is given in  Table 7.1a.
\mn
\centerline{Table 7.1a: Two dimensional irreducible level $p^\lambda$
                      representations}
\smallskip\noindent
\centerline{
\vbox{ \offinterlineskip
\def\Tablespace{ height2pt&\omit&&\omit&&\omit&&\omit&\cr }
\def\Tablerule{ \Tablespace
                \noalign{\hrule}
                \Tablespace      }
\hrule
\halign{&\vrule#&
  \strut\quad\hfil#\hfil\quad\cr
\Tablespace
& level && type of rep. && $\rho(S)$ && $\frac1{2\pi i}\log(\rho(T)) $  &\cr
\Tablerule
& 2 && $N_1(\chi_1)$
  &&  $\frac1{2} \pmatrix -1 &-\sqrt{3} \\ -\sqrt{3} &1 \endpmatrix$
  &&  $\diag( 0,\frac12)$
  &\cr \Tablerule
& $3$ && $N_1(\chi)$
  &&  $-\frac{i}{\sqrt{3}} \pmatrix 1 &\sqrt{2} \\ \sqrt{2} &-1 \endpmatrix$
  &&  $\diag(\frac13, \frac23)$
  &\cr \Tablerule
& $5$ && $R_1(1,\chi_{-1})$
  &&  $\frac{2i}{\sqrt{5}}
       \pmatrix  -\sin(\frac{\pi}5)  & \sin(\frac{2\pi}5)\\
                  \sin(\frac{2\pi}5) & \sin(\frac{\pi}5)
       \endpmatrix$
  &&  $\diag( \frac15, \frac45)$
  &\cr \Tablespace\Tablespace
& \omit && $R_1(2,\chi_{-1})$
  &&  $ \frac{2i}{\sqrt{5}}
      \pmatrix  -\sin(\frac{2\pi}5)  & -\sin(\frac{\pi}5) \\
                -\sin(\frac{\pi}5)   & \sin(\frac{2\pi}5)
      \endpmatrix$
  &&  $\diag( \frac25, \frac35)$
  &\cr\Tablerule
& $2^3$ && $N_3(\chi)_{+}$
  &&  $\frac{i}{\sqrt{2}} \pmatrix -1 &-1 \\ -1 &1 \endpmatrix$
  &&  $\diag( \frac38, \frac58)$
  &\cr \Tablespace
}
\hrule}
}
\mn\mn
Similarly, one obtains as a complete list of three dimensional
irreducible level $p^\lambda$ representations
$$ \align
 &p^\lambda=3^1,\qquad N_1(\chi_1) \\
 &p^\lambda=5^1,\qquad R_1(1,\chi_1),\ R_1(2,\chi_1) \\
 &p^\lambda=7^1,\qquad R_1(1,\chi_{-1}),\ R_1(2,\chi_{-1}) \\
 &p^\lambda=2^2, \qquad D_2(\chi)_+\otimes C_j \\
 &p^\lambda=2^3, \qquad R^0_3(1,3,\chi)_+ \otimes C_j,\
                        R^0_3(1,3,\chi)_- \otimes C_j\\
 &p^\lambda=2^4, \qquad R^0_4(1,1,\chi)_+ \otimes C_j,\
                        R^0_4(1,1,\chi)_- \otimes C_j,\\
 &\qquad\qquad\qquad    R^0_4(3,1,\chi)_+ \otimes C_j,\
                        R^0_4(3,1,\chi)_- \otimes C_j\\
 & \text{where}\ j=1,\dots,4.
\endalign $$
The explicit form of the representations which are not related by tensor
products with $C_j$ is given in Table 7.1b.

\vfill\eject
\centerline{Table 7.1b: Three dimensional irreducible level
 $p^\lambda$  representations}
\smallskip\noindent
\centerline{
\vbox{ \offinterlineskip
\def\Tablespace{ height2pt&\omit&&\omit&&\omit&&\omit&\cr }
\def\Tablerule{ \Tablespace
                \noalign{\hrule}
                \Tablespace      }
\hrule
\halign{&\vrule#&
  \strut\quad\hfil#\hfil\quad\cr
\Tablespace
& level && type of rep. && $\rho(S)$ && $\frac1{2\pi i}\log(\rho(T))$
 &\cr \Tablerule
& 3 && $N_1(1,\chi_1)$
  &&  $ \frac13\pmatrix  -1 &2 &2 \\ 2 &-1 &2 \\ 2 &2 &-1 \endpmatrix$
  &&  $\diag( \frac13, \frac23, 0)$
  &\cr \Tablerule\Tablerule
& 5 && $R_1(1,\chi_1)$
  &&  $ \frac2{\sqrt{5}}\pmatrix
              \frac12 &\frac1{\sqrt2}      &\frac1{\sqrt2} \\
       \frac1{\sqrt2} &-s_1  &s_2 \\
       \frac1{\sqrt2} & s_2 &-s_1
       \endpmatrix$
  &&  $\diag(0,\frac15, \frac45 )$
  &\cr \Tablespace\Tablespace
& \omit && $R_1(2,\chi_1)$
  &&  $ \frac2{\sqrt{5}}\pmatrix
              -\frac12 &-\frac1{\sqrt2}      &-\frac1{\sqrt2} \\
       -\frac1{\sqrt2} &-s_2   &s_1 \\
       -\frac1{\sqrt2} & s_1   &-s_2
       \endpmatrix$
  &&  $\diag( 0, \frac25, \frac35 )$
  &\cr \Tablespace
&\omit && \omit && $s_j = \cos(\frac{j\pi}5)$ && \omit
  &\cr \Tablerule\Tablerule
& 7 && $R_1(1,\chi_{-1})$
  &&  $\frac2{\sqrt{7}} \pmatrix
       s_1 &  s_2 &  s_3 \\
      s_2 & -s_3 &  s_1 \\
      s_3 &   s_1 & -s_2
      \endpmatrix$
  &&  $\diag( \frac27, \frac17, \frac47 )$
  &\cr \Tablespace\Tablespace
& \omit && $R_1(2,\chi_{-1})$
  &&  $ -- \text{"} -- $
  &&  $\diag( \frac57, \frac67, \frac37 )$
  &\cr \Tablespace
&\omit && \omit && $s_j = \sin(\frac{j\pi}7)$ && \omit
  &\cr \Tablerule\Tablerule
& $2^2$ && $D_2(\chi)_+$
  &&  $ \frac{i}2\pmatrix  0        &\sqrt{2} &\sqrt{2} \\
                           \sqrt{2} &-1        &1       \\
                           \sqrt{2} &1         &-1
       \endpmatrix$
  &&  $\diag(\frac14, \frac12, 0 )$
 &\cr \Tablerule\Tablerule
& $2^3$ && $R^0_3(1,3,\chi)_+$
  &&  $ \frac{i}2\pmatrix  0        &\sqrt{2} &\sqrt{2} \\
                           \sqrt{2} &1        &-1       \\
                           \sqrt{2} &-1       &1
       \endpmatrix$
  &&  $\diag( \frac12, \frac58, \frac18 )$
 &\cr\Tablespace\Tablespace
& \omit && $R^0_3(1,3,\chi)_-$
  &&   $ (-1)\ \cdot (--\text{"}--) $
  &&  $\diag( \frac12, \frac78,\frac38 )$
 &\cr\Tablerule\Tablerule
& $2^4$ && $R^0_4(1,1,\chi)_{+}$
  &&  $\frac{i}2\pmatrix  0        &\sqrt{2} &\sqrt{2} \\
                          \sqrt{2} &1        &-1       \\
                          \sqrt{2} &-1       &1
       \endpmatrix$
  &&  $\diag(\frac{5}{8}, \frac{1}{16}, \frac{9}{16} )$
&\cr\Tablespace\Tablespace
& \omit && $R^0_4(1,1,\chi)_{-}$
  &&  $-- \text{"} --$
  &&  $\diag( \frac{1}{8}, \frac{5}{16}, \frac{13}{8} )$
&\cr\Tablespace\Tablespace
& \omit && $R^0_4(3,1,\chi)_{+}$
  &&  $\frac{i}2\pmatrix  0        &\sqrt{2} &\sqrt{2} \\
                          \sqrt{2} &-1       &1        \\
                          \sqrt{2} &1        &-1
       \endpmatrix$
  &&  $\diag( \frac{7}{8}, \frac{3}{16}, \frac{11}{16} )$
&\cr\Tablespace\Tablespace
& \omit && $R^0_4(3,1,\chi)_{-}$
  &&  $ -- \text{"} -- $
  &&  $\diag( \frac{3}{8}, \frac{15}{16}, \frac{7}{16} )$
 &\cr\Tablespace
}
\hrule}
}
\vfill\eject
\centerline{Table 7.1c: Four dimensional irreducible level
                      $p^\lambda$ representations}
\smallskip\noindent
\centerline{
\vbox{ \offinterlineskip
\def\Tablespace{ height2pt&\omit&&\omit&&\omit&&\omit&\cr }
\def\Tablerule{ \Tablespace
                \noalign{\hrule}
                \Tablespace      }
\hrule
\halign{&\vrule#&
  \strut\quad\hfil#\hfil\quad\cr
\Tablespace
& level && type of rep. && $\rho(S)$ && $\frac1{2\pi i}\log(\rho(T))$  &\cr
\Tablerule
& 5 && $N_1(\chi),\ \chi^3 \not\equiv 1$
  && $\frac{2i}5
       \pmatrix  \eta_-       & \sqrt{3} s_2 & \eta_+       & \sqrt{3} s_4 \\
                 \sqrt{3} s_2 &  -\eta_+     & \sqrt{3} s_4 & \eta_- \\
                 \eta_+       & \sqrt{3} s_4 & -\eta_-      & -\sqrt{3} s_2 \\
                 \sqrt{3} s_4 & \eta_-       & -\sqrt{3} s_2& \eta_+  \\
       \endpmatrix$
  &&  $\diag( \frac35,\frac45,\frac25,\frac15 )$
 &\cr \Tablespace\Tablespace
&\omit && \omit
 && $ s_j = \sin(\frac{j\pi}5),\
      \eta_{\pm} = s_2 \pm s_4$
 && \omit
  &\cr \Tablespace\Tablespace
& \omit && $N_1(\chi),\ \chi^3 \equiv 1$
  && $-\frac25
       \pmatrix  \xi_1 & -\xi_2 & \xi_1 & -\xi_3  \\
                -\xi_2 & -\xi_1 & \xi_3 & \xi_1 \\
                 \xi_1 &  \xi_3 & \xi_1 & \xi_2 \\
                -\xi_3 &  \xi_1 & \xi_2 & -\xi_1  \\
       \endpmatrix$
  &&  $\diag( \frac35, \frac45,\frac25,\frac15 )$
  &\cr \Tablespace\Tablespace
&\omit && \omit
 && $  r_j = \cos(\frac{j\pi}5),\
      \xi_1 = r_1-r_4-\frac12,$
 && \omit
  &\cr \Tablespace
&\omit && \omit
 && $ \xi_2 = 3r_2+ 2r_4,\
      \xi_3 = 2r_2+ 3r_4$
 && \omit
  &\cr \Tablerule\Tablerule
& 7 && $R_1(1,\chi_1) $
  && $\sqrt{\frac{2}7}i \pmatrix
  -\frac1{\sqrt{2}} & -1     &   -1  &   -1 \\
  -1               &  \xi_1 & \xi_2 & \xi_3 \\
  -1               &  \xi_2 & \xi_3 & \xi_1 \\
  -1               &  \xi_3 & \xi_1 & \xi_2 \\
       \endpmatrix$
  &&  $ \diag( 0,\frac17,\frac47,\frac27 )$
  &\cr \Tablespace\Tablespace
& \omit && $R_1(2,\chi_1) $
  && $(-1) \cdot\qquad (-- \text{"} -- )$
  &&  $ \diag(0,\frac67,\frac37,\frac57)$
  &\cr \Tablespace\Tablespace
&\omit && \omit
 && $  s_j = \sqrt{\frac27}\sin(\frac{j\pi}7), $
&& \omit
  &\cr \Tablespace
&\omit && \omit
 && $ \xi_1 = 2s_2-s_4,\
      \xi_2 = 2s_4+s_6 $
&& \omit
  &\cr \Tablespace
&\omit && \omit
 &&  $\xi_2 = 2s_4+s_6,\
      \xi_3 = -2s_6-s_2 $
 && \omit
  &\cr \Tablerule\Tablerule
& $2^3$ && $N_3(\chi),\ \chi^3 \not\equiv 1 $
  && $ \frac{i}{\sqrt{8}} \pmatrix
   1 &  1 &  \sqrt{3} i & -s_1 \sqrt{3}i \\
   1 & -1 & -\sqrt{3} i & -s_1 \sqrt{3}i  \\
   -\sqrt{3} i  & \sqrt{3} i & 1 &  s_1 \\
  s_2 \sqrt{3}i & s_2 \sqrt{3} i & s_2 & -1 \\
       \endpmatrix$
  &&  $\diag(\frac38,\frac58,\frac18,\frac78) $
  &\cr \Tablespace\Tablespace
&\omit && \omit
 && $ s_j = e^{2\pi i \frac{j}3} $
 && \omit
  &\cr \Tablerule\Tablerule
& $3^2$ && $R^1_2(1,1,\chi),\ \chi^3 \equiv 1 $
  && $ \frac{2i}3\pmatrix
       -s_8 & -s_4 & -s_2 & -s_6 \\
       -s_4 &  s_2 & -s_8 &  s_6 \\
       -s_2 & -s_8 &  s_4 &  s_6 \\
       -s_6 &  s_6 &  s_6 &    0 \\
       \endpmatrix$
  &&  $\diag(\frac49,\frac19,\frac79,\frac13) $
  &\cr \Tablespace\Tablespace
& \omit && $R^1_2(2,1,\chi),\ \chi^3 \equiv 1 $
  && $ (-1)\ \cdot \quad (-- \text{"} --) $
  &&  $\diag( \frac29,\frac59,\frac89,\frac23 )$
  &\cr \Tablespace\Tablespace
& \omit && $R^1_2(1,1,\chi),\ \chi^3 \not\equiv 1 $
  && $ \frac{2}3\pmatrix
        s_1 &  s_5 &  s_7 &  s_6 \\
        s_5 & -s_7 & -s_1 &  s_6 \\
        s_7 & -s_1 &  s_5 & -s_6 \\
        s_6 &  s_6 &  -s_6 &    0 \\
       \endpmatrix$
  &&  $\diag( \frac49,\frac19,\frac79,\frac13 )$
  &\cr \Tablespace\Tablespace
& \omit && $R^1_2(2,1,\chi),\ \chi^3 \not\equiv 1 $
  &&  $ -- \text{"} -- $
  &&  $\diag( \frac59,\frac89,\frac29,\frac23) $
  &\cr \Tablespace\Tablespace
&\omit && \omit
 && $ s_j = \sin(\frac{\pi j}{18}) $
 && \omit &\cr\Tablespace
}
\hrule}
}
\vfill
\eject

Finally, one obtains as a complete list of four dimensional
irreducible level $p^\lambda$ representations
$$ \align
 &p^\lambda=5^1,\qquad N_1(\chi)\ (\chi^3\not\equiv 1),
                       N_1(\chi)\ (\chi^3\equiv 1),  \\
 &p^\lambda=7^1,\qquad R_1(1,\chi_1),\ R_1(2,\chi_1) \\
 &p^\lambda=2^3,\qquad N_3(\chi),\ C_4\otimes  N_3(\chi) \\
 &p^\lambda=3^2,\qquad B_i \otimes R^1_2(1,1,\chi),
                       B_i \otimes R^1_2(2,1,\chi)
\endalign $$
where $i=1,2,3$ and for $p^\lambda=3^2$ the character $\chi$ is a
primitive character of order 3 or 6 (so there are 12 four dimensional
irreducible level $3^2$ representations).

The explicit form of the representations which are not related by tensor
products with $C_j$ or $B_i$ is given in  Table 7.1c.

\vfill\eject

  \subhead
  7.2 The strongly-modular fusion algebras of dimension $\le 4$
  \endsubhead

In this appendix we give complete lists the simple strongly-modular
fusion algebras of dimension less than or equal to four.
\mn
\centerline{Table 7.2a: Two and three dimensional strongly-modular
            fusion algebras }
\smallskip\noindent
\centerline{
\vbox{ \offinterlineskip
\def\Tablespace{ height2pt&\omit&&\omit&&\omit&\cr }
\def\Tablerule{ \Tablespace
                \noalign{\hrule}
                \Tablespace      }
\hrule
\halign{&\vrule#&
  \strut\quad\hfil#\hfil\quad\cr
\Tablespace
& $\Cal F$ && $\rho(S)$ && $\frac1{2\pi i }\log(\rho(T))\bmod \Z $  &\cr
\Tablerule
& $\Phi_1\cdot\Phi_1 = \Phi_0$
  &&  $\frac1{\sqrt2} \pmatrix -1 &-1 \\ -1 &1 \endpmatrix$
  &&  $\cases
        \diag( \frac18, \frac38) &\\
        \diag( \frac78, \frac58) &\\
       \endcases$
   & \cr  \Tablespace\Tablespace
& ( $\Z_2$ ) &&\omit &&\omit &\cr \Tablerule\Tablerule
& $\Phi_1\cdot\Phi_1 = \Phi_0 + \Phi_1$
       &&  $\frac{2}{\sqrt{5}}
            \pmatrix -\sin(\frac{\pi}{5}) & -\sin(\frac{2\pi}{5}) \\
                     -\sin(\frac{2\pi}{5}) &  \sin(\frac{\pi}{5})
            \endpmatrix$
       && $\cases
            \diag( \frac{19}{20}, \frac{11}{20} ) &\\
            \diag( \frac{1}{20}, \frac{9}{20} ) &\\
            \endcases$
  &\cr  \Tablespace \Tablespace\Tablespace
& ( "$(2,5)$" ) &&  $\frac{2}{\sqrt{5}}
            \pmatrix  -\sin(\frac{2\pi}{5}) & \sin(\frac{\pi}{5}) \\
                       \sin(\frac{\pi}{5})  & \sin(\frac{2\pi}{5})
            \endpmatrix$
       && $\cases
            \diag( \frac{3}{20}, \frac{7}{20} )&\\
            \diag( \frac{ 17}{20}, \frac{13}{20} )&\\
           \endcases$
  &\cr  \Tablerule\Tablerule\Tablerule
& $\Phi_1\cdot\Phi_1 = \Phi_2$
  &&  $$
  &&  $$
  &\cr \Tablespace\Tablespace\Tablespace
& $\Phi_1\cdot\Phi_2 = \Phi_0$
  &&  $\frac1{\sqrt{3}}\pmatrix 1 & 1                  & 1 \\
                                1 & e^{2\pi i \frac13} & e^{2\pi i \frac23} \\
                                1 & e^{2\pi i \frac23} & e^{2\pi i \frac13}
       \endpmatrix$
  &&  $ \diag( \frac14, \frac7{12}, \frac7{12})$
  &\cr \Tablespace\Tablespace\Tablespace
& $\Phi_2\cdot\Phi_2 = \Phi_1$
  &&  $$
  &&  $$
  &\cr \Tablespace\Tablespace
& ( $\Z_3$ )
  &&  $$
  &&  $$
  &\cr \Tablerule\Tablerule
& $\Phi_1\cdot\Phi_1 = \Phi_0 + \Phi_2$
 &&  $\frac2{\sqrt{7}} \pmatrix
      -s_2 &  -s_1 &   s_3 \\
       -s_1 & -s_3 &  -s_2 \\
       s_3 & -s_2 &    s_1
      \endpmatrix$
  &&  $\cases
        \diag( \frac47, \frac17, \frac27 ) &\\
        \diag( \frac37, \frac67, \frac57 ) &\\
       \endcases$
  &\cr \Tablespace\Tablespace\Tablespace
& $\Phi_1\cdot\Phi_2 = \Phi_1 + \Phi_2$
  &&  $\frac2{\sqrt{7}} \pmatrix
       -s_3 &   -s_1 &   s_2 \\
       -s_1  &  -s_2 &  -s_3 \\
        s_2 &  -s_3 &    s_1
      \endpmatrix$
  &&  $\cases
        \diag( \frac17, \frac47, \frac27)&\\
        \diag( \frac67, \frac37, \frac57)&\\
       \endcases$
  &\cr \Tablespace\Tablespace\Tablespace
& $\Phi_2\cdot\Phi_2 = \Phi_0 +\Phi_1 +\Phi_2$
 &&  $\frac2{\sqrt{7}} \pmatrix
       s_1 &  s_2 &  s_3 \\
      s_2 & -s_3 &  s_1 \\
      s_3 &   s_1 & -s_2
      \endpmatrix$
  &&  $\cases
        \diag( \frac27, \frac17, \frac47 )&\\
        \diag( \frac57, \frac67, \frac37 )&\\
       \endcases$
  &\cr \Tablespace\Tablespace\Tablespace
& ( "$(2,7)$" )   && $ s_j = \sin(\frac{j \pi}7)$
  &&  \omit
  &\cr \Tablerule\Tablerule
& $\Phi_1\cdot\Phi_1 = \Phi_0$
  &&  $$
  &&  $$
  &\cr \Tablespace\Tablespace\Tablespace
& $\Phi_1\cdot\Phi_2 = \Phi_2$
  &&  $\frac12\pmatrix 1         &1         & \sqrt{2}\\
                       1         &1         &-\sqrt{2}\\
                       \sqrt{2}  &-\sqrt{2} &0
       \endpmatrix$
  &&  $\cases
        \diag( \frac{8-n}{16}, \frac{16-n}{16}, \frac{n}8 ) &\\
        \diag( \frac{16-n}{16}, \frac{8-n}{16}, \frac{n}8 ) &\\
        n=0,\dots, 7 &\\
          \endcases$
  &\cr \Tablespace\Tablespace\Tablespace
& $\Phi_2\cdot\Phi_2 = \Phi_0 +\Phi_1$
  &&  $$
  &&  $$
  &\cr \Tablespace
& ( "$(3,4)$" )
  &&  $$
  &&  $$
  &\cr \Tablespace
}
\hrule}
}
\vfill
\eject
\centerline{Table 7.2b: Four dimensional simple strongly-modular fusion
algebras}
\smallskip\noindent
\centerline{
\vbox{ \offinterlineskip
\def\Tablespace{ height2pt&\omit&&\omit&&\omit&\cr }
\def\Tablerule{ \Tablespace
                \noalign{\hrule}
                \Tablespace      }
\hrule
\halign{&\vrule#&
  \strut\quad\hfil#\hfil\quad\cr
\Tablespace
& $\Cal F$ && $\rho(S)$ && $\frac1{2\pi i }\log(\rho(T)) \bmod \Z$  &\cr
\Tablerule
& ${\displaystyle{ \Phi_1^2 = \Phi_2,\ \ \Phi_1\cdot\Phi_2 = \Phi_3,}
      \atop \displaystyle{} }
   \atop
   {\displaystyle{} \atop
    \displaystyle {
    \Phi_2^2 = \Phi_0,\ \  \Phi_1\cdot\Phi_3 = \Phi_0,} }$
 && $\frac12 \pmatrix 1&1&1&1\\ 1&i&-1&-i\\
                      1&-1&-1& -1\\ 1&-i& -1&i
             \endpmatrix$
 && $\cases
      \diag(\frac78, \frac14, \frac 38, \frac 14) &\\
      \diag(\frac38, \frac14, \frac 78, \frac 14) &\\
      \endcases$
 &\cr \Tablespace\Tablespace
& ${\displaystyle{ \Phi_3^2 = \Phi_2,\ \ \Phi_2\cdot\Phi_3 = \Phi_1,}
      \atop \displaystyle{} }
   \atop
   {\displaystyle{} \atop
    \displaystyle { (\ \Z_4 \ )} }$
 && $\frac12 \pmatrix 1&1&1&1\\ 1&-i&-1&i\\
                      1&-1&-1& -1\\ 1&i& -1&-i
             \endpmatrix$
 && $\cases
      \diag(\frac58, \frac34, \frac 18, \frac 34) &\\
      \diag(\frac18, \frac34, \frac 58, \frac 34) &\\
      \endcases$
 &\cr \Tablerule\Tablerule
& ${\displaystyle{ \Phi_1^2 = \Phi_0,\ \ \Phi_1\cdot\Phi_2 = \Phi_3,}
      \atop \displaystyle{\Phi_2^2 = \Phi_0,\ \  \Phi_1\cdot\Phi_3 = \Phi_2,} }
   \atop
   {\displaystyle{\Phi_3^2 = \Phi_0,\ \  \Phi_2\cdot\Phi_3 = \Phi_1 \,} \atop
    \displaystyle{} }$
 &&$\frac12 \pmatrix 1&-1&-1&-1 \\ -1&1&-1&-1 \\ -1&-1&1&-1 \\ -1&-1&-1&1
           \endpmatrix$
 &&$ \cases
   \diag(0,0,0,\frac12) &\\
   \diag(\frac12,0,0,0) &\\
   \endcases$
 &\cr \Tablespace\Tablespace
& ( $\Z_2 \otimes \Z_2$ ) && \omit && \omit
 &\cr \Tablerule\Tablerule
& ${\displaystyle{ \Phi_1^2 = \Phi_0+\Phi_3}  \atop \displaystyle{} }
   \atop
   {\displaystyle{} \atop
    \displaystyle {\Phi_1 \cdot \Phi_2 = \Phi_1+\Phi_3} }$
 && $ \frac23 \pmatrix
       -s_4 &  s_1 &  s_3 & -s_2 \\
        s_1 &  s_2 &  s_3 &  s_4 \\
        s_3 &  s_3 &    0 & -s_3 \\
       -s_2 &  s_4 & -s_3 &  s_1 \\
       \endpmatrix$
  && $\cases  \diag(\frac7{36},\frac{19}{36},\frac1{12},\frac{31}{36})  &\\
              \diag(\frac{29}{36},\frac{17}{36},\frac{11}{12},\frac5{36})  &\\
      \endcases $
 &\cr \Tablespace\Tablespace
& ${\displaystyle{ \Phi_1\cdot\Phi_3 = \Phi_2+\Phi_3}  \atop
    \displaystyle{} }
   \atop
   {\displaystyle{} \atop
    \displaystyle {  \Phi_2^2 = \Phi_0+\Phi_2+\Phi_3 } }$
 && $ \frac23 \pmatrix
       s_1 & s_2 & s_3 & s_4 \\
       s_2 &-s_4 & s_3 & -s1 \\
       s_3 & s_3 &   0 &-s_3 \\
       s_4 &-s_1 & -s_3 & s_2 \\
       \endpmatrix$
  && $\cases  \diag(\frac{31}{36},\frac7{36},\frac{1}{12},\frac{19}{36})  &\\
              \diag(\frac5{36},\frac{29}{36},\frac{11}{12},\frac{17}{36})  &\\
      \endcases $
 &\cr \Tablespace\Tablespace
& ${\displaystyle{ \Phi_2\cdot\Phi_3 = \Phi_1+\Phi_2+\Phi_3 }  \atop
    \displaystyle{} }
   \atop
   {\displaystyle{} \atop
    \displaystyle {  \Phi_3^2 = \Phi_0+\Phi_1+\Phi_2+\Phi_3 } }$
 && $ \frac23 \pmatrix
       s_2 &-s_4 & s_3 &-s_1 \\
      -s_4 & s_1 & s_3 &-s_2 \\
       s_3 & s_3 &   0 &-s_3 \\
      -s_1 &-s_2 &-s_3 &-s4 \\
       \endpmatrix$
  && $\cases \diag(\frac{19}{36},\frac{31}{36},\frac1{12},\frac7{36})  &\\
             \diag(\frac{17}{36},\frac{5}{36},\frac{11}{12},\frac{29}{12})  &\\
      \endcases $
 &\cr \Tablespace\Tablespace
& ( "$(2,9)$" )  && $ s_j = \sin(\frac{j\pi}9) $ && \omit
 &\cr \Tablespace
}
\hrule}
}

\vfill\eject

  \subhead
  7.3 The strongly-modular fusion algebras of dimension less than 24:
      Representations $\rho$, fusion matrices and graphs
  \endsubhead

In this appendix we present the representations $\rho$ of the modular
group, the fusion matrices and the fusion graphs related to the
nondegenerate strongly-modular fusion algebras of dimension less than 24.
\mn
\centerline{Table 7.3: Simple nondegenerate strongly-modular fusion of
dimension
                      less than 24}
\centerline{ ($q$ is a prime satisfying $q<47$)}
\smallskip\noindent
\centerline{
\vbox{ \offinterlineskip
\def\Tablespace{ height2pt&\omit&&\omit&&\omit&\cr }
\def\Tablerule{ \Tablespace
                \noalign{\hrule}
                \Tablespace      }
\hrule
\halign{&\vrule#&
  \strut\quad\hfil#\hfil\quad\cr
\Tablespace
&  fusion && dim && $\rho$
   &\cr \Tablerule
& $\Z_2$
   && $2$
   && $C_4\otimes N_3(\chi)_{\pm},\ (p^\lambda = 2^3)$
   &\cr \Tablerule
& "$c(3,4)$"
   && $3$
   && $ C_4\otimes D_2(\chi)_+,\quad (p^\lambda = 2^2) $
  &\cr \Tablespace
&  \omit    && \omit
   && $C_4\otimes R^0_3(1,3,\chi)_{\pm}, \quad (p^\lambda = 2^3)$
   &\cr \Tablespace
&  Ising      && \omit
   && $C_4\otimes R^0_4(r,3,\chi)_{\pm}, \ (r=1,2; p^\lambda = 2^4)$
   &\cr \Tablerule
&    "$(2,q)$"
   && $\frac12 (q-1)$
   && $ C_4^{\frac{q+1}2}\otimes
        R_1(r,\chi_{-1}),\ ( \bigl({r\over p}\bigr) = \pm 1;
                                p^\lambda = q  )$
   &\cr \Tablerule
&  "$(2,9)$"
   &&  4
   &&  $C_4\otimes R^1_2(r,1,\chi),\quad
        (r= 1,2; \chi^3\equiv 1; p^\lambda=3^2)$
   &\cr \Tablerule
& $\text{B}_{9}$
   &&  6
   && $N_2(\chi),\quad (\chi^3 \equiv 1; p^\lambda = 3^2)$
   &\cr \Tablerule
& $\text{B}_{11}$
   &&  10
   &&  $N_1(\chi),\quad (\chi^3\equiv 1; p^\lambda =11)$
   &\cr \Tablerule
& $\text{G}_9$
   && 12
   && $C_4\otimes R_3^1(r,1,\chi),\quad (r=1,2; \chi^3\equiv 1;p^\lambda=3^3)$
   &\cr \Tablerule
& $\text{G}_{17}$
   &&  16
   &&  $N_1(\chi),\quad (\chi^3\equiv 1; p^\lambda =17)$
   &\cr \Tablerule
& $\text{E}_{23}$
   && 22
   &&  $N_1(\chi),\quad (\chi^3\equiv 1; p^\lambda =23)$
   &\cr \Tablespace
}
\hrule
}
}
\mn
The fusion matrices ${\Cal N}_1$ which define the distinguished basis
of the simple nondegenerate strongly-modular fusion algebras of
dimension less than 24 are given by:
\mn
$$\align
   &\Z_2:\qquad\quad
    {\Cal N}_1 = \pmatrix 0 & 1 \\ 1 & 0 \endpmatrix  \\
   &\text{"} (3,4) \text{"}: \quad
    {\Cal N}_1 = \pmatrix 0 & 0 & 1 \\ 0 & 0 & 1 \\ 1 & 1 & 0 \endpmatrix \\
   &\text{"} (2,q) \text{"}:\quad
    {\Cal N}_1 = \left. \pmatrix  0 & 1       &         &     \\
                           1 & \ddots  &  \ddots &     \\
                             & \ddots  &   0     &  1  \\
                             &         &   1     &  1  \\
                \endpmatrix  \right\} \scriptstyle{\frac{q-1}{2}}\\
  & \text{B}_{9}:\ \qquad
    {\Cal N}_1 =  \pmatrix 0&1&0& & & \\
                           1&0&1&1& & \\
                           0&1&0&0&1& \\
                            &1&0&1&1&0\\
                            & &1&1&1&1\\
                            & & &0&1&1
                  \endpmatrix \\
\endalign
$$
\vfill\eject
$$\align
 &\text{B}_{11}:\qquad
    {\Cal N}_1  = \left(\smallmatrix 0&1&0&0& & & & & & \\
                           1&0&1&0&0& & & & & \\
                           0&1&0&0&1&0& & & & \\
                           0&1&0&0&1&0&1& & & \\
                            &0&1&1&0&1&0&1& & \\
                            & &0&0&1&0&0&0&1& \\
                            & & &1&0&0&0&1&0&0\\
                            & & & &1&0&1&1&1&0\\
                            & & & & &1&0&1&1&1\\
                            & & & & & &0&0&1&1\\
                 \endsmallmatrix\right) \\
&\text{G}_9:\qquad
  {\Cal N}_1 =
   \left(\smallmatrix
0& 1& 0& 0& 0&  &  &  &  &  &  &  \\
1& 1& 1& 1& 0& 0&  &  &  &  &  &  \\
0& 1& 1& 1& 1& 1& 0&  &  &  &  &  \\
0& 1& 1& 0& 0& 1& 0& 0&  &  &  &  \\
0& 0& 1& 0& 1& 1& 1& 1& 0&  &  &  \\
 & 0& 1& 1& 1& 1& 0& 1& 1& 0&  &  \\
 &  & 0& 0& 1& 0& 1& 1& 0& 1& 1&  \\
 &  &  & 0& 1& 1& 1& 1& 1& 0& 1& 1\\
 &  &  &  & 0& 1& 0& 1& 0& 0& 0& 1\\
 &  &  &  &  & 0& 1& 0& 0& 1& 1& 0\\
 &  &  &  &  &  & 1& 1& 0& 1& 1& 1\\
 &  &  &  &  &  &  & 1& 1& 0& 1& 1\\
   \endsmallmatrix\right) \\
 &\text{G}_{17}:\qquad
{\Cal N}_1 =
 \left(\smallmatrix
0&1&0&0&0&0&0&0&0& & & & & & & \\
1&1&0&1&1&0&0&0&0&0& & & & & & \\
0&0&0&0&0&1&0&0&0&1&0& & & & & \\
0&1&0&0&1&0&0&0&0&0&1&0& & & & \\
0&1&0&1&1&0&0&0&0&0&1&1&0& & & \\
0&0&1&0&0&0&0&0&0&1&0&0&0&1& & \\
0&0&0&0&0&0&0&0&1&1&0&0&0&0&1& \\
0&0&0&0&0&0&0&0&0&0&1&0&0&1&0&1\\
0&0&0&0&0&0&1&0&1&0&0&0&1&0&1&0\\
 &0&1&0&0&1&1&0&0&1&0&0&0&1&1&0\\
 & &0&1&1&0&0&1&0&0&1&1&0&0&0&1\\
 & & &0&1&0&0&0&0&0&1&1&1&0&0&1\\
 & & & &0&0&0&0&1&0&0&1&1&0&1&1\\
 & & & & &1&0&1&0&1&0&0&0&1&1&1\\
 & & & & & &1&0&1&1&0&0&1&1&1&1\\
 & & & & & & &1&0&0&1&1&1&1&1&1\\
\endsmallmatrix\right) \\
&\text{E}_{23}:\qquad
{\Cal N}_1 =
\left( \smallmatrix
0&1&0&0&0&0&0&0&0&0&0&0&0&0&0& & & & & & & \\
1&0&1&1&0&0&1&0&0&0&0&0&0&0&0&0& & & & & & \\
0&1&0&0&1&0&0&0&0&1&0&0&0&0&0&0&0& & & & & \\
0&1&0&1&0&0&0&0&0&1&0&0&0&0&0&1&0&0& & & & \\
0&0&1&0&0&0&1&1&0&0&0&0&0&0&0&0&0&0&1& & & \\
0&0&0&0&0&0&0&0&0&1&0&1&0&0&1&0&0&0&0&0& & \\
0&1&0&0&1&0&0&0&0&1&0&0&0&1&0&1&0&0&0&0&0& \\
0&0&0&0&1&0&0&0&0&1&0&0&0&1&1&0&0&0&0&0&0&1\\
0&0&0&0&0&0&0&0&1&0&0&0&1&0&1&0&0&0&0&0&1&0\\
0&0&1&1&0&1&1&1&0&0&0&0&0&0&0&1&0&0&1&1&0&0\\
0&0&0&0&0&0&0&0&0&0&0&1&1&0&1&0&0&1&0&0&0&1\\
0&0&0&0&0&1&0&0&0&0&1&0&0&0&0&1&0&0&0&1&1&0\\
0&0&0&0&0&0&0&0&1&0&1&0&1&0&0&0&1&1&0&0&1&0\\
0&0&0&0&0&0&1&1&0&0&0&0&0&0&0&0&1&0&1&1&0&1\\
0&0&0&0&0&1&0&1&1&0&1&0&0&0&0&0&1&0&1&1&1&0\\
 &0&0&1&0&0&1&0&0&1&0&1&0&0&0&1&0&0&1&1&0&1\\
 & &0&0&0&0&0&0&0&0&0&0&1&1&1&0&1&1&0&0&1&1\\
 & & &0&0&0&0&0&0&0&1&0&1&0&0&0&1&1&0&1&1&1\\
 & & & &1&0&0&0&0&1&0&0&0&1&1&1&0&0&1&0&1&1\\
 & & & & &0&0&0&0&1&0&1&0&1&1&1&0&1&0&1&1&1\\
 & & & & & &0&0&1&0&0&1&1&0&1&0&1&1&1&1&1&1\\
 & & & & & & &1&0&0&1&0&0&1&0&1&1&1&1&1&1&1\\
\endsmallmatrix
\right).
\endalign
$$
The corresponding fusion graphs can be found in a seperate postscript file.

\vfill\eject

\subhead
7.4 Minimal models of Casimir $\w$-algebras
\endsubhead

In this last appendix we give some data related to the rational models
of Casimir $\w$-algebras.

Let ${\Cal K}$ be a simple Lie algebra of rank $l$ over $\C$.
Then the rational models of the Casimir $\w$-algebra
related to this Lie algebra have central charge
$$c_{\Cal K}(p,q) = l -
{\textstyle{12\over pq}}(q\,\rho - p\,{\rho^\vee})^2
\quad p,q\ \hbox{coprime}, \quad {h^\vee} \le p \quad h\le q
$$
where $p$ and $q$ have to be chosen minimal,
$h$ ($h^\vee$) denotes the (dual) Coxeter number of
${\Cal K}$ and $\rho$ ($\rho^\vee$) denotes the sum of
its (dual) fundamental weights $\lambda_i$ ($\lambda_i^\vee$).
The conformal dimensions and conformal characters of the minimal model
are given by \cite{FKW}:
$$\align
   h_{\lambda,\nu^\vee} &= {1\over{2pq}}
     \left( (q\lambda-p\nu^\vee)^2-(q\rho-p\rho^\vee)^2\right) \\
  \chi_{\lambda,{\nu^\vee}}(q)  &=
      \eta(q)^{-l} \sum_{w \in W}
      \sum_{ t \in {\Lambda^\vee}} \epsilon (w)
      q^{{1\over2pq}{(q w(\lambda+\rho) -
                      p ({\nu^\vee}+{\rho^\vee}) + pqt)}^2}
  \endalign$$
where $\lambda$ ($\nu^\vee$) lies in the (dual) weight lattice
so that
$\lambda = \sum_{i=1}^l l_i \lambda_i$ and
$\nu^\vee = \sum_{i=1}^l l_i^\vee \lambda_i^\vee.$
$\lambda$ and $\nu^\vee$ have to satisfy
$\sum_{i=1}^l l_i m_i \le p-1,
\ \sum_{i=1}^l l_i^\vee m_i^\vee \le q-1$
where $m_i$ are the normalized components of the highest root $\psi$ in
the directions of the simple roots $\alpha_i$, i.e.\
${{\psi}\over{\psi^2}} = \sum_{i=1}^l m_i {{\alpha_i}\over{\alpha_i^2}}$.
$m_i^\vee$ is given by $m_i^\vee  = {2\over{\alpha_i^2}} m_i$.
Note that the set of conformal
dimensions given by this condition has a symmetry so that all conformal
dimensions of the minimal model occur with the same multiplicity in it
(in the nonsimply laced cases the multiplicity is just 2). For more details
see \cite{FKW,B,GO,FL}.
\mn
\centerline{
\vbox{
\hbox{Table 7.4: Values of $m_i,m_i^\vee$ for all simple
      Lie algebras \cite{GO}.}
\hbox{\vrule \hskip 1pt\vbox{\offinterlineskip
\def\tablespace{height2pt&\omit&&\omit&&\omit&\cr}
\def\tablerule{\tablespace\noalign{\hrule}\tablespace}
\hrule\halign{&\vrule#&\strut\quad\hfil#\hfil\quad\cr
\tablespace\tablespace
& {\it Lie algebra}  && $(m_i)$       && $(m_i^\vee)$ &\cr
\tablerule
& ${\Cal A}_l$ && $(1,\dots,1)$       && $(1,\dots,1)$ &\cr
\tablespace
& ${\Cal B}_l$ && $(1,2,\dots,2,1)$   && $(1,2,\dots,2)$ &\cr
\tablespace
& ${\Cal C}_l$ && $(1,\dots,1)$       && $(2,\dots,2,1)$ &\cr
\tablespace
& ${\Cal D}_l$ && $(1,2,\dots,2,1,1)$ && $(1,2,\dots,2,1,1)$ &\cr
\tablespace
& ${\Cal E}_6$ && $(1,2,2,3,2,1)$     && $(1,2,2,3,2,1)$ &\cr
\tablespace
& ${\Cal E}_7$ && $(2,2,3,4,3,2,1)$   && $(2,2,3,4,3,2,1)$ &\cr
\tablespace
& ${\Cal E}_8$ && $(2,3,4,6,5,4,3,2)$ && $(2,3,4,6,5,4,3,2)$ &\cr
\tablespace
& ${\Cal F}_4$ && $(1,2,3,2)$         && $(2,4,3,2)$ &\cr
\tablespace
& ${\Cal G}_2$ && $(2,1)$             && $(2,3)$ &\cr
\tablespace}\hrule}\hskip 1pt \vrule}
}}
\vfill\eject


\head
\endhead


\Refs
\refstyle{A}

\ref\key
\by W.\ Eholzer
\paper Fusion Algebras Induced by Representations of the
       Modular Group
\jour  Int. J. Mod. Phys. {\bf A}
\vol 8 \yr 1993 \pages 3495-3507 (see \S3)
\endref

\ref\key
\by W.\ Eholzer
\paper On the Classification of Modular Fusion Algebras
\jour  preprint BONN-TH-94-18, MPI-94-91, Commun. Math. Phys.
(to appear) (see \S2.2, \S2.3, \S3, \S7.1-3)
\endref

\ref \key
\by W.\ Eholzer, N.\ -P.\ Skoruppa
\paper Modular Invariance and Uniqueness of Conformal Characters
\jour preprint BONN-TH-94-16, MPI-94-67, Commun. Math. Phys. (to appear)
 (see \S2.1, \S4)
\endref

\ref\key
\by W.\ Eholzer, N.-P.\ Skoruppa
\paper  Conformal Characters and Theta Series
\jour preprint MSRI No. 012-95, BONN-TH-94-24, Lett. Math. Phys. (to appear)
(see \S5)
\endref

\ref\key
\by  R. Blumenhagen, W. Eholzer, A. Honecker, K. Hornfeck,
     R. H{\"u}bel
\paper Coset Realization of Unifying $\w$-Algebras
\jour preprint BONN-TH-94-11, DFTT-25/94, Int. Jour. Mod. Phys. A
      (to appear)  (see \S7.4)
\endref

\endRefs

\Refs
\refstyle{A}
\widestnumber\key{xxxxxxxxx}


\ref \key AM
\by G.\ Anderson, G.\ Moore
\paper  Rationality in Conformal Field Theory
\jour Commun. Math. Phys.
\vol 117  \yr 1988   \pages  441-450
\endref

\ref \key B
\by  P.~Bouwknegt
\paper Extended Conformal Algebras from Kac-Moody Algebras
\jour Proceedings of the meeting `Infinite dimensional Lie algebras and
Groups',
      CIRM, Luminy, Marseille
\yr 1988 \pages  527-554
\endref

\ref\key BEH$^3$
\by  R. Blumenhagen, W. Eholzer, A. Honecker, K. Hornfeck,
     R. H{\"u}bel
\paper Coset Realization of Unifying $\w$-Algebras
\jour preprint BONN-TH-94-11, DFTT-25/94, hep-th/9406203,
      Int. Jour. Mod. Phys. A
\toappear
\endref

\ref \key BFKNRV
\by R.\ Blumenhagen, M.\ Flohr, A.\ Kliem,
    W.\ Nahm, A.\ Recknagel, R.\ Varnhagen
\paper $\w$-Algebras with Two and Three Generators
\jour    Nucl. Phys. B
\vol 361  \yr 1991  \pages  255-289
\endref

\ref\key BPZ
\by A.A.\ Belavin, A.M.\ Polyakov, A.B.\ Zamolodchikov
\paper Infinite Conformal Symmetry in Two-Dimensional
       Quantum Field Theory
\jour Nucl. Phys. B
\vol 241 \yr 1984 \pages 333-380
\endref

\ref\key BS
\by P.\ Bouwknegt, K.\ Schoutens
\paper $\w$-Symmetry in Conformal Field Theory
\jour Phys. Rep.
\vol  223 \yr 1993 \pages 183-276
\endref


\ref\key C
\by J.L.\ Cardy
\paper Operator Content of Two-Dimensional
       Conformally Invariant Theories
\jour Nucl. Phys. B
\vol 270 \yr 1986 \pages 186-204
\endref

\ref\key CIZ
\by A.\ Cappelli, C.\ Itzykson, J.B.\ Zuber
\paper The A-D-E Classification of Minimal and $A_1^{(1)}$
       Conformal Invariant Theories
\jour  Commun. Math. Phys.
\vol 113 \yr 1987 \pages 1-26
\endref

\ref\key CPR
\by M.\ Caselle, G.\ Ponzano, F.\ Ravanini
\paper Towards a Classification of Fusion Rule Algebras in Rational
       Conformal Field Theories
\jour Int. J. Mod. Phys. B
\vol 6 \yr 1992 \pages 2075-2090
\endref

\ref \key De
\by P.\ Degiovanni
\paper $\Z/N\Z$ Conformal Field Theories
\jour Commun. Math. Phys. \vol 127 \yr 1990 \pages 71-99
\endref

\ref\key Do
\by L.\ Dornhoff
\book Group Representation Theory
\publ Marcel Dekker Inc.
\publaddr New York
\yr 1971
\endref

\ref\key E$1$
\by W.\ Eholzer
\paper Exzeptionelle und Supersymmetrische $\w$-Algebren
       in Konformer Quantenfeldtheorie
\jour Diplomarbeit BONN-IR-92-10
\endref

\ref\key E$2$
\by W.\ Eholzer
\paper Fusion Algebras Induced by Representations of the
       Modular Group
\jour  Int. J. Mod. Phys. A
\vol 8 \yr 1993 \pages 3495-3507
\endref


\ref\key E$3$
\by W.\ Eholzer
\paper On the Classification of Modular Fusion Algebras
\jour  preprint BONN-TH-94-18, MPI/94-91,hep-th/9408160,
       Commun. Math. Phys.
\toappear
\endref

\ref\key EFH$^2$NV
\by W.\ Eholzer, M.\ Flohr, A.\ Honecker, R.\ H{\"u}bel, W.\ Nahm,
    R.\ Varnhagen
\paper Representations of $\w$-Algebras with Two Generators
       and New Rational Models
\jour Nucl. Phys. B
\vol 383 \yr 1992 \pages 249-288
\endref

\ref \key ES$1$
\by W.\ Eholzer, N.\ -P.\ Skoruppa
\paper Modular Invariance and Uniqueness of Conformal Characters
\jour preprint BONN-TH-94-16, MPI/94-67, hep-th/9407074, Commun. Math. Phys.
\toappear
\endref

\ref\key ES$2$
\by W.\ Eholzer, N.-P.\ Skoruppa
\paper  Conformal Characters and Theta Series
\jour preprint MSRI No. 012-95, BONN-TH-94-24, hep-th/9410077, Lett. Math.
Phys.
\toappear
\endref

\ref \key F
\by J.\ Fuchs
\paper Fusion Rules in Conformal Field Theory
\jour Fortschr. Phys.
\vol 42 \yr 1994 \pages  1-48
\endref

\ref \key FHL
\by I.B.\ Frenkel, Y.\ Huang, J.\ Lepowsky
\book  On Axiomatic Approaches to Vertex Operator Algebras and Modules
\bookinfo Memoirs of the American Mathematical Society, Volume 104, Number 494
\publ American Mathematical Society
\publaddr Providence, Rhode Island
\yr 1993
\endref

\ref\key FKW
\by E.\ Frenkel, V.\ Kac, M.\ Wakimoto
\paper Characters and Fusion Rules for $\w$-Algebras
       via Quantized Drinfeld-Sokolov Reduction
\jour Commun. Math. Phys.
\vol 147 \yr 1992  \pages 295-328
\endref

\ref \key FL
\by  V.A.\ Fateev, S.L.\ Luk'yanov
\paper Additional Symmetries and Exactly-Soluble Models in Two-Dimensional
       Conformal Field Theory
\jour  Sov. Sci. Rev. A. Phys. {\bf 15/2} \yr 1990
\endref

\ref\key FRT
\by L.\ F\'eher, L.\ O'Raifeartaigh, I.\ Tsutsui
\paper The Vacuum Preserving Lie Algebra of a Classical $\w$-algebra
\jour Phys. Lett. B
\vol 316 \yr 1993 \pages 275-281
\endref

\ref \key FZ
\by I.B.\ Frenkel, Y.\ Zhu
\paper Vertex Operator Algebras Associated to Representations
       of Affine and Virasoro Algebras
\jour Duke Math. J.
\vol 66(1) \yr 1992  \pages 123-168
\endref

\ref \key G
\by R.\ C.\ Gunnings
\book  Lectures on Modular Forms
\publ Princeton University Press
\publaddr Princeton, New Yersey
\yr 1962
\endref

\ref \key Gi
\by P.\ Ginsparg
\paper Applied Conformal Field Theory
\jour proceedings of the `Les Houches Summer School 1988'
\yr 1988 \pages 1-168
\endref

\ref \key GO
\by P.\ Goddard, D.\ Olive
\paper Kac-Moody Algebras and Virasoro Algebras in Relation to Quantum Physics
\jour Int. Jour. Mod. Phys. A
\vol 1 \yr 1986  \pages 303-414
\endref

\ref \key GP
\by C.\ Batut, D.\ Bernardi, H.\ Cohen, M.\ Olivier
\paper PARI-GP
\publ Universit\'e Bordeaux 1
\publaddr Bordeaux
\yr 1989
\endref

\ref \key GSW
\by M.\ Green, J.H.\ Schwarz, E.\ Witten
\book Superstring Theory I, II
\publ Cambridge University Press
\publaddr Cambridge
\yr 1987
\endref

\ref\key Ha
\by R.\ Haag
\book Local Quantum Physics
\publ Springer
\publaddr Berlin - Heidelberg - New York
\yr 1992
\endref

\ref\key HL
\by   Y.-Z.\ Huang, J.\ Lepowsky
\paper A Theory of Tensor Products for Module Categories for a
       Vertex Operator Algebra I, II
\jour preprints, hep-th/9309076, hep-th/9309159
\endref

\ref
\by  \phantom{xxxxxxxxxx} Y.-Z.\ Huang, J.\ Lepowsky
\paper Tensor Products of Modules for a Vertex Operator Algebra and
       Vertex Tensor Categories
\jour preprint hep-th/9401119
\endref

\ref\key Hu
\by   Y.-Z.\ Huang
\paper A Theory of Tensor Products for Module Categories for a
       Vertex Operator Algebra IV
\jour private communication
\endref

\ref\key Kac
\by V.\ Kac
\book Infinite Dimensional Lie Algebras and Groups
\publ World Sientific
\publaddr Singapore
\yr 1989
\endref

\ref \key Ki
\by E.\ B.\ Kiritsis
\paper Fuchsian Differential Equations for Characters on the Torus:
       A Classification
\jour Nucl. Phys. B
\vol 324  \yr 1989  \pages 475-494
\endref

\ref\key KRV
\by J.\ Kellendonk, M.\ R\"osgen, R.\ Varnhagen
\paper Path Spaces and $\w$-Fusion in Minimal Models
\jour   Int. Jour. Mod. Phys. A
\vol 9 \yr 1994 \pages 1009-1023
\endref

\ref\key KW
\by H.G.\ Kausch, G.M.T.\ Watts
\paper A Study of $\w$-Algebras Using Jacobi Identities
\jour Nucl. Phys. B
\vol 354  \yr 1991  \pages 740-768
\endref

\ref \key L
\by H.\ Li
\paper Symmetric Invariant Bilinear Forms on Vertex Operator Algebras
\jour preprint Rutgers University
\endref

\ref \key M
\by G.\ Moore
\paper Atkin-Lehner Symmetry
\jour Nucl. Phys. B
\vol 293  \yr 1987 \pages 139-188
\endref

\ref\key MH
\by J\. Milnor and D\. Husemoller
\book Symmetric bilinear forms
\publ Springer
\publaddr Berlin - Heidelberg - New York
\yr 1973
\endref

\ref\key MMS
\by S.\ Mathur, S.\ Mukhi, A.\ Sen
\paper On the Classification of Rational Conformal Field Theories
\jour Phys. Lett. B
\vol 213 \yr 1988 \pages 303-308
\endref

\ref\key Na
\by W.\ Nahm
\book Chiral Algebras of Two-Dimensional Chiral
       Field Theories and Their Normal Ordered Products
\bookinfo proceedings of the Trieste Conference on
          ``Recent Developments in Conformal Field Theories''
\publ World Scientific
\publaddr Singapore
\yr 1989
\endref

\ref\key NRT
\by W.\ Nahm, A.\ Recknagel, M.\ Terhoeven
\paper Dilogarithm Identities in Conformal Field Theory
\jour  Mod. Phys. Lett. A
\vol 8 \yr 1993 \pages 1835-1847
\endref

\ref \key NW
\by A.\ Nobs
\paper Die irreduziblen Darstellungen der Gruppen $SL_2(Z_p)$
       insbesondere $\qquad$ $SL_2(Z_2)$  I
\jour Comment. Math. Helvetici
\vol 51 \yr 1976 \pages  465-489
\endref

\ref
\by \phantom{xxxxxxxxxx} A.\ Nobs, J.\ Wolfart
\paper Die irreduziblen Darstellungen der Gruppen $SL_2(Z_p)$
       insbesondere $SL_2(Z_2)$  II
\jour Comment. Math. Helvetici
\vol 51 \yr 1976 \pages  491-526
\endref

\ref\key RC
\by A.\ Rocha-Caridi
\book Vacuum Vector Representations of the Virasoro Algebra
\bookinfo in 'Vertex Operatos in Mathematics and Physics',
          S. Mandelstam and I.M. Singer
\publ MSRI Publications Nr. 3, Springer
\publaddr Berlin - Heidelberg - New York
\yr 1984
\endref

\ref\key Sche
\by A.\ N.\  Schellekens
\paper Meromorphic $c=24$ Conformal Field Theories
\jour Commun. Math. Phys.
\vol 153 \yr 1993 \pages 159-185
\endref

\ref\key Scho
\by B\. Schoeneberg
\book Elliptic Modular Functions
\bookinfo Die Grundlehren der mathematischen Wissenschaften
in Einzeldarstellungen, Bd 204
\publ Springer
\publaddr Berlin - Heidelberg - New York
\yr 1974
\endref

\ref\key Sh
\by G.\ Shimura
\book Introduction to the Arithmetic Theory of Automorphic Functions
\publ $\qquad$ Iwanami Sholten, Princeton Press
\publaddr Tokyo, Japan
\yr 1971
\endref

\ref\key Sk
\by N.-P.\ Skoruppa
\paper \"Uber den Zusammenhang zwischen Jacobiformen
       und Modulformen halbganzen Gewichts
\jour Bonner Mathematische Schriften
\vol 159 \yr 1985
\endref

\ref\key Va
\by C.\ Vafa
\paper Toward Classification of Conformal Theories
\jour Phys. Lett. B
\vol 206 \yr 1988  \pages 421-426
\endref

\ref\key Ve
\by E.\ Verlinde
\paper Fusion Rules and Modular Transformations in 2d Conformal Field Theory
\jour Nucl. Phys. B
\vol 300 \yr 1988  \pages 360-376
\endref

\ref\key Vi
\by M-F\. Vign\,eras
\book Arithm\'etique des alg\`ebres de quaternions (Lecture Notes in
Mathematics 800)
\publ Springer
\publaddr Berlin - Heidelberg - New York
\yr 1980
\endref

\ref \key Wa
\by W.\ Wang
\paper  Rationality of Virasoro Vertex Operator Algebras
\jour Int. Research Notices (in Duke Math. J.)
\vol 7 \yr 1993   \pages  197-211
\endref

\ref\key Wi
\by E.\ Witten
\paper  Quantum Field Theory and the Jones Polynomial
\jour Commun. Math. Phys.
\vol 121 \yr 1989 \pages 351-399
\endref

\ref\key Wo
\by K.\ Wohlfahrt
\paper An extension of F.\ Klein's level concept
\jour Illinois J. Math.
\vol 8 \yr 1964 \pages 529-535
\endref

\ref \key Za$1$
\by D.\ Zagier
\paper private communication
\endref

\ref \key Za$2$
\by D.\ Zagier
\paper Modular Forms and Differential Operators
\jour preprint MPI/94-13
\endref

\ref \key Zh
\by Y.\ Zhu
\paper  Vertex Operator Algebras, Elliptic Functions, and Modular Forms
\jour Ph.D. thesis, Yale University
\yr  1990
\endref

\endRefs
\enddocument